\documentclass[rmp,aps,nofootinbib,preprint,endfloats*]{revtex4}

\usepackage[]{graphics}
\usepackage{epsfig}
\usepackage{amsmath}
\usepackage{longtable}

\begin{document}
\title{Inelastic X-ray Scattering by Electronic Excitations under High Pressure}
\author{Jean-Pascal Rueff}
\email{jean-pascal.rueff@synchrotron-soleil.fr}
\affiliation{Synchrotron SOLEIL, L'Orme des Merisiers, BP 48, Saint Aubin, 91192 Gif sur Yvette, France}
\affiliation{Universit\'{e} Pierre et Marie Curie - Paris 6, CNRS-UMR7614, Laboratoire de Chimie Physique - Mati\`{e}re et Rayonnement, 11 Rue Pierre et Marie Curie, 75005 Paris, France}
\author{Abhay Shukla}
\affiliation{Universit\'{e} Pierre et Marie Curie - Paris 6, CNRS-UMR7590, Institut de Min\'{e}ralogie et de Physique des Milieux condens\'{e}s, 140 rue de Lourmel, 75015 Paris, France}

\begin{abstract}  
Investigating electronic structure and excitations under extreme conditions gives access to a rich variety of phenomena. High pressure typically induces behavior such as magnetic collapse and the insulator-metal transition in $3d$ transition metals compounds, valence fluctuations or Kondo-like characteristics in $f$-electron systems, and coordination and bonding changes in molecular solids and glasses.
This article reviews research concerning electronic excitations in materials under extreme conditions using inelastic x-ray scattering (IXS). IXS is a spectroscopic probe of choice for this study because of its chemical and orbital selectivity and the richness of information it provides. Being an all-photon technique, IXS has a penetration depth compatible with high pressure requirements. Electronic transitions under pressure in $3d$ transition metals compounds and $f$-electron systems, most of them strongly correlated, are reviewed. Implications for geophysics are mentioned. Since the incident X-ray energy can easily be tuned to absorption edges, resonant IXS, often employed, is discussed at length. Finally studies involving local structure changes and electronic transitions under pressure in materials containing light elements are briefly reviewed.     
\end{abstract}                                                                 

\date{\today}
\maketitle
\tableofcontents

\section{Introduction}
\subsection{Why this work}
Pressure is an effective means to alter electronic density, and thereby electronic structure, hybridization and magnetic properties. Applying pressure therefore can lead to phenomena of importance from the physical point of view such as magnetic collapse, metal-insulator transitions (MIT), valence changes, or the emergence of superconducting phases.

Probing the electronic properties of materials under high pressure conditions, however, remains a formidable task, the  sample environment preventing easy access to the embedded material. With the exception of optical absorption which provides information about low energy excitations, the experimental difficulties mean that high-pressure studies have been mostly restricted so far to structural refinement, to study of the Raman modes or to the characterization of transport properties.

The availability of extremely intense and focused x-ray sources through the latest generation of synchrotrons has opened new perspectives for spectroscopic studies at high pressure. Though standard spectroscopic techniques such as X-ray absorption have been in use in high-pressure studies for quite some time, newer methods like nuclear forward scattering, the synchrotron-based equivalent of M\"{o}ssbauer spectroscopy are fast becoming a choice tool to investigate magnetism in selected elements.  
On the other hand, inelastic x-ray scattering with hard x-rays, used in tandem with high pressure, is a powerful spectroscopic tool for a variety of physical and chemical applications. It is an all-photon technique fully compatible with high-pressure environments and applicable to a vast range of materials. In the resonant regime, it ensures that the electronic properties of the element under scrutiny are selectively observed. Standard focalization of x-rays below of 100 microns and micro-focusing to a few microns ensures that small sample size in a pressure cell is not a problem. This also corresponds approximately to the scattering volume in the hard x-ray region. Given these conditions, we can expect maximum throughput with IXS-derived techniques.

Though IXS techniques have been used for some time now, the combination of these with high pressure has opened a new line of research which is now rapidly reaching maturity. Our aim in this manuscript is to provide an overview of this field in two classes of materials which have been at the heart of research efforts in ``condensed matter'' physics: strongly correlated transition metal oxides and rare-earth compounds. These materials are not yet well understood from a fundamental point of view but are also found in many technologically advanced products, such as in recording media based on GMR (Giant Magneto Resistance) materials, spintronics or magnetic structures.  In the introductory materials to the relevant sections, we restrict ourselves to useful concepts for understanding the nature of $d$ and $f$ electronic states, and more specifically their behavior under high pressure. An extensive theoretical description of these is beyond the scope of this work and in particular, magnetic structure and interactions are not discussed, except in close connection with the electronic properties. Because it is a method with which the reader might not be very familiar, we will start off by discussing theoretical and experimental basics of inelastic x-ray scattering in some detail. The following sections are devoted to an extensive review of experimental results under high pressure with a main focus on magnetic transitions in transition metals in combination with metal insulator transition, electron delocalization in mixed valent materials and finally bonding changes in light elements that uncovers a marginal, yet unique aspect of IXS. It will be followed by conclusions and perspectives.

\subsection{Historical context}
Before discussing IXS as a probe of the electronic properties of materials under pressure it can be useful to place this new spectroscopy in the wider historical context of research carried out over the past thirty years in high pressure physics. These studies, though also focusing on electronic transitions and in particular on the metal insulator transition, valence changes and magnetic collapse, used very different and complementary techniques. 

Resistivity and optical spectroscopy under pressure were initiated soon after the development of pressure cells, starting from the earlier pressure apparatus of Bridgman and Drickamer and later with diamond anvil cells  (see ~\textcite{Jayaraman1983} for a complete review). Both techniques can probe pressure-induced metal insulator transitions, and have been extensively applied to elemental systems, semi conductors, wide gap insulators (see e.g. \textcite{Syassen1985,Syassen1987,Chen1993}) and correlated systems~\cite{Tokura1992}. 

Magnetometric measurements are more difficult under pressure because of the weakness of the magnetic signal coming from the sample. But several groups have reported successful experiments in specially designed pressure cells. Magnetic susceptibility is particularly efficient in detecting superconducting phases under pressure such as in Li and S (cf.~\textcite{Struzhkin2004} for a recent review). No spin state transition has been reported so far with this technique. In contrast, M\"{o}ssbauer spectroscopy is a widespread method of investigation of the magnetic state of transition metals and rare earths under high pressure.  The measurements require isotope substitution which sets some constraints on the possible range of detected elements. But M\"{o}ssbauer research has been very active in high pressure physics owing largely to its high sensitivity to Fe magnetism. Magnetic transitions have been observed in elemental Fe and several compounds and minerals up to the megabar pressure range \cite{Pipkorn1964,Abd-Elmeguid1988,Taylor1991, Pasternak1997, Speziale2005}. The more recent development of synchrotron-based nuclear forward scattering has augmented the M\"{o}ssbauer capacities to smaller or more diluted samples coupled to the laser heating technique~\cite{Lin2006}. To complete this brief overview of pressure compatible magnetometric probes, one should mention x-ray magnetic circular dichroism (XMCD) and neutron magnetic scattering (for a more extensive comparison, cf.~\textcite{Astuto2006}), which both are well established magnetic probes. Neutron scattering is usually restricted to moderate pressures as the large beam eventually limits the sample dimensions and therefore the maximum attainable pressure. It however allows a full determination of the magnetic structure, as recently shown up to 20 GPa~\cite{Goncharenko1998}. XMCD benefits on the other hand from the x-ray brilliance and chemical selectivity just as IXS, while the polarization of the light provides the magnetic sensitivity. Following the seminal work of \textcite{Odin1998}, a handful of XMCD experiments have been performed at the K-edges of transition metals under pressure up to the megabar range~\cite{Iota2007}. We will refer to  some of these results while discussing the spin state transitions of $3d$ metals.

Finally, the sensitivity of x-ray absorption spectroscopy to the valence state has been long used for studying materials under high pressure. Although, XAS is closely related to IXS and will be discussed later, it is worth mentioning the pioneering work of \textcite{Syassen1982} and \textcite{Roehler1988} on the valence change of rare earth systems under pressure. On the contrary measuring the K-edges of the light elements under high pressure conditions is a unique and recent possibility thanks to IXS.

\subsection{Energy scales}
Exploring the phase (structural, magnetic or electronic) diagram of materials requires tuning key external parameters. Among them temperature and pressure are equally important to explore the free energy landscape of the system. A temperature ($T$) induced phase transition is driven by entropy. More simply, the temperature effects in terms of energy scale can be expressed by considering electronic excitations from the ground state via the Boltzmann constant and the approximate relationship 1000~K $\cong$ 86.17 meV. 

On the other hand, pressure-energy conversion can be obtained through the Gibbs free energy for a closed system, defined as  $dG=-SdT+\mathcal{V}dP$. At constant temperature, the expression of the total energy change (for a given pressure variation $\Delta P$) reduces to a simple integration of the $\mathcal{V}dP$ term. Although solids are not easily compressible, the volume variation at very high pressure regime, as envisaged in this study, is far from being negligible. Using the compressibility $\kappa=-1/\mathcal{V}\left(\partial \mathcal{V}/\partial P\right)_T$, one can estimate the energy variation from Eq.~\ref{Gibbs}.
\begin{equation}
\Delta G=\frac{\mathcal{V}_0}{\kappa}\left(1-e^{-\kappa\Delta P}\right)
\label{Gibbs}
\end{equation}
with $\mathcal{V}_0$, the molar volume at ambient pressure. At low pressure, this expression can be approximated by $\Delta G\sim \mathcal{V}_0\Delta P$, which can be derived directly from the Gibbs free energy supposing $\mathcal{V}$ independent of $P$. Let us estimate the internal energy change in a system for a $\Delta P$ of 100 GPa ($\equiv$ 1 Mbar) in the two classes of materials of main interest here: transition metals and rare earths. Transition metals are poorly compressible metals. Their isothermal bulk modulus ($K_T=1/\kappa$) falls within the megabar range. Application of Eq.~\ref{Gibbs} to Fe ($K_T=170$~GPa) yields a variation $\Delta G\sim 5.3$ eV for the considered $\Delta P$. Rare earths have lower $K_T$ values and in Ce for instance ($K_T=22$~GPa) this implies a somewhat smaller $\Delta G\sim 4.6$ eV for $\Delta P$=100 GPa with respect to transition metals. 

Independently of the materials under consideration, $T$ and $P$ variations map onto totally different energy scales in the free energy landscape of the system. Temperatures of several thousand Kelvin can be achieved with resistive or laser-assisted setups but this still corresponds to a modest amount on an energy scale. At least one order of magnitude in energy can be gained by using pressure as an external parameter if we consider that megabar pressure can be achieved. Pressure induced phase transitions may also lead to new types of ordering, since entropy is not involved. The existence of a quantum critical point (QCP) in strongly correlated materials is such a manifestation of a new state of matter. 

\section{Basics of inelastic x-ray scattering}
\subsection{A question of terminology}
Appropriate naming of new spectroscopic techniques is always useful but rarely easy and like many recent techniques the terminology for inelastic x-ray scattering (IXS) has gone through a maze of mutations.  

\textcite{Sparks1974} first showed the ``inelastic resonance emission of x rays'' using a laboratory x-ray source. The new experimental finding, here correctly designated as an emission process in the resonant conditions, differs from early results obtained in the Compton regime for which the photon energy is chosen far from any resonances, and at high momentum transfer. The perfect suitability of synchrotron radiation for inelastic x-ray scattering was demonstrated a few years later by~\textcite{Eisenberger1975a} who first performed ``resonant x-ray Raman scattering'' at the Cu K-edge, and simultaneously adopted Raman terminology for an x-ray based process. Though historically justified, this widespread terminology is somewhat confusing. 
In this work, we will limit ourselves to the use of \emph{resonant inelastic x-ray scattering} (RIXS). An exception will be made for \emph{resonant x-ray emission spectroscopy} (RXES) or \emph{x-ray emission spectroscopy} (XES) as a sub-category of RIXS, when it clearly applies.

Non-resonant IXS (nrIXS)\footnote{The acronym for non-resonant inelastic scattering (nrIXS) should not be confused with that of nuclear resonant inelastic x-ray scattering} is historically the older technique. Non-resonant experiments of DuMond and coworkers on ``x-ray Compton scattering'' precede resonant measurements by several decades. This was followed by pioneer work of M. Cooper and W. Sch\"{u}lke with x-ray rotating anodes (cf.\ Refs.\ in \textcite{Schulke1991}) and of G. Loupias with synchrotron light \cite{Loupias1980}. \textcite{Susuki1967} later measured the K-edge of Be using ``X-ray Raman scattering'' (XRS) by extending the energy loss region away from the Compton region. This terminology is still in use to distinguish the measurements of the absorption edges of light elements in the x-ray scattering mode from that of other types of non-resonant scattering events, such as scattering from phonons. In our manuscript we refer to \emph{inelastic x-ray scattering} (IXS) as the general scattering process from which both RIXS and nrIXS originate.

\subsection{IXS cross section}
\begin{figure}[htbp]
\centering
\includegraphics[width=0.90\linewidth]{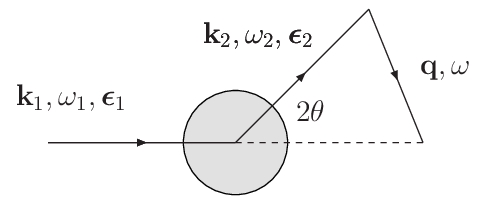}
\caption[Scattering process]{Scattering process of a photon by an electron system (gray area).}
\label{fig:scattering}
\end{figure}

The general inelastic x-ray scattering process is illustrated in Fig.~\ref{fig:scattering}. An incident photon defined by its wave vector, energy and polarization ($\mathbf{k}_1$, $\hbar\omega_1$, $\boldsymbol{\epsilon}_1$) is scattered by the system through an angle $2\theta$, the scattered photon being characterized by $\mathbf{k}_2$, $\hbar\omega_2$, $\boldsymbol{\epsilon}_2$. $\mathbf{q}=\mathbf{k}_1-\mathbf{k}_2$ and $\hbar\omega=\hbar\omega_1-\hbar\omega_2$ define the momentum and energy respectively transferred during the scattering process. For x-rays, $k_1\approx k_2$, so that 
\begin{equation}
q\approx 2k_1\sin(\theta).
\end{equation}
The momentum transfer depends only on the scattering angle and incoming wavelength.

The starting point for describing the scattering process theoretically is the photon-electron interaction Hamiltonian $\mathcal{H}$. For perturbation treatment, $\mathcal{H}$ is conventionally separated into a term $\mathcal{H}_i$ describing the interaction between the electrons and the incident electromagnetic field and a term $\mathcal{H}_0$ corresponding to the non-interacting electron system:
\begin{equation}
    \mathcal{H}=\mathcal{H}_0+\mathcal{H}_i. 
\end{equation}
The non-interacting term reads
\begin{equation}
    \mathcal{H}_0=\sum_j\frac{1}{2m}\mathbf{p}^2_j+\sum_{jj'}V(r_{jj'}),
\end{equation}
and
\begin{equation}
\mathcal{H}_i=
\sum_j\frac{e^2}{2mc^2}\mathbf{A}^2(\mathbf{r}_j)
-\frac{e}{mc}\mathbf{A}(\mathbf{r}_j)\cdot\mathbf{p}_j,
\label{eq:hamiltonian_int}
\end{equation}
$\mathbf{A}$  and $V$ are the vector and scalar potentials of the interacting electromagnetic field and the electrons are defined by their momentum $\mathbf{p}$ and position $\mathbf{r}$. The sum is over all the electrons of the scattering system. We use the Coulomb gauge ($\nabla\mathbf{A}=0$). The spin-dependent terms in $\mathcal{H}_i$  are smaller by a factor $\hbar/mc^2$ and are not considered in this study. 

\begin{figure}[htbp]
\centering
\includegraphics[width=0.90\linewidth]{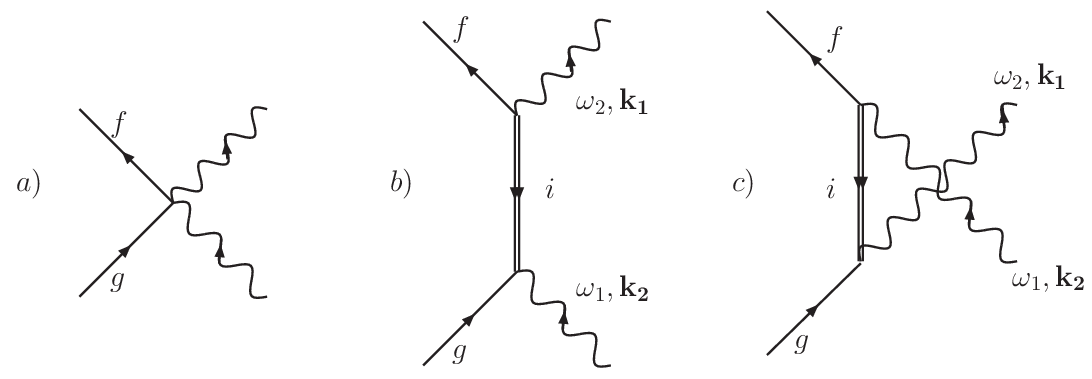}
\caption[Scattering diagrams]{Terms in the IXS cross section : a) non-resonant, b) and c) resonant scattering. Wavy (straight) lines represent the photon (electron) wave functions; double lines are inner-shell core-hole.}
\label{fig:diag_1}
\end{figure}

The double differential scattering cross section can derived from the interaction Hamiltonian using the Fermi Golden rule in the sudden approximation. For a second order process, this is known as the Kramers-Heisenberg formula~\cite{Kramers1925}. It consists of the sum of three terms, represented as Feynman diagrams in Fig~\ref{fig:diag_1}, which we now discuss in some more detail. 

\subsubsection{Non-resonant scattering}
\label{sec:nrIXS}
The first term (Fig.~\ref{fig:diag_1}(a)) arises from the $\mathbf{A}^2$ term in the interaction Hamiltonian~(\ref{eq:hamiltonian_int}), in first order of perturbation, which dominates far from any resonances. The non-resonant scattering cross section depends on the dynamical structure factor $S(\mathbf{q},\omega)$. Using the notation of Fig.~\ref{fig:scattering}, the non-resonant scattering cross section reads: 
\begin{equation}
\frac{d^2\sigma}{d\Omega_2{}d\omega_2}=r_0^2
\left(
\frac{\omega_2}{\omega_1}
\right)
|\boldsymbol\epsilon_1\cdot\boldsymbol\epsilon_2^*|^2 
S(\mathbf{q},\omega),
\label{eq:nrIXS}
\end{equation} 
where $r_0$ is the classical electron radius, $r_0=e^2/mc^2$. The pre-factor in expression~(\ref{eq:nrIXS}) represents the Thomson scattering by free electrons. 
\begin{equation}
\left(\frac{d\sigma}{d\Omega_2}\right)_{Th}=r_0^2\left(\frac{\omega_2}{\omega_1}\right)|
\boldsymbol\epsilon_1\cdot\boldsymbol\epsilon_2^*|^2,
\label{eq:thomson}
\end{equation} 

The dynamical structure factor 
\begin{equation}
S(\mathbf{q},\omega)=\sum_f|\langle f|\sum_j e^{i\mathbf{q}\cdot\mathbf{r_j}}|g\rangle|^2\delta(E_g-E_f+\hbar\omega).
\label{eq:dyn_str_factor}
\end{equation}
contains the main information on the system. It relates the non-resonant scattering process to the excitations of the electron system allowed by energy and momentum conservation. Following Van~Hove~\cite{VanHove1954}, $S(\mathbf{q},\omega)$ can be further written as the Fourier transform of the electron pair-correlation function: 
\begin{equation}
S(\mathbf{q},\omega)=
\frac{1}{2\pi}\int_{-\infty}^{\infty}dt\,e^{-i\omega t}\langle g|\sum_{jj'}e^{-i\mathbf{q}\cdot\mathbf{r}_{j'}(t)}e^{i\mathbf{q}
\cdot\mathbf{r}_{j}(0)}|g\rangle
\label{eq:VH}
\end{equation}
where $\left|g\right\rangle$ is the ground state and the sum is carried over the positions $(\mathbf{r}_j,\mathbf{r}_{j'})$ of the electron pairs. The two notations~(\ref{eq:dyn_str_factor}) and (\ref{eq:VH}) of the dynamical structure factor reflect the fluctuation-dissipation theorem: In the non-resonant regime, the system excitations (dissipation) are connected to the scattering due to density fluctuation in the ground state, i.e. in absence of perturbation 
Depending on how $q$ compares with the characteristic length scale of the system in the probed energy transfer regime, $\lambda_c$, equation~(\ref{eq:VH}) describes phenomena ranging from dynamics of collective modes ($q\lambda_c\ll 1$) to single particle excitations ($q\lambda_c\gg 1$). 

In the case of a homogeneous electron system, $S(\mathbf{q},\omega)$ can be related to the dielectric function $\varepsilon$ through 
\begin{equation}
S({\mathbf q},\omega)=
(1+\eta_B)\frac{q^2}{4\pi e^2}\,\mbox{Im}
\left[\frac{-1}{\varepsilon({\bf q},\omega)}\right],
\end{equation}
where $\eta_B=1/[\exp(\hbar\omega/k_BT)-1]$ is the Bose factor. This equation is similar to the electron energy loss spectroscopy (EELS) cross section when the pre-factor is replaced with an appropriate cross section for electron-electron scattering. nrIXS can probe a wide domain in the ($\mathbf{q}$,$\omega$) phase space because of the high photon energy, and the absence of kinematic limitations which allows $\mathbf{q}$ to vary independently of $\omega$. In this respect it is different from neutron scattering and also very complementary to EELS from the experimental point of view. The energy transfer is limited by the best achievable energy resolution. 

\subsubsection{X-ray Raman scattering}
\label{sec:XRS}
We have considered so far excitations of the valence electrons. A particular case of nrIXS, x-ray Raman scattering (XRS) is the excitation of core electrons into unoccupied states. As we will see in section~\ref{sec:light_element}, this technique is relevant primarily to light elements whose binding energy falls in the soft x-ray region.

Substituting in Eq.~(\ref{eq:nrIXS}) the dynamical structure factor $S({\mathbf q},\omega)$ by its expression~(\ref{eq:dyn_str_factor}) and using Eq.~(\ref{eq:thomson}), the non-resonant scattering reads : 
\begin{eqnarray}
\frac{d^2\sigma}{d\Omega_2d\omega_2}&=&
\left(\frac{d\sigma}{d\Omega_2}\right)_{Th} 
\sum_{g,f}|\langle f|\sum_j e^{i\mathbf{q}\cdot\mathbf{r}_j}|g\rangle|^2 \nonumber \\   &\times&\delta(E_g-E_f+\hbar\omega) 
\label{eq:XRS}
\end{eqnarray} 

In this form, equation~(\ref{eq:XRS}) is equivalent to an absorption cross section but with  $e^{i\mathbf{q}\cdot\mathbf{r}}$ playing the role of the transition operator. The dependence of XRS on the momentum transfer can be better visualized by expanding the transition operator in:
\begin{equation}
e^{i\mathbf{q}\cdot\mathbf{r}}=
1+
i\mathbf{q}\cdot\mathbf{r}+(i\mathbf{q}\cdot\mathbf{r})^2/2+
\ldots
\label{eq:qr_expansion}
\end{equation}

In the low $q$ limit, the second term $\mathbf{q}\cdot\mathbf{r}$ in Eq.~(\ref{eq:qr_expansion}) dominates; the constant term normally does not contribute to the cross section providing the initial and final states are orthogonal. This can be compared to the conventional absorption transition operator $(\boldsymbol\epsilon\cdot\mathbf{r})e^{i\mathbf{k}\cdot\mathbf{r}}$ which simplifies into $\boldsymbol\epsilon\cdot\mathbf{r}$ in the dipolar approximation ($e^{i\mathbf{k}\cdot\mathbf{r}}\approx 1$). Thus in XRS, $\mathbf{q}$ plays a role comparable to the polarization vector $\boldsymbol\epsilon$ in x-ray absorption spectroscopy.
 
The equivalence with the absorption cross section has been more strictly formalized by~\textcite{Mizuno1967} in a one electron approximation.  Using the scattering tensor $\mathbf{T}(\omega)$:
\begin{equation}
\mathbf{T}(\omega)=\sum_{g,f}\langle f|\sum_j\mathbf{r}_j|g\rangle\langle
g|\sum_j\mathbf{r}_j|f\rangle \times\delta(E_g-E_f+\hbar\omega),
\label{eq:IXS_2}
\end{equation}
the authors showed that equation~(\ref{eq:XRS}) is equivalent to:
\begin{equation}
\frac{d^2\sigma}{d\Omega_2d\omega_2} = 
\left(
\frac{d\sigma}{d\Omega_2}
\right)_{Th}
\mathbf{q}\cdot\mathbf{T}(\omega)\cdot\mathbf{q}.
\label{eq:mizuno}
\end{equation}
With the same formalism, the soft x-ray absorption cross section is found proportional to $\boldsymbol\epsilon\cdot\mathbf{T}(\omega)\cdot\boldsymbol\epsilon$ which has the same form as expression~(\ref{eq:mizuno}) except for the substitution of the momentum transfer by the polarization vector.

Since $\mathbf{q}\cdot\mathbf{r}$ varies with the scattering angle, the dipolar approximation may not be valid in certain scattering configurations where in particular the monopolar term can be dominant. The respective weight of the multipolar expansion was studied by~\textcite{Doniach1971} in the case of  Li metal. The many body interaction due to the core-hole potential was taken into account in the XRS cross-section by the Mahan-Nozi\`{e}re-de Dominicis (MND) theory of edge singularity close to an energy threshold $\hbar\omega_0$. In this framework, the dynamical structure factor can be expressed as:
\begin{equation}
S({\mathbf q},\omega)=\sum_l A_l(q)\overline{R}_l(\omega),
\label{eq:Doniach}
\end{equation}
with $A_l(q)$, the generalized matrix element and $\overline{R}_l(\omega)$ a quantity which diverges as $1/(\omega-\omega_0)^{\alpha_l}$; ${\alpha_l}$ is the MND threshold exponent. 
\begin{figure}[h!]
\centering
\includegraphics[width=0.90\linewidth]{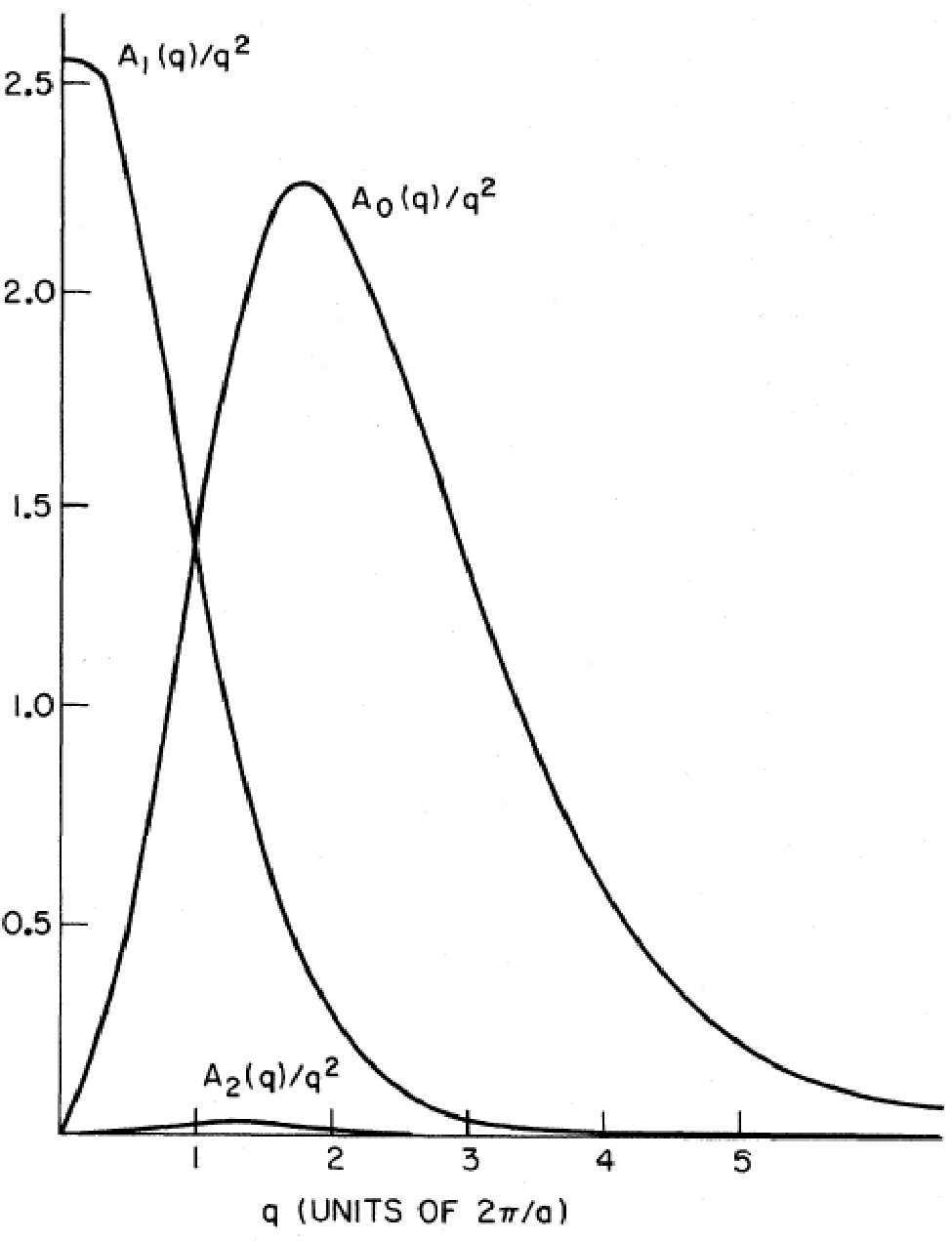}
\caption[X-ray Raman scattering matrix element]{$q$-dependence of the matrix element $A_l(q)$ in the x-ray Raman scattering cross section. The momentum transfer $q$ is given in unit of $2\pi/a$ where $a=3.50$ \AA\ is the Li unit cell parameter. From~\textcite{Doniach1971}.}
\label{fig:Doniach}
\end{figure}
Eq.~(\ref{eq:Doniach}) is illustrated graphically in Fig.~\ref{fig:Doniach}. In the low $q$ limit in forward scattering, the dipolar $A_1(q)$ term dominates and XRS is equivalent to an absorption process in the soft x-ray region. At larger scattering angle, the cross section is dominated by the monopolar contribution $A_0(q)$ while the $A_2(q)$ quadrupolar terms can be neglected.    

Unfortunately, the MND approach is no longer valid in the cases of insulator or semiconductors which require a more accurate treatment of the core-hole electron interaction. Based on the Bethe-Salpeter formalism, \textcite{Soininen2000} have proposed an expression of the dynamical structure factor in terms of an effective Hamiltonian $\mathcal{H}_{eff}$ which carries the many body interactions in the excited state and $\rho(\mathbf{q})$ the Fourier transform of the density-fluctuation operator:
\begin{equation}
S({\mathbf q},\omega)=-\frac{1}{\pi}\mbox{Im}
\langle
g|\rho(\mathbf{q})\frac{1}{\omega-\mathcal{H}_{eff}+
i\Gamma(\omega)}\rho(\mathbf{q})^+|g
\rangle; 
\end{equation}
$\Gamma(\omega)$ accounts for lifetime broadening effects. The method was proven effective to describe the $q$-dependence of the Li K-edge in the wide gap insulator LiF as measured by XRS~\cite{Hamalainen2002}. 

\subsubsection{Resonant scattering}
\begin{figure}[htbp]
\centering
\includegraphics[width=0.90\linewidth]{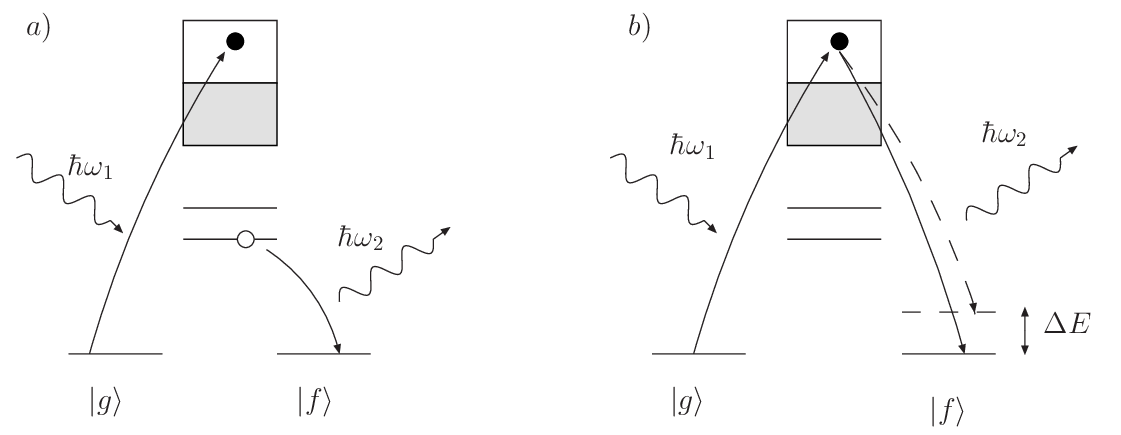}
\caption[RIXS process]{One electron picture of two types of RIXS process : resonant emission (a), and direct recombination (b). The shaded areas are occupied states. $\Delta E$ stands for the energy of one particular excited state relatively to the ground state.}
\label{fig:rixs}
\end{figure}

When the incident photon energy is tuned to the vicinity of an absorption edge, the non-resonant contribution ($\mathbf{A}^2$ term) is no longer the leading term of the interaction Hamiltonian which is now dominated by the $\mathbf{A}\cdot\mathbf{p}$ term (Fig~\ref{fig:diag_1}(b,c)). Since the scattering process is described by an incoming as well as an outgoing photon the $\mathbf{A}\cdot\mathbf{p}$ term has to be considered to the second order. In a one electron picture the resonant inelastic x-ray scattering (RIXS) process can be described by the absorption of an incident photon followed by the emission of a secondary photon as shown in Fig.~\ref{fig:rixs}(a). In reality, the absorption and emission interfere and the resonant scattering process has to be treated as a unique event. Within the limits of the second-order perturbation approach and neglecting the spin-dependent terms, the total double differential cross section is expressed the Kramers-Heisenberg formula:
\begin{widetext}
\begin{eqnarray}
\label{eq:KH}
\frac{d^2\sigma}{d\Omega d\hbar\omega_2} & = & 
r_0^2
\left(\frac{\omega_2}{\omega_1}\right)
\sum_f
\left|
\langle f |\sum_j e^{i\mathbf{q}\cdot\mathbf{r}_j}|i\rangle
(\boldsymbol\epsilon_1\cdot\boldsymbol\epsilon^*_2)\right.\nonumber\\
&+& 
\left.
\left(\frac{\hbar}{m}\right)
\sum_i\sum_{jj'}\left[
\frac{
\left\langle
f \left|
(\boldsymbol\epsilon^*_2\cdot\mathbf{p}_j)e^{-i\mathbf{k}_2\cdot\mathbf{r}_j})
\right|i
\right\rangle
\left\langle 
i\left|
(\boldsymbol\epsilon_1\cdot\mathbf{p}_{j'})e^{i\mathbf{k}_1\cdot\mathbf{r}_{j'}})
\right|g\right\rangle
}
{E_g-E_i+\hbar\omega_1-i\Gamma_i/2}\right.\right.\nonumber\\
& + &
\left.\left.
\frac{
\left\langle
f \left|
(\boldsymbol\epsilon_1\cdot\mathbf{p}_j)e^{i\mathbf{k}_1\cdot\mathbf{r}_j})
\right|i
\right\rangle
\left\langle 
i\left|
(\boldsymbol\epsilon^*_2\cdot\mathbf{p}_{j'})e^{-i\mathbf{k}_2\cdot\mathbf{r}_{j'}})
\right|g\right\rangle
}
{E_g-E_i-\hbar\omega_2}\right]\right|^2\nonumber\\
&\times&
\delta(E_g-E_f+\hbar\omega_1-\hbar\omega_2),
\end{eqnarray}
\end{widetext}
where $|g\rangle$, $|f\rangle$ and $|i\rangle$ stand for the ground state, final state, and intermediate state with energies $E_g$, $E_f$ and $E_i$ respectively. We use the standard notation for the incident and outgoing photon wave vector, energy and polarization ($\mathbf{k}_1$, $\hbar\omega_1$, $\boldsymbol\epsilon_1$) and ($\mathbf{k}_2$, $\hbar\omega_2$, $\boldsymbol\epsilon_2$); $\Gamma_i$ is the lifetime broadening of the core-excited state. The sums are carried over the intermediate and final states, and over the electronic positions $\mathbf{r}$. In a RIXS experiment, the incident photon energy is chosen close to an absorption edge such that $\hbar\omega_1\approx E_g-E_i$. Keeping only the leading term in Eq.~(\ref{eq:KH}), the Kramers-Heisenberg formula then simplifies into:
\begin{widetext}
\begin{equation}
\label{eq:KH_small}
\frac{d^2\sigma}{d\Omega d\hbar\omega_2} =  
r_0^2
\left(\frac{\omega_2}{\omega_1}\right)
\sum_f
\left|\left(\frac{\hbar}{m}\right)
\sum_i
\frac{
\left\langle
f \left|
(\boldsymbol\epsilon^*_2\cdot\mathbf{p}_j)e^{-i\mathbf{k}_2\cdot\mathbf{r}_j})
\right|i
\right\rangle
\left\langle 
i\left|
(\boldsymbol\epsilon_1\cdot\mathbf{p}_{j'})e^{i\mathbf{k}_1\cdot\mathbf{r}_{j'}})
\right|g\right\rangle
}
{E_g-E_i+\hbar\omega_1-i\Gamma_i/2}\right|^2
\times
\delta(E_g-E_f+\hbar\omega),
\end{equation}
\end{widetext}
with $\hbar\omega$ the transferred energy and omitting the implicit sum over $jj'$. Energy conservation is reflected by the argument of the $\delta$-function in Eq.~(\ref{eq:KH_small}) and applies to the overall scattering process. It is not binding for the $|g\rangle\rightarrow|i\rangle$ transition due to  the short lifetime of the intermediate state. The energy conservation condition gives rise to the so-called Raman shift of the scattered photon energy $\hbar\omega_2$, which varies linearly as a function of the incident photon energy. The resonant denominator and interference terms in the corresponding cross section (Eq.~(\ref{eq:KH_small})) characterize the regime of resonant inelastic x-ray scattering. 
 
If one takes into account the finite lifetime in the final state $\Gamma_f$, the $\delta$-function in Eq.~(\ref{eq:KH_small}) has to be replaced by a Lorentzian $\Delta(\omega)=\Gamma_f/(\omega^2+\Gamma_f^2)$. This in turn plays a fundamental role in the asymmetry of the RIXS profile on resonance~\cite{Agren2000} (cf.~\ref{sec:narrowing}). Other correction terms would include convolution by Gaussian functions to account for the experimental resolution and incident energy bandwidth. 

\subsubsection{Resonant emission and Direct recombination}
\label{sec:RIXS}
After the absorption of the primary photon, the system is left in an excited intermediate state.  The decay of the RIXS intermediate state can either leave a spectator electron in the final state (Fig.~\ref{fig:rixs}(a)), or involve the participant electron (Fig.~\ref{fig:rixs}(b)). We will discuss the former in the next section~\ref{sec:narrowing} in terms of resonant x-ray emission spectroscopy (RXES). For clarity, we distinguish RXES from the radiative decay process (denoted RIXS by default) where the excited electron recombines with the core-hole, thus returning the system either to the ground state or to an excited configuration (cf.\ Fig.~\ref{fig:rixs}(b)). The energy difference from the ground state $\Delta E$ is transferred to the electron system, and an excitation spectrum for the system is thus measured through the resonant cross section. In section~\ref{sec:nrIXS}, we remarked that information about the single particle excitation spectrum is contained in the dynamical structure factor $S(\mathbf{q},\omega)$ which is directly related to the non-resonant scattering cross section. A non-resonant experiment can thus be interpreted directly, but in practice the cross-section is weak which is problematic for studying heavy elements as the non resonant cross section approximately falls with $\approx 1/Z^2$ (with a jump at $Z\sim40$, cf.\ Fig.~1 in \textcite{Scopigno2005}). In addition the non-resonant measurement lacks chemical selectivity.  

Thanks to the resonant enhancement, the RIXS direct recombination process allows low energy excitations to be probed in complex materials in the absence of the core-hole in the final state~\cite{Kao1996,Hill1998}. Compared to other spectroscopic techniques such as EELS or optical absorption, RIXS presents several advantages: i) the momentum transfer can be varied on a larger scale and the dispersion of the excitation studied over multiple Brillouin zones; ii) the scattering cross section benefits from the resonant enhancement including the chemical selectivity; iii) the penetration depth is significantly larger than for electron scattering; iv) the energy-loss spectra are not contaminated by multiple scattering contributions.

We will see in section~\ref{sec:MIT} that RIXS is a powerful method when dealing with metal-insulator transitions under pressure. The theoretical treatment of RIXS however is not a trivial task as the excitonic pair formed in the intermediate state may interact with the valence electrons, requiring an ad-hoc treatment beyond the second-order perturbation theory discussed in section~\ref{sec:third_order}. 

\subsubsection{Fluorescence}
Far above the absorption edge resonant processes still exist but coherence between the absorption and emission is lost. It is no longer possible to determine when the photon is absorbed or emitted. Time-permuted events such as described by diagram Fig~\ref{fig:diag_1}(c) contribute to the scattering process. This situation corresponds to the fluorescence regime or x-ray emission spectroscopy (XES), where the emitted photon energy no longer depends on the choice of the energy of the incident photon.  The fluorescence cross-section is well approximated by using a two-step model (absorption followed by emission) by multiplying the x-ray-absorption cross section with the emission cross section. This applies to the K$\alpha$ ($2p\rightarrow 1s$) and K$\beta$ ($3p\rightarrow 1s$) emission that we will study later in this review. Such an approximation is however limited to ionic systems where configuration interactions in the intermediate state can be neglected. For covalent systems, a coherent second-order model gives a more accurate description when relaxation in the intermediate state can occur.

\subsection{Narrowing effects}
\label{sec:narrowing}
\begin{figure*}[htbp]
\begin{tabular}{cc}
\includegraphics[width=0.45\linewidth]{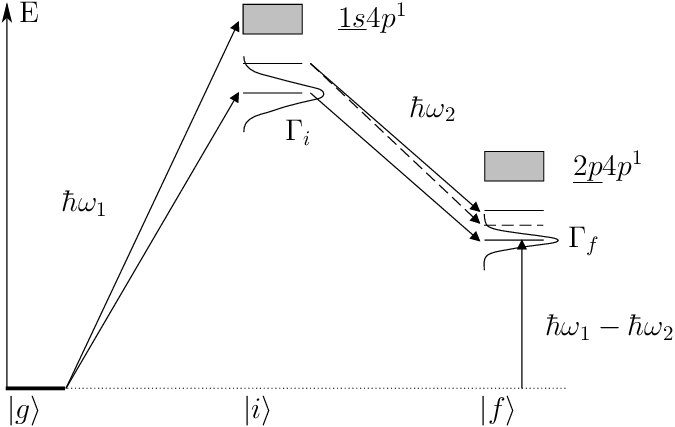} &
\includegraphics[width=0.45\linewidth]{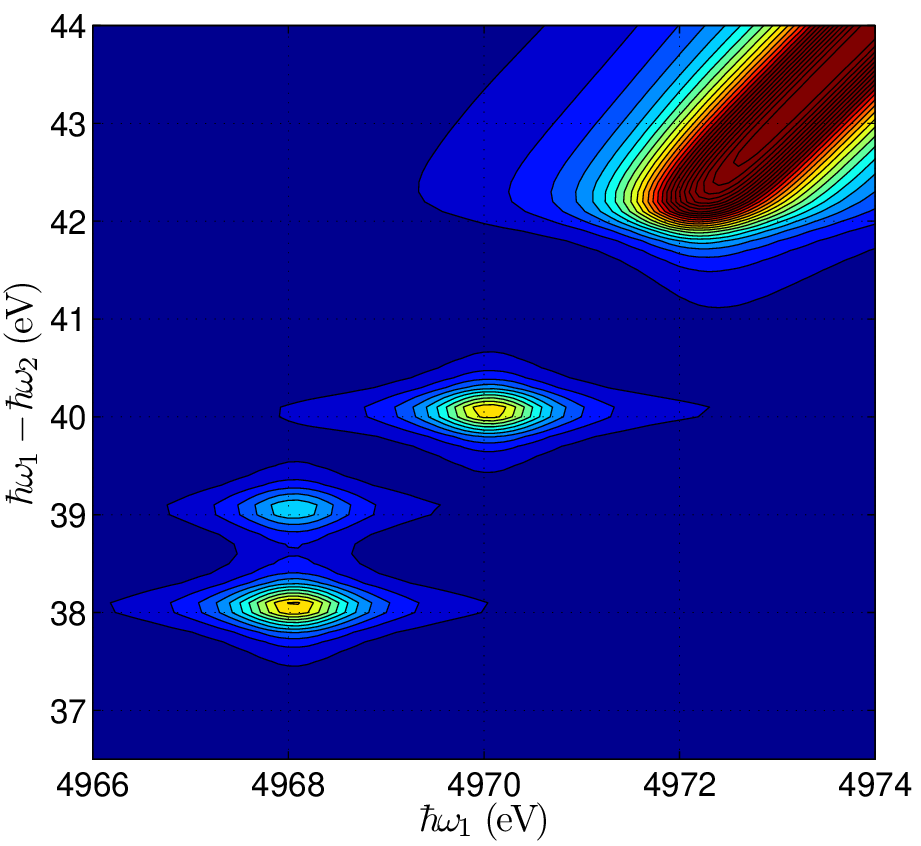} \\
{\small a)} & {\small b)}\\
\\
\includegraphics[width=0.45\linewidth]{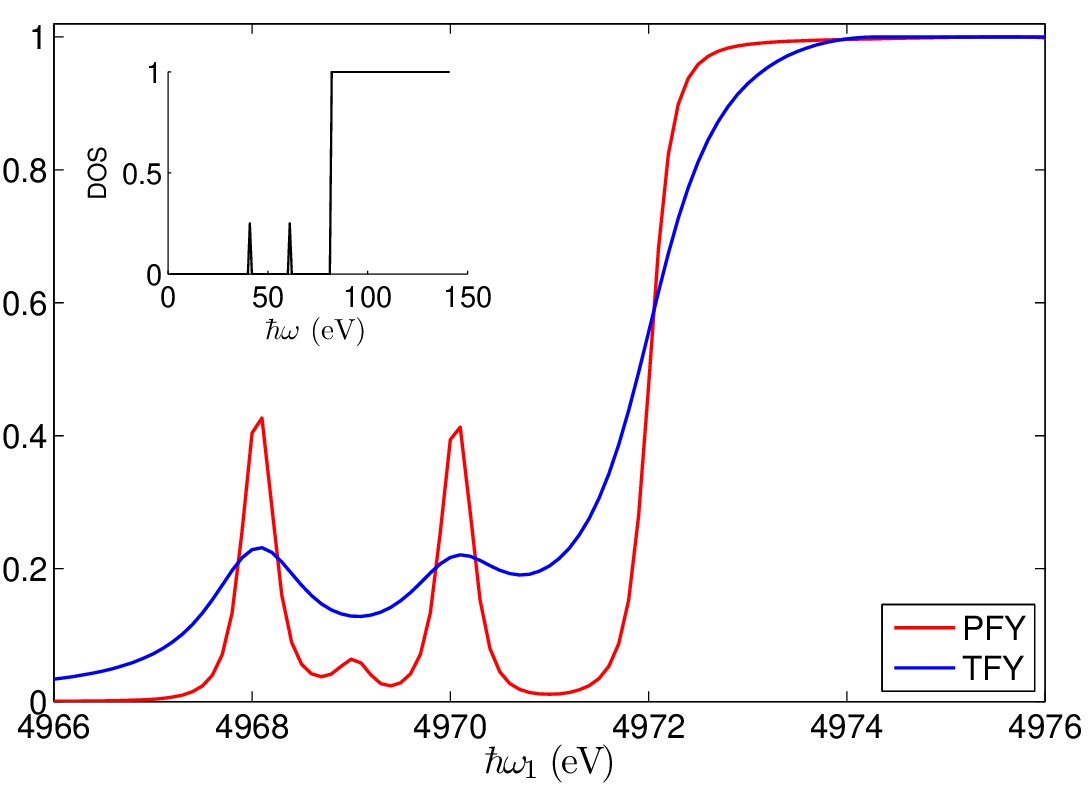} &
\includegraphics[width=0.45\linewidth]{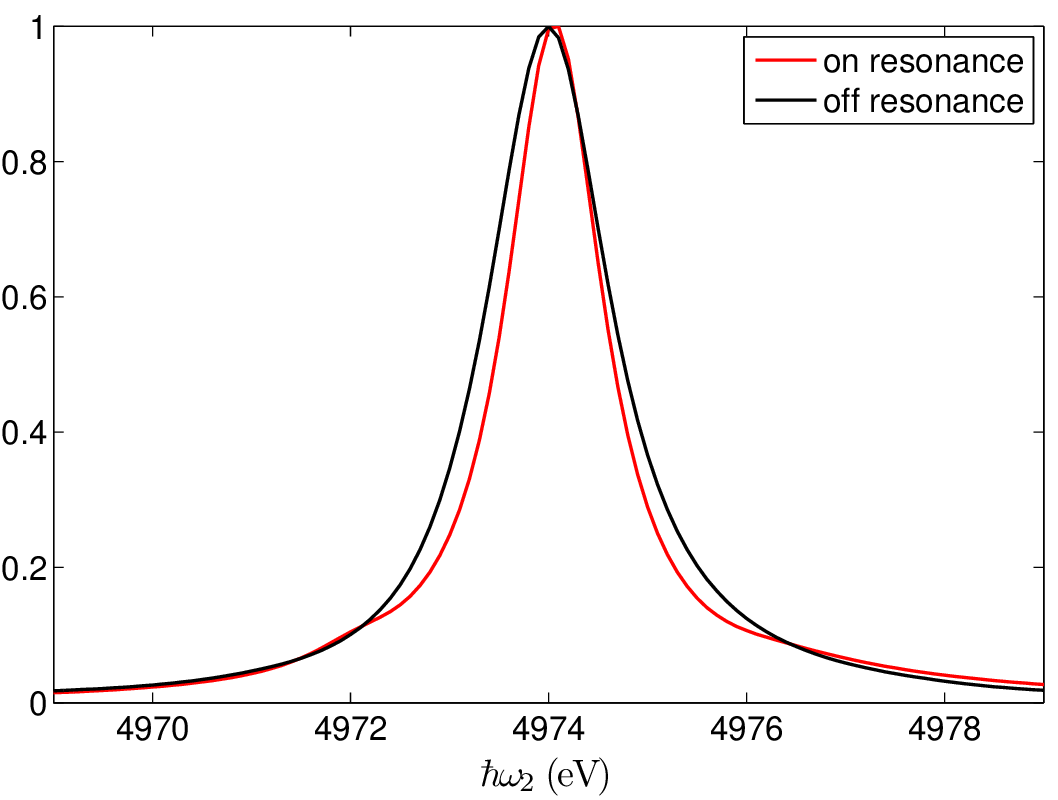} \\
{\small c)} & {\small d)}\\
\end{tabular}
\caption[Narrowing effects]{(Color online) (a) $1s2p$-RXES process in a configuration scheme; (b) map of the cross section in the incident vs.\ transfer energy plane; (c) Partial and total fluorescence yield absorption spectra - inset indicates the model density of unoccupied states; (d) Comparison of RXES spectra on resonance to fluorescence regime (off resonance).}
\label{fig:Narrowing}
\end{figure*}
In the resonant regime the energy resolution of the measured line is limited by the core hole lifetime but this broadening $\Gamma$ can be partly overcome, depending on the detuning of the incident photon energy with respect to the resonance energy. Such a narrowing effect was first observed at the Cu K edge \cite{Eisenberger1976}. As explained below, narrowing is mostly effective when the 
hole created in the intermediate state belongs to a narrow and shallow level such as in RXES which we will describe extensively in section~\ref{sec:RE_HP} of this manuscript.

\subsubsection{Resonant X-ray emission} 
The RXES process consists of the absorption of an incident photon ($\hbar\omega_1$) which provokes the transition of a core electron to empty states followed by the emission of a secondary photon ($\hbar\omega_2$) upon recombination of another electron to the primary vacancy. To illustrate the narrowing effects in RXES, we consider the case where the intermediate states are delocalized states with little overlap with the core-hole wavefunction. Because the primary electron is ejected into a continuum level, the sum over discrete intermediate states in Eq.~(\ref{eq:KH_small}) has to be substituted by an integration over a continuous density of unoccupied states $\eta(\varepsilon)$, namely $\sum_i\mapsto\sum_i\int d\varepsilon\eta(\varepsilon)$~\cite{Tulkki1982,Aberg1985,Taguchi2000}. Omitting interference effects, the cross section then reads:
%
%
%
\begin{widetext}
\begin{equation}
\frac{d^2\sigma}{d\Omega d\hbar\omega_2} =  
\sum_f
\sum_i
\int d\varepsilon\eta(\varepsilon)
\frac{
\left\langle f \left|T_2\right|i \right\rangle^2
\left\langle i\left|T_1\right|g \right\rangle^2
}{(E_g-E_i-\varepsilon+\hbar\omega_1)^2+\Gamma_i^2/4}
\times
\frac{\Gamma_f/2\pi}{(E_g-E_f-\varepsilon+\hbar\omega_1-\hbar\omega_2)^2+\Gamma_f^2/4},
\label{eq:KH1s2p}
\end{equation}
\end{widetext}
where $T_1$ and $T_2$ are the transition operators for the incident and emitted photons. In this simplified form, the cross section merely reduces to a product of two  Lorentzian functions of width proportional to $\Gamma_{i}$ and $\Gamma_{f}$ and centered at two different energies, respectively function of $\omega_1$ and $\omega_1-\omega_2$. 

Following the description made in \textcite{Hayashi2003,Glatzel2009}, we have computed the RXES cross section in the case of a $1s2p$-RXES using the simplified expression Eq.~(\ref{eq:KH1s2p}). As schematized in Fig.~\ref{fig:Narrowing}(a) in a configuration scheme, the RIXS process involves the creation of successively a $1s$ and $2p$ core-holes. We used $\Gamma_{i}=$7 eV, $\Gamma_{f}=$2 eV for lifetime broadening effects and considered a model empty density of states shown in the inset to Fig.~\ref{fig:Narrowing}(c). The continuum states are represented by a step function. In the pre-edge region, the Dirac peaks mimic the presence of localized $3d$ states. The results is shown in Fig.~\ref{fig:Narrowing}(b) as a function of incident and transfer energy. In this plane, emission from localized states appears at constant transfer energy, while fluorescence emission disperses along the main diagonal. 

Cuts of this surface at fixed $\hbar\omega_1$ probe the final states with a $\Gamma_{f}$ resolution. In the opposite direction, at fixed transfer energy, one is able to scan through the intermediate state but with a enlarged resolution $\Gamma_{i}>\Gamma_{f}$. The differential resolving power is clearly observed in the pre-edge region, which stretches further in the direction parallel to incident energy axis (cf.\ Fig.~\ref{fig:Narrowing}(b)). At the resonance, a narrowing of the emission below the lifetime broadening is observed (Fig.~\ref{fig:Narrowing}(c,d)). 

\subsubsection{Partial Fluorescence Yield X-ray absorption}
Instead of measuring the emitted spectra at fixed incident energy as in RXES one can measure scattered intensity at fixed emission energy while the incident energy is varied across an absorption edge. This corresponds to cuts along the diagonal in Fig.~\ref{fig:Narrowing}(b). As demonstrated originally by \textcite{Hamalainen1991}, the resulting spectrum in this so-called partial fluorescence yield (PFY) mode is interesting because it resembles a standard x-ray absorption (or total fluorescence yield (TFY)) spectrum but with better resolution. The sharpening effect results from the absence of a deep core-hole in the final state. As opposed to measurements in the TFY mode however, the PFY spectra is not strictly equivalent to an absorption process~\cite{Carra1995}, since it depends on the choice of the emitted energy. Multiplet effects in the RXES final state can also distorts the PFY lineshape. 

The sharpening effect is exemplified in Fig.~\ref{fig:Narrowing}(c) where PFY and TFY spectra calculated from Eq.~(\ref{eq:KH1s2p}) are superimposed and compared to our model density of states. In the PFY mode, the lifetime broadening $\Gamma_\text{PFY}$ can be approximated  by : 
\begin{equation}
\label{Eq:PFY}
\frac{1}{\Gamma_\text{PFY}^2}=\frac{1}{\Gamma_i^2}+\frac{1}{\Gamma_f^2}
\end{equation}
%
In general, the lifetime broadening of the final state is considerably smaller than that of core excited state ($\Gamma_f\ll\Gamma_i$), thus giving the possibility of performing x-ray absorption spectroscopy below the natural width of the core excited state. The sharpening effect is especially marked in the pre-edge region as shown in Fig.~\ref{fig:Narrowing}(c). 

\subsection{Third-order terms}
\label{sec:third_order}
\begin{figure}[htbp]
\centering
\includegraphics[width=0.90\linewidth]{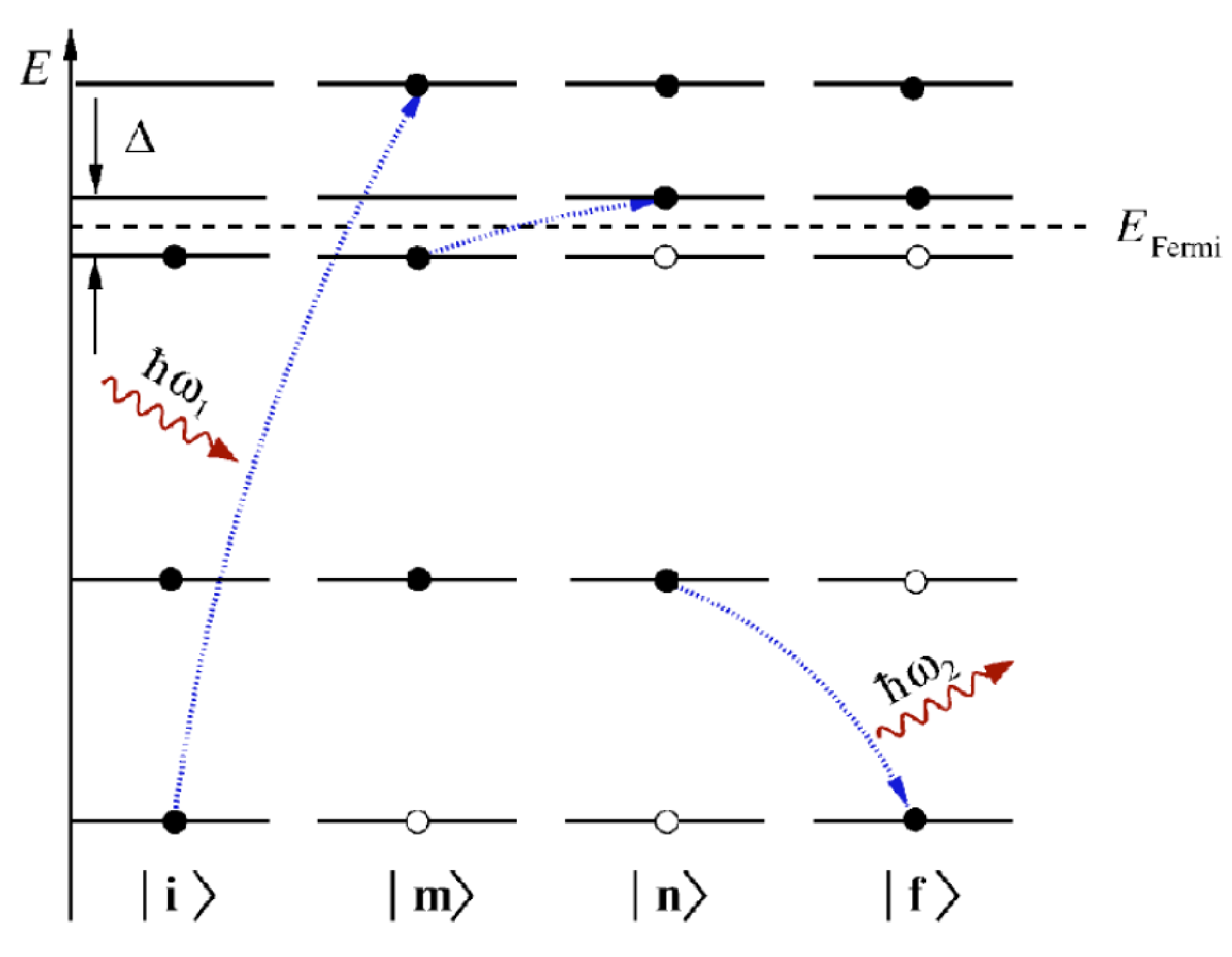}
\caption[shakeup process]{(Color online) Shakeup process in the intermediate state of a RIXS process. $|i\rangle$ and $|f\rangle$ are the initial and final state, and $|m\rangle$ and $|n\rangle$ two intermediate state. From~\textcite{Doring2004}.}
\label{fig:shakeup}
\end{figure}

\begin{figure}[htbp]
\centering
\includegraphics[width=0.90\linewidth]{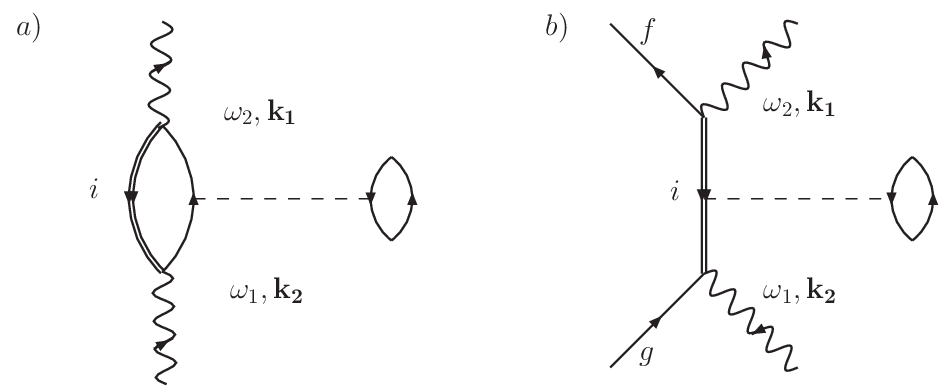}
\caption[First order Coulomb corrections]{First order Coulomb corrections (dotted line) to RIXS. In diagram a), an excitonic pair is formed in the intermediate state.}
\label{fig:diag_3}
\end{figure}

The Kramers-Heisenberg equation which we have used so far to describe the RIXS process is derived in the so-called sudden approximation. It relies on the implicit assumption that the core hole left in the RIXS intermediate state, which can be considered to form a virtual excitonic pair with the excited electron, is short-lived enough not to perturbate the rest of the electronic system. What it fails, the Coulomb interaction of the newly formed exciton may act as an extra potential that could scatter off valence electrons (Fig.~\ref{fig:shakeup}). The occurrence of such a shake up event requires an ad-hoc treatment beyond the Kramers-Heisenberg formulation. 
The shake up process has been described by a third order perturbation treatment of the scattering cross section~\cite{Platzman1998,Doring2004}, inspired by the Raman cross-section for light scattering by phonons. The Coulomb interaction $\mathcal{H}_C$ between the virtual exciton and the rest of the valence electrons is singled out from the total interacting Hamiltonian and treated in the perturbation theory (cf. Fig.~\ref{fig:diag_3}(a)). The resulting Kramers-Heisenberg cross section possesses an additional term which contains two intermediate states $|m\rangle$ and $|n\rangle$~\cite{Doring2004} as follows:
\begin{widetext}
\begin{eqnarray}
\label{eq:KH_third}
\frac{d^2\sigma}{d\Omega d\hbar\omega_2} &=&  
r_0^2
\left(\frac{\omega_2}{\omega_1}\right)
\sum_f
\left|
\sum_i
\frac{
\left\langle f \left|T_2\right|i\right\rangle
\left\langle i\left|T_1\right|g\right\rangle
}
{E_g-E_i+\hbar\omega_1-i\Gamma_i/2}\right.\nonumber\\
&+&
\left.
\sum_{i,n}
\frac{
\left\langle f \left|T_2\right|n\right\rangle
\left\langle n \left|\mathcal{H}_C\right|i\right\rangle
\left\langle i\left|T_1\right|g\right\rangle
}
{(E_g-E_i+\hbar\omega_1-i\Gamma_i/2)(E_g-E_n+\hbar\omega_1-i\Gamma_n/2)}
\right|^2\nonumber\\
&\times&
\delta(E_g-E_f+\hbar\omega),
\end{eqnarray}
\end{widetext}
The second term in Eq.~(\ref{eq:KH_third}) describes  the three-step scattering process of Fig.~\ref{fig:shakeup}.

In the shake up description, the RIXS cross section is explicitly related to the dynamical structure factor $S(\mathbf{q},\omega)$, weighed by a resonant denominator~\cite{Abbamonte1999,Brink2005}. Though it is a matter of debate whether this treatment is necessary, it was suggested that third order corrections could explain the deviation from linear Raman shift observed in cuprates. 

\section{Instrumentation}
IXS is a second order process of weak intensity. Even though the very first experiments were performed on laboratory  and second generation sources, the flowering of IXS as a spectroscopic probe coincides with the development of insertion devices on third generation synchrotrons. Simultaneously, new x-ray optics based on the Rowland circle geometry have provided relatively large acceptance angles while maintaining an excellent energy resolution. 

\subsection{IXS Spectrometer}
\subsubsection{Energy selection}
Energy discrimination is achieved through Bragg reflection with a crystal analyzer. Because perfect crystal quality is required to attain the best resolving power, Si or Ge analyzers are preferentially used. Another key point is to adapt the Bragg angle to the photon energy in order to minimize the geometrical contribution to the resolution. This quantity is given by equation~(\ref{eq:geom_contrib}) where $\Delta\theta$ is the source size (including the beam divergence) and $\theta_B$ the Bragg angle of a given reflection. Thus, the higher the Bragg angle, the smaller the geometrical term.  
\begin{equation}
\left.\frac{\Delta E}{E}\right|_g=\Delta\theta\cot\theta_B
\label{eq:geom_contrib}
\end{equation}

Typical analyzers are indicated in Table \ref{tab:analyzers} for selected transition metals, rare earths and actinides emission energies. 
\begin{table}[htbp]
\caption[Analyzer crystals]{Analyzer crystals and Bragg angles sorted by increasing emission energies for selected transition metals, rare-earths and actinides.}
\begin{tabular}{lccc}
\hline
\hline
Emission Line & Energy (eV) & Analyzer & Bragg angle (deg)\\
\hline
Mn-K$\alpha_{1}$ & 5900.4 & Si(440) & 71.40$^\circ$\\
Fe-K$\alpha_1$ & 6405.2 & Si(333) & 67.82$^\circ$\\
Mn-K$\beta_{1,3}$ & 6490.4 & Si(440) & 84.10$^\circ$\\
Co-K$\alpha_1$ & 6930.9 & Si(531) & 76.99$^\circ$\\
Fe-K$\beta_{1,3}$ & 7059.3 & Si(531) & 73.06$^\circ$\\
Ni-K$\alpha_1$ & 7480.3 & Si(620) & 74.82$^\circ$\\
Co-K$\beta_{1,3}$ & 7649.1 & Si(620) & 70.70$^\circ$\\
Cu-K$\alpha_1$ & 8046.3 & Si(444) & 79.38$^\circ$\\
Ni-K$\beta_{1,3}$ & 8264.6 & Si(551) & 80.4$^\circ$\\
Cu-K$\beta_{1,3}$ & 8903.9 & Si(553) & 79.97$^\circ$\\
\hline
Ce-L$\alpha_1$ & 4840.2 & Si(400)& 70.62$^\circ$\\
Yb-L$\alpha_{1}$ & 7416.0 & Si(620) & 76.78$^\circ$\\
U-L$\alpha_{1}$ & 13614.7 & Ge(777) & 77.40$^\circ$\\
\hline
\hline
\end{tabular}
\label{tab:analyzers}
\end{table}

\subsubsection{Rowland circle}
\begin{figure}[htbp]
\includegraphics[width=0.90\linewidth]{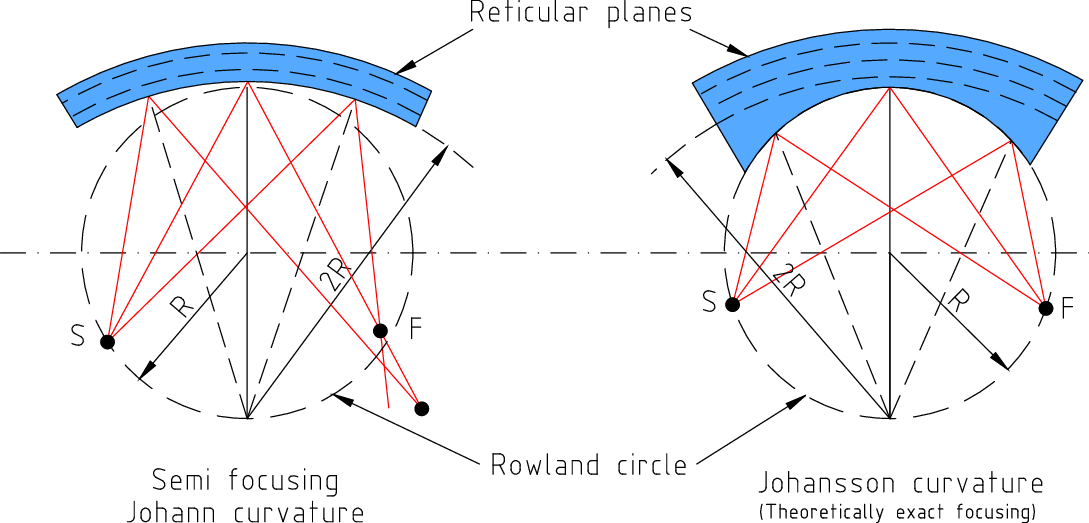}
\caption[Johann's vs.\ Johanssonn's geometry]{(Color online) Johann geometry is an approximation of the Johanssonn geometry for large $\Delta E$.}
\label{fig:Johann}
\end{figure}

IXS has largely benefited from the technological developments concerning x-ray spectrometers. A major step towards high resolution - high flux spectrometers was to adapt the Rowland circle in the Johann geometry to x-ray optics. In this approximate geometry, the sample, the analyzer and the detector sit on a circle whose diameter corresponds to the analyzer bending radius $R$. The Johann geometry departs from the exact focusing or Johansson geometry by a different curvature of the analyzer surface, as illustrated in figure \ref{fig:Johann}. In the latter, the analyzer surface entirely matches the Rowland circle, while in the former the focusing condition is only fulfilled at a single point. 

Grinding the analyzer surface to fulfill the Johansson condition is a difficult task, and the Johann geometry is usually preferred. The consequent Johann error can be expressed by equation~(\ref{eq:Johann}), where $r$ is the distance from the analyzer center. This contribution to the overall resolution is negligible as long as the analyzer diameter is small compared to the bending radius. 
\begin{equation}
\left.\frac{\Delta E}{E}\right|_J=\frac{1}{2}\left(\frac{r}{R}\right)^2\cot^2\theta_B
\label{eq:Johann}
\end{equation}

\subsubsection{Analyzer bending} 
\begin{figure}[htbp]
\includegraphics[width=0.90\linewidth]{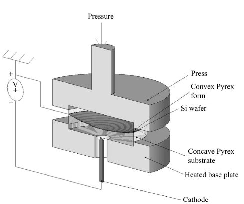}
\caption[Bent analyzer]{(Color online) (left) Press for anodic bonding. From~\textcite{Collart2005}.}
\label{fig:analyzer}
\end{figure}

The spherical crystal analyzer is the key element of the spectrometer and must be optimized for the best trade-off between resolution and count-rate. It collects scattered photons from a large solid angle, selects the required photon energy and focuses the beam onto the detector. Among different possible focusing setups and corresponding analyzer design, we will describe here spherically bent analyzers for they combine several advantages including large solid-angles and relatively high-resolution. Static bending can be realized by pressing the analyzer wafers onto a spherical glass substrate. The two pieces are bonded together either by gluing the analyzer backface with a resin, or by anodic-bonding method which was recently applied to the fabrication of Si analyzers~\cite{Collart2005}. In this technique, bonding is ensured by migration of Na$^+$ ions in the glass at high temperature and in presence of a high electric field, away from the glass/Si interface. The fixed O$^{2-}$ ions at the interface exert a very strong Coulomb force on the Si wafer which irreversibly adheres to the substrate due to the formation of Si-O bonds. Figure \ref{fig:analyzer} shows a press developed for anodic bonding at IMPMC (Paris). 

Bending a crystal results in elastic deformations that affect the energy resolution according to 
\begin{equation}
\left.\frac{\Delta E}{E}\right|_P=\frac{l}{R}\left|\cot^2(\theta_B)-\nu\right|
\label{eq:Poisson}
\end{equation}
\noindent where $l$ is the effective thickness of the crystal and $\nu$ the material Poisson ratio. This term is normally small, but a conventional gluing process generally introduces extra strain locally due to an inhomogeneous layer of glue which further contributes to enlarge the energy bandwidth. Because of the absence of interfacial gluing resin, the anodic bonding technique provides better resolution. An intrinsic resolution of the order of 200 meV was obtained at 8.979 keV with a 2-m radius Si(553) analyzer prepared at IMPMC. Another possibility is to use diced Si analyzers~\cite{Masciovecchio1996} which are more suitable for applications requiring very high resolutions, below the 100 meV level but one generally pays a price associated with a correspondingly lower count-rate.

\subsection{Pressure setups for the spectroscopist}
\subsubsection{Scattering geometries at high pressure}
Diamond anvil cells (DAC) are easily the most widely used pressure cells in x-ray spectroscopy (cf.~\textcite{Jayaraman1983} for an extensive though somewhat outdated description of the DAC technique). Let alone the exceptional hardness of diamonds which has pushed the highest achievable pressures to the megabar region, diamonds are transparent in a broad spectral range covering infrared, visible light, and x-ray (mostly above 5 keV). DAC are small devices which can be easily mounted on a goniometer head, in a vacuum chamber or in a cryostat for low-temperature measurements, or coupled to a power-laser source such as in laser-heating technique. 
\begin{figure}[htbp]
\includegraphics[width=0.90\linewidth]{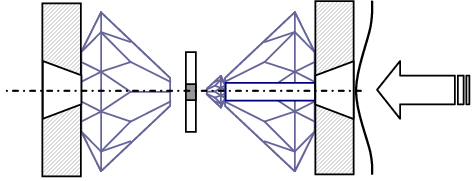}
\caption[Diamond anvil cell]{(Color online) Diamond anvil cell equipped with a solid (left) and  perforated diamond (right). The sample shown in dark gray is contained in a gasket that is compressed between the anvil and the piston. Pressure can be realized by inflating a metallic membrane schematized by the curvy line.}
\label{fig:DAC}
\end{figure}

\begin{figure}[htbp]
\includegraphics[width=0.90\linewidth]{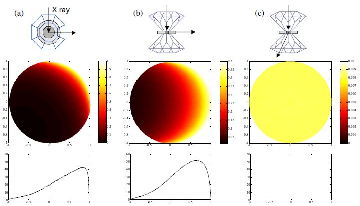}
\caption[Self absorption effect]{(Color online) Self absorption effect in a model transition-metal sample contained in a pressure cell. From top to bottom, sample geometry (a) in-plane scattering (b) transverse geometry, (c) full transmission (the sample and the gasket are shown in gray); 2D emission profile; integrated intensity. The sample diameter is 100 $\mu m$, and we considered an attenuation length of 30 $\mu m$, typical of metal oxides around 8 keV. From \textcite{Rueff2005}.}
\label{fig:DAC_geom}
\end{figure}

The standard procedure is to load the sample in a chamber drilled in a gasket that serves to limit the pressure gradient during compression by the two diamonds. To ensure hydrostaticity, the gasket chamber is normally filled with a pressure transmitting medium. Ruby chips are also inserted for pressure calibration. Different geometries can be envisaged depending on the experimental needs. Fig.~\ref{fig:DAC_geom} illustrates more particularly the setups used in x-ray spectroscopy with in-plane, transverse or transmission geometries. Both in-plane and transverse geometries require x-ray transparent gasket material such as high-strength Be. Because Be is lighter than C, in-plane detection through Be gasket seems to be the most efficient geometry while the absorption of the diamonds, particularly strong along the exit path -- the scattered energies typically fall within the 5--10 keV energy range -- makes it difficult to work in full transmission geometry. However determining the optimum geometry requires self-absorption of the scattered x-rays to be taken into consideration.

The self-absorption strength depends on the total sample length projected along the detection direction, here the sample-analyzer axis, and the x-ray attenuation length for the considered material. The 2D intensity profile emitted by the sample is simulated in figure~\ref{fig:DAC_geom} for different geometries : (a) in-plane scattering through a Be gasket, (b) transverse geometry (the incident x-ray enters the cell through diamond and exits through a Be gasket), or (c) full transmission through the diamonds. The simulation was carried out by considering a sample of diameter 100 $\mu m$ placed in an incident x-ray beam of 15 keV, and an attenuation length of 30 $\mu m$ typical of   transition-metal and rare earth compounds. As expected, the highest peak-intensity is obtained for the in-plane configuration, but because both incident and emitted x-ray are strongly absorbed, only a portion covering about one-third of the sample surface is visible from the analyzer point of view. In the transverse configuration, the fluorescence comes approximately from one-half of the sample. Even if the first diamond absorbs part of the incident beam, the integrated intensity is comparable to the in-plane configuration thanks to the wider emitting area. Finally, a homogeneous sample can be obtained in the full transmission mode, but then the emitted intensity is strongly absorbed by the exit diamond, resulting in a loss of intensity by a factor of $\approx$30. 

The latter limitation can be avoided to a large extent by using perforated diamonds as recently proposed \cite{Dadashev2001}. The diamonds can be either partially emptied leaving simply a thin but opaque back-wall (down to 200 microns) or fully drilled as illustrated in Fig.~\ref{fig:DAC}; in such a case, a small diamond (typically of 500 microns height) is glued onto the tip of the perforated (bigger) diamond with the advantage of an optical access to the sample chamber. 
In the very high pressure regime, solid diamonds are preferable and the transverse geometry therefore appears as the best compromise between integrated intensity and sample homogeneity, as far as the sample size is kept small compared the attenuation length and hydrostaticity preserved throughout the entire pressure range. 
Finally, EXAFS measurements down to the S K-edge (2.47 keV) under high-pressure were recently made possible by using Be gaskets in the in-plane geometry where part of the gasket material was hollowed-out along the scattering path. 

\subsubsection{Combined pressure / temperature}
To explore the complete phase diagram of electronic transitions, it is essential to be able to apply high-pressure while simultaneously varying  temperature. At one extreme, combined high-temperature and high-pressure permits the description of, for instance, magnetism in transition metals or materials of geophysical interest. High temperature at high pressure can be reached by resistive oven or laser heating techniques. Especially, double sided laser heating enables a homogeneous and constant temperature over the illuminated sample area in the DAC~\cite{Lin2005a,Schultz2005}. At the other extreme, low temperature allows one to explore for instance the rich phenomena related to quantum criticality in heavy fermions which we will discuss further in section~\ref{sec:QCP}. For low temperature applications the pressure cell can be mounted in a cryostat and put in thermal contact with the cold finger.\\

The next sections will be devoted to experimental results. We will discuss successively excitations of $d$ and $f$ electrons before addressing those in light elements. 

\section{Local magnetism of transition metal compounds}
\begin{figure}[htbp]
\centering
\includegraphics[width=0.90\linewidth]{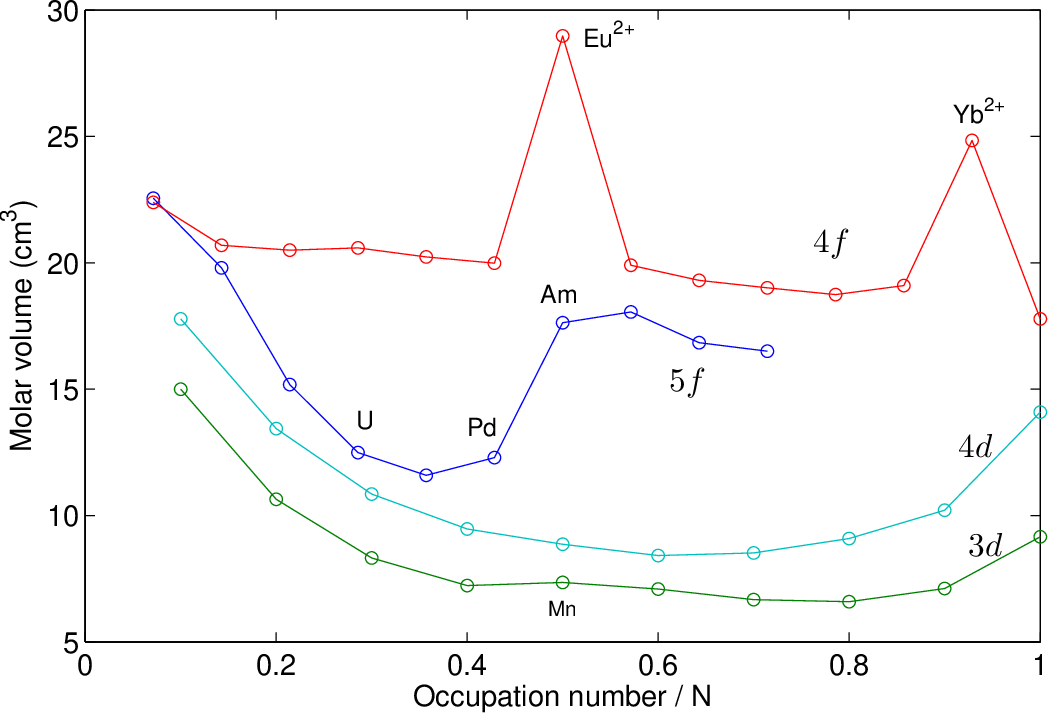}
\caption[Occupation number]{(Color online) Molar volume as a function of the occupation number for the $3d$, $4d$, $4f$, and $5f$ series.}
\label{fig:molar_vol}
\end{figure}

The general behavior of $d$ electrons suggests a delocalized character. In transition metals, they form a band, located in the vicinity of the Fermi energy, of large width when compared to the other characteristic energy scales. A further proof of the band like behavior of $d$ electron is found in the dependence of the molar volume as a function of band filling shown in Fig.~\ref{fig:molar_vol}. The quasi-parabolic behavior observed in the $d$ series is indicative of a simple band state.
However this view is simplistic and the localized or itinerant behavior of $d$ electrons has in fact been the subject of a long-standing controversy which originally goes back to Van Vleck and Slater's study of magnetism. An emblematic example of the apparent dual behavior of $d$ electrons is Fe. The metallic character of Fe indicates itinerant $d$ electrons, while its magnetic properties are well described by an assembly of localized spins. Another striking contradiction appears in transition metal oxides, such as NiO, as revealed in the early work of de~Boer and Verwey. NiO has a partially filled $d$ band and should be metallic. Instead, NiO is a wide gap insulator as are most transition metal oxides.  
%

\subsection{Electron correlations in the compressed lattice}
\subsubsection{Mott-Hubbard approach}
In many transition metal compounds, the Coulomb repulsion $U$ between $d$ electrons is of the same order of magnitude as the $d$ bandwidth. The Mott-Hubbard Hamiltonian is a simplified, yet effective, approach for dealing with electron correlations. The correlated system is described by a single band model in which the $d$ electrons experience a Coulomb $U$ interaction when two of them occupy the same site. In the Mott Hubbard framework, the contribution of $U$ is formalized as an extra term added to the kinetic energy $t$ in the Hamiltonian:
\begin{equation}
\mathcal{H}=\sum_{i,j,\sigma}t_{ij}a_{i\sigma}^{+}a_{j\sigma}
+U\sum_{i}n_{i\uparrow}n_{i\downarrow}
\label{eq:MH}
\end{equation}
$a_{i\sigma}^{+}$ ($a_{i\sigma}$) creates (annihilates) an electron of spin $\sigma$ at site $i$, and $n_{i\sigma}=a_{i\sigma}^{+}a_{i\sigma}$. In the strongly correlated picture, the relative magnitude of $U$ and the $d$-bandwidth $W$ governs the tendency toward localized ($U/W>1$) or itinerant ($U/W<1$) behavior of the $d$ electrons. Correlations thus provide an explanation for the non-metallic character of several transition metal compounds: in the case of half (or less) filling, hopping of $d$ electrons through the lattice is energetically unfavorable because of the strong on-site Coulomb repulsion, leading to the splitting of the associated $d$-band through the opening of a correlation gap and the consequent characteristic insulating state. 
This theoretical understanding was later extended and refined by Zaanen, Sawatzky, and Allen (ZSA)~\cite{Zaanen1985} to account for large discrepancies observed between the estimated and measured band gap in some transition metal insulators and also explain their photoemission spectra. In addition to the on-site $d$-$d$ Coulomb interaction $U$ employed in the original Mott-Hubbard theory, the ligand-valence bandwidth, the ligand-to-metal charge-transfer energy ($\Delta$), and the ligand-metal hybridization interaction are explicitly included as parameters in the model Hamiltonian. Systems where $U<\Delta$ are dubbed Mott insulators while $U>\Delta$ characterizes so-called charge-transfer insulators. In particular, it is now well established that the  correlation energy $U$ is relatively high in NiO and the band gap is of the charge-transfer type that is primarily O-$2p$ to Ni-$3d$ character, because the correlation gap is actually larger than the charge transfer gap.
Correlations are also a necessary ingredient in transition metals to derive the correct magnetic anisotropy~\cite{Yang2001} or charge density~\cite{Dudarev2000}.

This classification scheme has been very successful in describing the diverse properties and some seemingly contradicting behavior of a large number of these compounds. However, these high-energy-scale charge fluctuations are primarily characteristic of the elements involved, and thus cannot be freely adjusted for systematic study of their effects, although they can be varied somewhat by external temperature and magnetic field. On the other hand, pressure can introduce much larger perturbations of these parameters than can either temperature or magnetic field. Hence, it is of great interest to study the high-pressure behavior of these systems, and specifically, to correlate observed transformations with changes in electronic structure.

\subsubsection{Pressure induced metal-insulator transition}
\begin{figure}[htbp]
\centering
\includegraphics[width=0.90\linewidth]{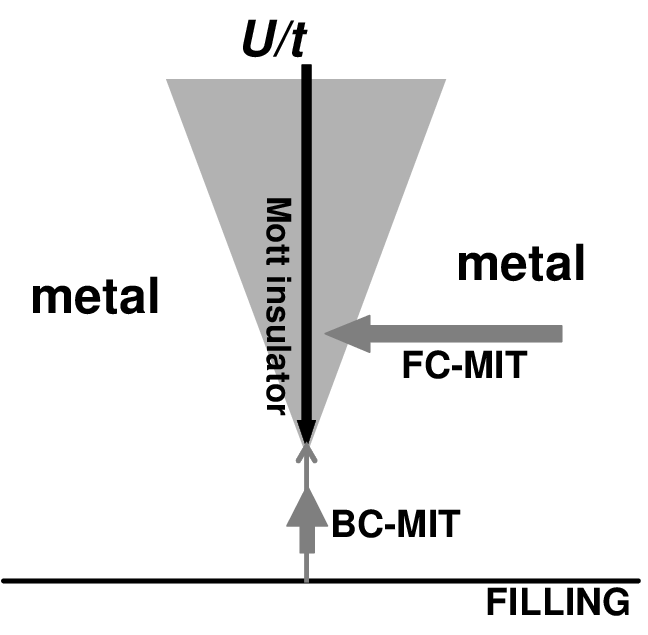}
\caption[Metal-insulator transition]{Metal-insulator transition in correlated transition metal. From \textcite{Imada1998}.}
\label{fig:Imada}
\end{figure}
One important aspect of pressure-induced electronic changes are metal-insulator transitions. According to the classification proposed by~\textcite{Imada1998}, pressure deals with bandwidth-control (BC) MIT as it affects the interatomic distances, hence the orbital overlap and the related bandwidth. In this picture, the control parameter $U/t$ (or equivalently $U/W$) determines the transition from a Mott insulator to a metallic state. V$_2$O$_3$ is a prototypical example of a BC-type insulator to metal transition by application of pressure. In correlated materials, the metallic state (gray area in Fig.~\ref{fig:Imada}) in the immediate vicinity of the insulator state shows an anomalous behavior: the carriers are on the verge of localization, and the system is subject to strong spin, charge and orbital fluctuations. This is the case for example in V$_2$O$_3$ which is characterized by anomalous specific heat and susceptibility near the MIT region. 

The strength of resonant spectroscopy lies in its ability to decouple these different degrees of freedom while applying pressure. The change in the charge transfer and electronic correlations through the MIT will be more specifically discussed in sections~\ref{sec:MIT}.

\subsection{Magnetic collapse}
\subsubsection{Stoner picture}
In his pioneering work, N.~Mott already pointed out the close relationship between the insulating state of transition-metal compounds and electronic density \cite{Mott1968}. In Mott's picture, the insulating character persists upon increasing density (i.e. pressure) until screening becomes effective enough to destroy the electronic correlation that maintains the insulating state, while the $d$-bandwidth increases due to the growing band overlap. At high pressure, the system is therefore expected to undergo a first-order insulator-metal transition, which is usually accompanied by the disappearance of the local $d$ magnetic moment (and not only the long-range magnetization). Using the Hubbard description of the itinerant magnetism (Eq.~\ref{eq:MH}), the stability of the $d$ magnetism can be formalized by the Stoner criterion (Eq.~\ref{eq:Stoner}). Depending on the strength of the on-site Coulomb repulsion $U$, the electron system will behave as a Pauli paramagnet at small $U$ while turning ferromagnetic when $U$ exceeds a critical value $U_c$ defined by:
\begin{equation}
U_c\times n(\varepsilon_F)=1,
\label{eq:Stoner}
\end{equation}
with $n(\varepsilon_F)$ the density of the paramagnetic states at the Fermi energy. The Stoner criterion expresses the balance between exchange and kinetic energies. It has  straightforward implications for the high pressure electronic behavior. As the $d$ bandwidth increases, $n(\varepsilon_F)$ decreases, eventually leading to a state where the Stoner criterion is no longer fulfilled, with a significant loss of the magnetic moment. Here, this \em magnetic collapse \em is understood as a direct consequence of the progressive delocalization of the $d$ electrons under pressure. 
\citet{Krasko1987} has proposed an extended Stoner criterion where $n(\varepsilon_F)$ is replaced by the averaged density of state $\overline{n}$ which explicitly depends both on the spin and magnetic moments. The extended Stoner calculations bridge the gap between the localized approach (crystal field-induced) and the conventional Stoner theory and was especially applied to magnetic collapse in transition metal oxides~\cite{Cohen1997}. 

\begin{table*}[htbp]
\caption[$d$-compound properties]{Summary of main properties of the studied transition metal compounds under pressure. $S$ is the metal spin state obtained from XES. The magnetic state is either paramagnetic (PM), ferromagnetic (FM), antiferromagnetic (AF) or non magnetic (NM); I(M) corresponds to insulating (metallic) state; LS, IS, and HS stand for low spin, intermediate and high spin states. \\
($^*$debated structure; $^\dagger$or semi-conducting).}
\begin{tabular}{lcccccc}
\hline\hline
Sample & Formal Valence & P (GPa) & T (K) & Structure & Properties & $S$\\
\hline
MnO & 2+ & 0 & 300 & NaCl & (AF)I & HS\\
& & 100 & 300 & NiAs & (NM)M  & LS\footnote{\textcite{Mattila2007}} \\
\hline
Fe & 2+ & 0 & 300 & bcc & (FM)M & HS \\
& & 13 & 300 & hcp & (NM)M & LS\footnote{\textcite{Rueff1999a}}\\
& & 20 & 1400 & fcc & (PM)M & LS\footnote{\textcite{Rueff2008}}\\
FeS & 2+ & 0 & 300 & NiAs & (AF)I & HS\\
& & 10 & 300 & Monoclinic & (NM)M$^\dagger$ & LS\footnote{\textcite{Rueff1999}}\\
FeO& 2+ & 0 & 300 & NaCl & (AF)I & HS \\
& & 140 & 300 & NiAs$^*$ & (NM)M & LS\footnote{\textcite{Badro1999}}\\
Fe$_2$O$_3$ & 3+ & 0 & 300 & Corundum & (AF)I & HS\\
& & 60 & 300 & Corundum$^*$ & (NM)M & LS\footnote{\textcite{Badro2002}}\\
Fe$_3$C & 3+ & 0 & 300 & Orthorhombic & (FM)M & HS\\
& & 10--25 & 300 & Orthorhombic & (NM)M & LS\footnote{\textcite{Lin2004a}}\\
Fe-Ni (Invar) & 2+ & 0 & 300 & fcc & (FM)M & HS \\
& & 20 & 300 & fcc & (NM)M & LS\footnote{\textcite{Rueff2001}}\\
(Mg,Fe)O & 2+ & 0 & 300 & NaCl & (PM)I & HS\\
& & 60  & 300 & NaCl & (NM)M & LS\footnote{\textcite{Badro2003,Lin2005,Kantor2006}}\\
& & 80  & 2000 & NaCl & (NM)M & LS\footnote{\textcite{Lin2007}}\\
(Mg,Fe)SiO$_3$ & 2+/3+ & 0 & 300 & Perovskite & PM(I) & HS/HS\\
& & 120 & 300 & Perovskite & (NM)I & LS/LS\footnote{\textcite{Badro2004}}\\
& & 138 & 2500 & Post-perovskite & (NM)I & IS/-\footnote{\textcite{Lin2008}}\\
\hline
CoO & 2+ & 0 & 300 & NaCl & (AF)I & HS \\
& & 100 & 300 & NaCl & (NM)M & LS\footnotemark[1] \\
LaCoO$_3$ & 3+ & 0 & 300 & Perovskite & (PM)I & IS  \\
& & 10 & 300 & Perovskite & (NM)M  & LS\footnote{\textcite{Vanko2006}}\\
La$_{0.72}$Sr$_{0.18}$CoO$_3$ & 3+/4+ & 0 & 34--300 & Perovskite & (PM)M & IS/LS \\
& & 14 & 34--300 & Perovskite & (NM)I & LS/LS\footnote{\textcite{Lengsdorf2007}}\\
\hline
NiO & 2+ & 0 & 300 & NaCl & (AF)I & HS\\
& & (140) & 300 & NaCl & (AF)I & HS\footnotemark[1] \\
\hline\hline
\end{tabular}
\label{table:d-sample}
\end{table*}

\subsubsection{Description in the atomic multiplet approach}
Alternatively, magnetic collapse can be discussed within the atomic multiplet picture which retains the localized $3d$ aspects. The multiplet approach is mostly useful when discussing core-hole spectroscopic data as the core-hole wave-function overlaps strongly with the valence orbitals, leading to strong Coulomb interaction (cf.~\ref{sec:Kbeta}). An extensive description of the  multiplet approach can be found in \textcite{Cowan1981} while its application to core hole spectroscopy is the object of a recent work by \textcite{Groot2008}. Let us consider the case of a free atom with $N$ electrons. The atomic interaction  Hamiltonian is expressed by:
\begin{equation}
\mathcal{H}_{ATOM}=\sum_{pairs}\frac{e^2}{r_ij}+\sum_N\zeta(r_i)l_i\cdot s_i
\label{eq:H_Multiplet}
\end{equation}
It contains the effective electron repulsion and spin-orbit coupling. We have omitted the kinetic energy term of the electrons, the Coulomb interaction with the nucleus and the spherical part of the electronic repulsion which are equivalent for all the electrons. They define the average energy of the electronic configuration while Eq.~(\ref{eq:H_Multiplet}) gives the relative energy of the different states within a given configuration. The configurational energy can be estimated by computing the $\mathcal{H}_{ATOM}$ matrix element. For a $3d^N$ ion with a $^{2S+1}L_J$ term symbol, the Coulomb part is usually expressed in terms of Slater-Condon direct and exchange integrals $F^k,G^k$ ($f_k,g_k$) for the radial (angular) part: 
\begin{equation}
\left\langle^{2S+1}L_J\left|\frac{e^2}{r_ij}\right|{^{2S+1}L}_J\right\rangle
=\sum_k f_kF^k+\sum_k g_kG^k
\end{equation}

The presence of a crystal electric field (CEF) potential $\phi(r)$ is treated as a perturbation to the atomic Hamiltonian. In $d$ electron systems, the CEF strength is larger than the spin-orbit coupling and will strongly affect the energy levels by lifting their degeneracy.  In octahedral ($O_h$) symmetry, the crystal field depends on a unique parameter $10Dq$, defined as the average energy separation between the two crystal field split $d$ orbitals, $t_{2g}$ and $e_g$ (cf.\ Fig.~\ref{fig:hsls}). The energy level splitting as a function of the crystal field is known as the Tanabe-Tsugano diagram.
\begin{figure}[htbp]
\centering
\includegraphics[width=0.90\linewidth]{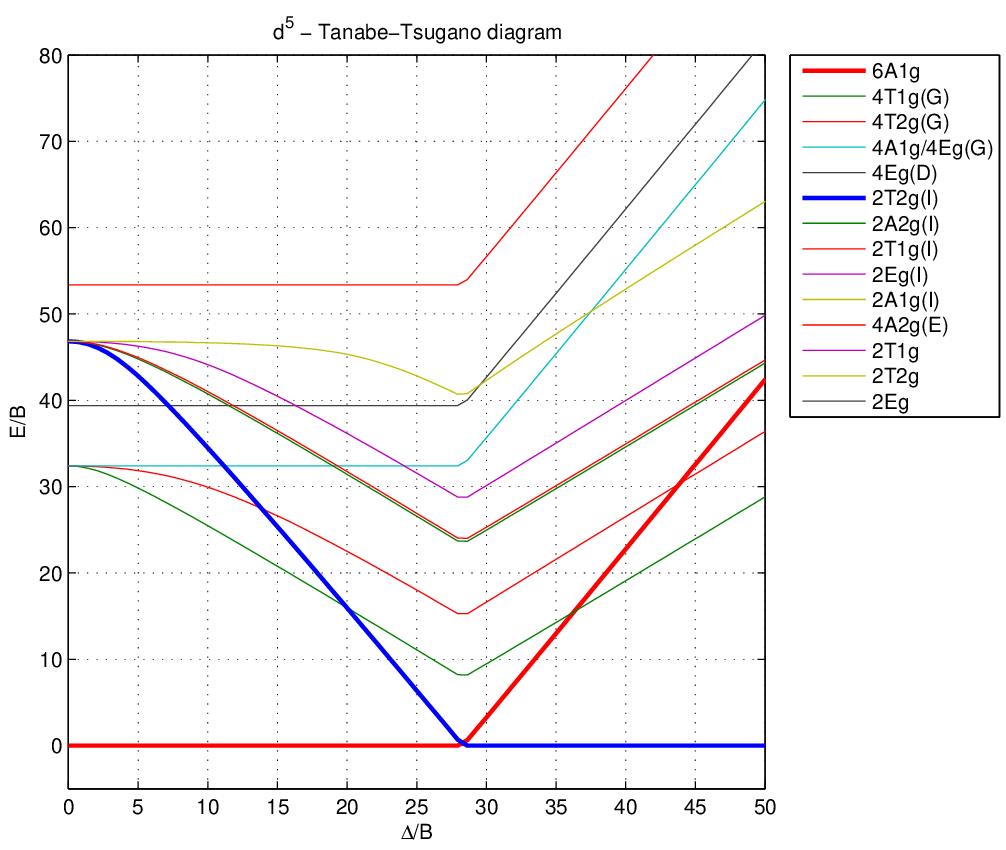}
\caption{(Color online) Tanabe-Tsugano diagram of a $3d^5$ ion in octahedral symmetry. The diagram was computed with the CAMMAG program~\cite{Cruse1979}. The configuration energy ($E$) and crystal field strength ($\Delta$) are normalized by the Racah parameter $B$. Thick lines indicate the ground state in weak and high field limits.}
\label{fig:Tanabe_Tsugano}
\end{figure}
Fig.~\ref{fig:Tanabe_Tsugano} shows as an example the energy level splitting for $3d^5$ ion. The diagram offers a rationale for the magnetic collapse under pressure as the CEF strength depends sensitively on the interatomic distance and therefore on pressure. In our example, the ground state in weak field (low pressure) limit has a $^6A_{1g}$ symmetry with all the five electrons spin up ($S=5/2$). When pressure is applied, the crystal field strength increases as a result of the metal-ligand distance shortening, eventually resulting in a high spin (HS) to low spin (LS) transition and spin pairing. The ground state term changes to $^2T_{2g}$ and the spin moment diminishes to $S=1/2$. In this picture, the magnetic collapse therefore results from a competition between the crystal field and the exchange interaction. As an intra-atomic property, the latter is barely affected by pressure in contrast to the CEF, and the magnetic collapse occurs when the CEF strength overcomes the magnetic exchange. The picture is summarized in Fig.~\ref{fig:hsls}.
\begin{figure}[htbp]
\centering
\includegraphics[width=0.90\linewidth]{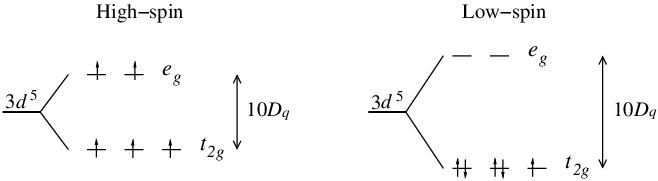}
\caption[HS-LS in $3d^5$ ion]{Example of high-spin ($S=5/2$) and low spin ($S=1/2$) configurations for a $3d^5$ metal ion in octahedral symmetry. $10D_q$ is the crystal field strength.}
\label{fig:hsls}
\end{figure}

Experimental results of pressure-induced magnetic collapse in transition-metal as well as the relative influence of crystal field vs.\ bandwidth will be extensively discussed in sections~\ref{sec:MIT} to \ref{sec:thermal_excitation}. Table~\ref{table:d-sample} gives a summary of the results obtained so far in transition metal compounds under pressure along with their main physical properties.

\subsection{XES at the K$\beta$ line}
\label{sec:Kbeta}
Besides other techniques conventionally devoted to magnetism, x-ray emission spectroscopy can be used as an alternative probe of the transition-metal magnetism. XES is well suited to high-pressure studies thanks to the intense fluorescence yield in the hard x-ray energy range, especially when combined with bright and focused x-ray beams provided by third-generation synchrotron sources. More particularly, the K$\beta$ (${3p}\rightarrow{1s}$) emission line from the transition metal atom (and to a less extent the K$\alpha$ (${2p}\rightarrow{1s}$) line) turns out to be extremely sensitive to the transition metal spin state. 

As we will discuss below in details, the overall spectral lineshape of the K$\beta$ line in transition metal consists of an intense main line (K$\beta_{1,3}$) and a satellite structure (K$\beta'$) located on the low energy side. The satellite has been successively proposed to arise from exchange interaction~\cite{Tsutsumi1976}, shake-up or plasmon phenomena, and charge transfer effects~\cite{Kawai1990}, before being attributed correctly to the multiplet structure~\cite{Groot1994,Peng1994a,Wang1997}.\\

To illustrate our purpose, we now consider the case of a Fe trivalent ion ($3d^5$) in octahedral ($O_h$) symmetry which exhibits particularly clear spectral changes.

\subsubsection{Example of a $3d^5$ ion}
\begin{figure}[htbp]
\includegraphics[width=0.90\linewidth]{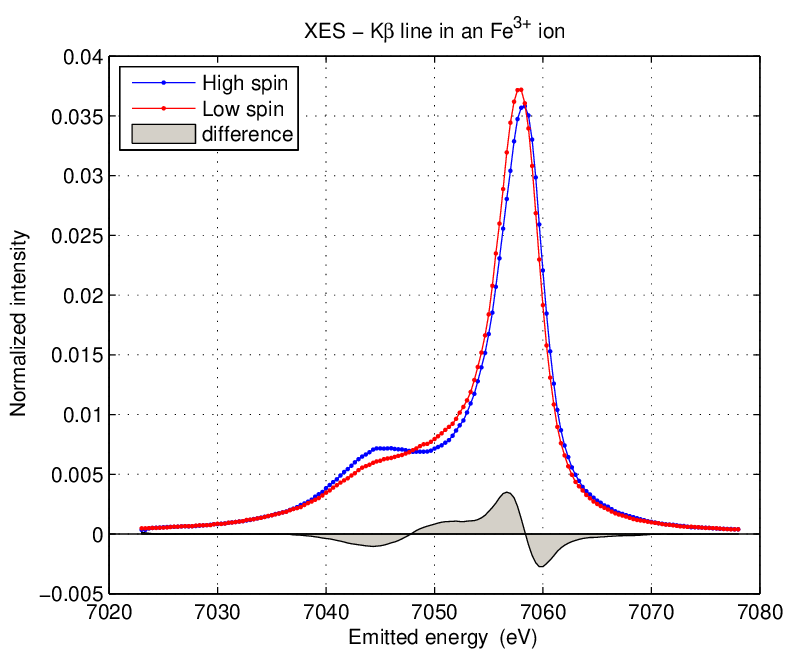}
\caption[XES in Fe]{(Color online) Fe K${\beta}$ emission line through a pressure-induced high-spin to low-spin transition in a $3d^5$ Fe-based molecular complex. The spectra are normalized to the integral. HS-LS difference spectrum is shown in gray.}
\label{fig:Fe_XES}
\end{figure}

The Fe$^{3+}$ K$\beta$ emission line from \textcite{Vanko2006a} is shown in Fig.~\ref{fig:Fe_XES} for both high spin and low spin configurations. They were measured in a Fe$^{3+}$ spin-crossover compound where the spin transition is driven by temperature change. The spectra are normalized to the integral. At the spin state transition, the low energy satellite decreases while its spectral weight is transferred to the main peak which increases. The modification of the lineshape is correlated with a slight energy shift which ensures that the spectrum center of mass stays constant.

\begin{figure}[htbp]
\includegraphics[width=0.90\linewidth]{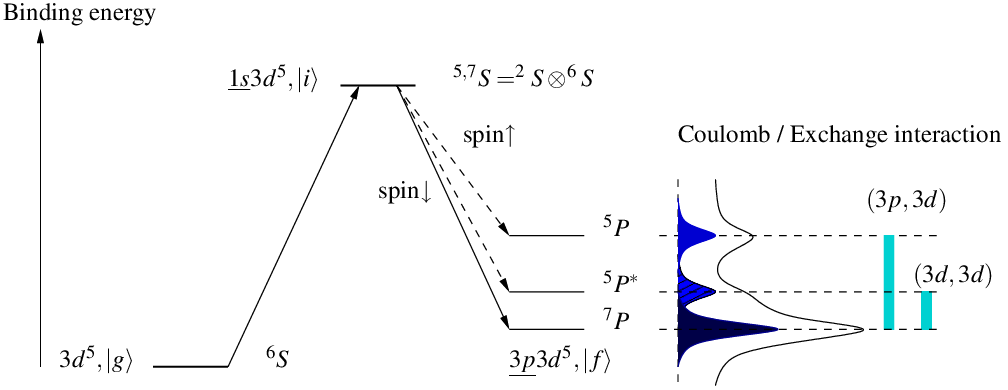}
\caption[Kbeta process]{(Color online) K${\beta}$ process on a configuration level scheme in the case of a $3d^5$ ion. $\left|g\right\rangle,\left|i\right\rangle,\left|f\right\rangle$ denote the ground state, intermediate and final states. Spectroscopic terms are indicated.}
\label{fig:Kbeta_process}
\end{figure}

Fig.~\ref{fig:Kbeta_process} illustrates formally the K$\beta$ XES process in a configuration scheme for a $3d^5$ ion. The initial state is formed by a $1s$ core-hole ($^2S$ configuration) which couples to the $3d$ states. For a $3d^n$ ion with a $^{2S+1}L$ configuration,  the $1s$-$3d$ exchange interaction splits the degenerate ground state into two configurations of high-spin ($^{2S+2}L$) and low-spin symmetry ($^{2S}L$). Thus in a $3d^5$ ($^6S$) configuration, one expects two intermediate states of $^5S$ and $^7S$ symmetry by application of the $^2S\otimes^6S$ cross product. In the final state, the $3p$-$3d$ exchange interaction leads to two ``spin-polarized'' configurations,  ($\underline{3p^{\uparrow}}3d^{\uparrow}$) and  ($\underline{3p^{\downarrow}}3d^{\uparrow}$), according to the two possible spin orientations for the $3p$ hole with respect to spin-up $d$ electrons.
In our example, two final states of $^7P$ and $^5P$ symmetry are formed leading to a main peak (K$\beta_{1,3}$) and a satellite (K$\beta'$) structure that characterizes the emission spectrum. An additional feature ($^5P^*$) is found in the spin-up channel. It is due to a spin-flip excitation in the $3d$ band and shows up as a shoulder to the main peak. Configuration interaction both in the initial and final states may lead to mixing of states, ending in a complicated multiplet structure. But the spread of the multiplet terms is nevertheless dominated by the $3p$-$3d$ exchange interaction because of the strong $3p$ overlap with the $3d$ states. As originally proposed by~\textcite{Tsutsumi1976}, the energy difference between the main peak and the satellite is, to a crude approximation, proportional to $G^{1,3}(3p,3d)(2S+1)$, where $G^{1,3}(3p,3d)$ is the Slater exchange parameter between the electrons in the $3p$ and $3d$ shells (of the order of 15 eV). The $3p$ spin orbit splits the states further within $\sim$1~eV. Through a magnetic collapse transition, the $3d$ magnetic moment abruptly changes and so does the K$\beta$ lineshape (see Fig.~\ref{fig:Fe_XES}). Note that the final state splitting is less clear in the K$\alpha$ XES because of the weaker $2p$-$3d$ overlap. 

Thus, XES appears as a local probe of the $3d$ magnetism. No external magnetic field is required since XES benefits from the intrinsic spin-polarization of the $d$ electrons. In the following, we will review XES results obtained in various transition metal compounds under high pressure (and temperature) conditions. Starting from a purely phenomenological approach, we will see that the variation of the XES lineshape across a magnetic transition is well accounted for by full multiplet calculations including ligand field and charge transfer effects providing valuable information concerning $d$ electrons properties under extreme conditions.

\subsubsection{Integrated absolute difference}
\begin{figure}[htbp]
\includegraphics[width=0.90\linewidth]{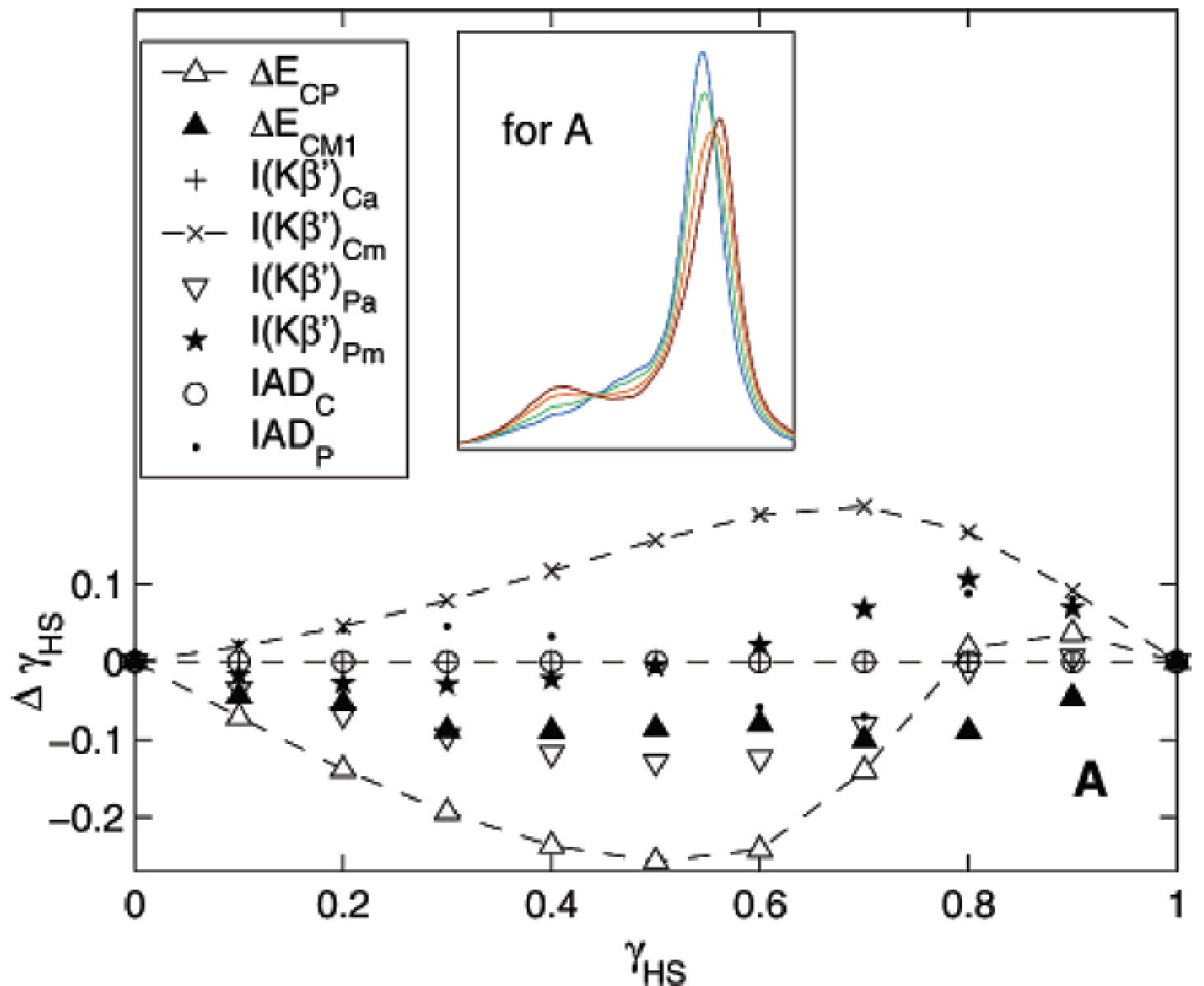}	
\caption[IAD vs.\ HS fraction]{(Color online) Deviations of the HS fraction $\gamma_{HS}$ as determined from different extraction techniques (see details in the cited reference) in simulated XES spectra of known spin state. The spectra are constructed from a linear superposition of theoretical HS and LS spectra in a Fe$^{2+}$ ion with a HS weight $\gamma_{HS}$ (inset). The IAD values (open circles and dots) show the best results. Adapted from~\textcite{Vanko2006a}.}
\label{fig:IAD_Vanko}
\end{figure}

Unfortunately, the magnetic information contained in the XES spectra is not immediately available. In the absence of formal sum rules such as in x-ray magnetic circular dichroism (XMCD), one is restricted to using a more approximate approach. The changes of the local magnetic moment can be estimated from the integrated absolute difference (IAD)~\cite{Rueff1999a,Vanko2006,Vanko2006a} which relates the spectral lineshape to the $3d$ spin-state as follows :
\begin{equation}
\mathrm{IAD}=\int\left|I_{XES}(\omega,P)-I_{XES}(\omega,P_0)\right|d\omega,
\label{eq:IAD}
\end{equation}
where $I_{XES}(P)$ is the intensity of the x-ray emission at a given pressure $P$ and $P_0$ a reference pressure point. The IAD is a phenomenological analysis but shows a remarkable agreement when compared to model systems. \textcite{Vanko2006a} have applied the IAD analysis to Fe$^{2+}$ spectra of known spin state (Fig.~\ref{fig:IAD_Vanko}). The spectra were constructed from a linear combination of $\gamma_{HS}$ HS and $(1-\gamma_{HS})$ LS XES spectra. The deviation $\Delta\gamma_{HS}$ of the extracted high spin fraction compared to the nominal values is negligible. 

\subsubsection{Temperature effect}
\label{sec:thermal_broadening}
\begin{figure}[ht]
\includegraphics[width=0.90\linewidth]{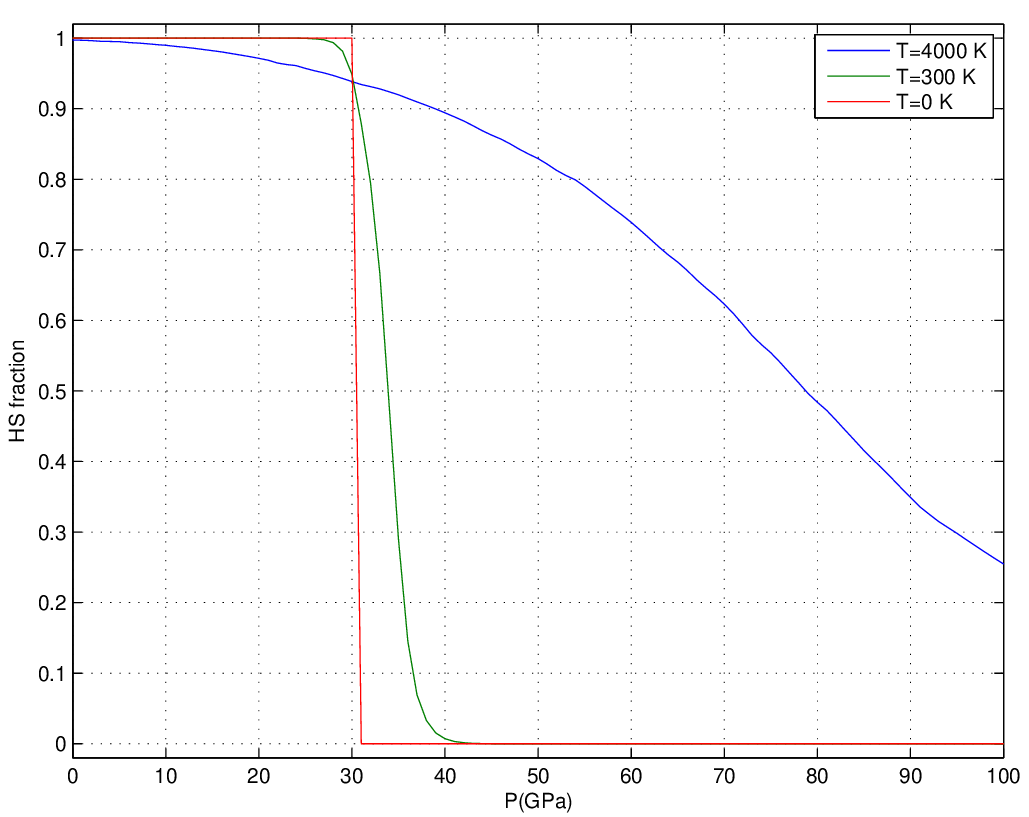}	
\caption[HS fraction in Fe$^{2+}$]{(Color online) HS fraction in Fe$^{2+}$. Pressure dependence of enthalpies was borrowed from~\textcite{Tsuchiya2006}.}
\label{fig:HS_frac}
\end{figure}

At a given pressure, excited spin states of energies within $k_BT$ from ground state will mix. At equilibrium, the spin population in a pure atomic approach can be described by considering an assembly of ions in a series of spin states $i$ defined by their enthalpy $H_i$ and degeneracy $g_i$. The fraction of ions in the $i$-state is expressed by Eq.~(\ref{eq:spin_frac}), assuming a Boltzmann statistics ($\beta=1/k_BT$). The main effect of temperature is to broaden the spin transition as illustrated in Fig.~\ref{fig:HS_frac}. According to this simulation, a broadening of the transition is already observed at room temperature, where most of the measurements were performed. However, the smearing due to thermal excitations becomes dominant only in the very high temperature region. 
\begin{equation}
n_i=\frac{1}{1+\sum_{j\neq i}\frac{g_j}{g_i}\exp[-\beta(H_j-H_i)]}	
\label{eq:spin_frac}
\end{equation}

\subsection{Connection to Metal-Insulator transition} 
\label{sec:MIT}
We now turn to experimental results about high-pressure magnetic properties of strongly correlated electrons with an emphasis on transition metal oxides. Because these are usually wide gap insulators with antiferromagnetic correlations, we expect magnetic collapse occurs at very high pressure in close connection with metal-insulator transition.
 
\subsubsection{Transition-metal monoxides}
\label{sec:mono_MO_HP}
\begin{figure}[htbp]
\includegraphics[width=0.90\linewidth]{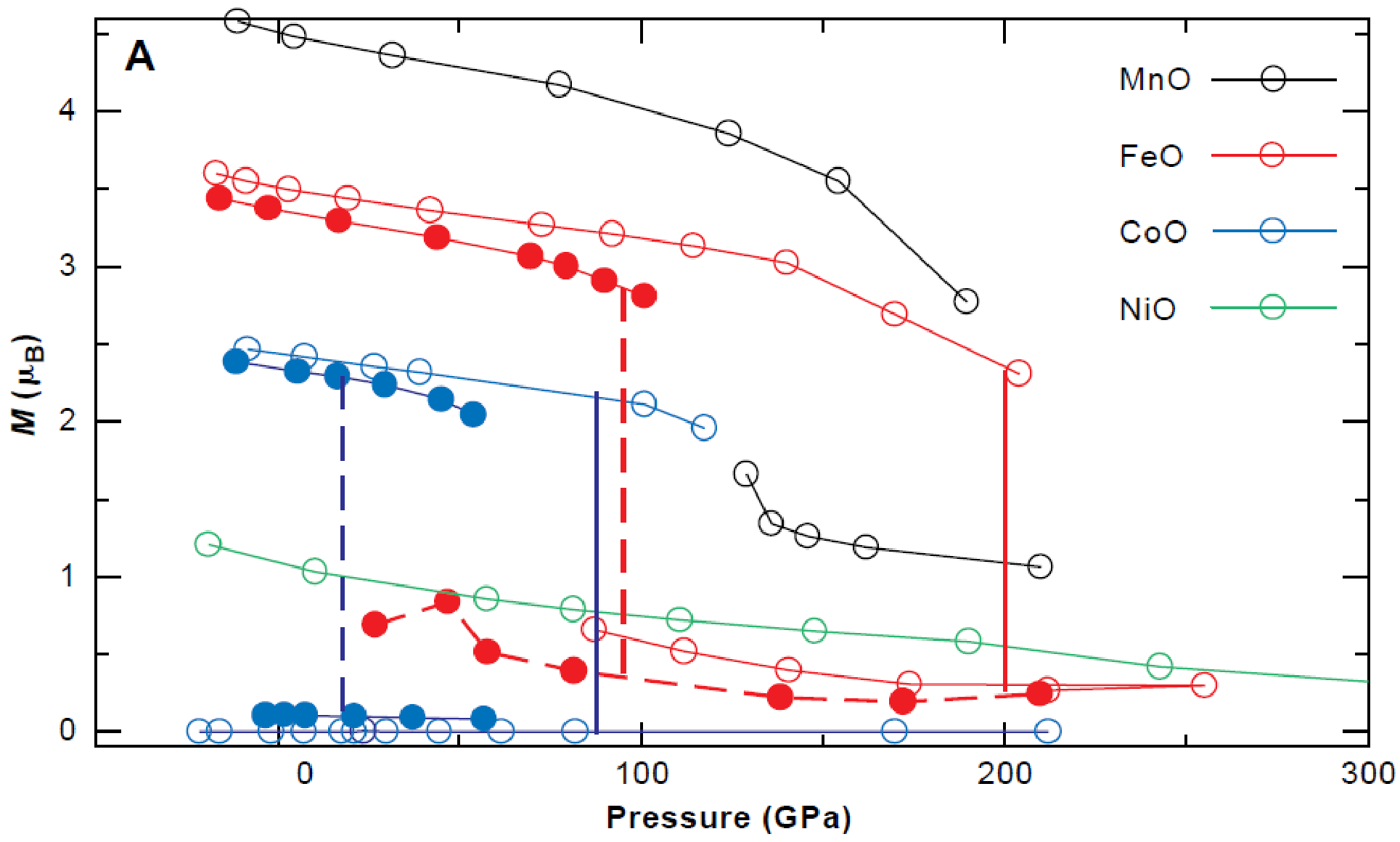}
\caption[MO Calculated magnetic moment]{(Color online) Calculated magnetic moment in transition metal oxide as a function of pressure; GGA (open circles) and LDA (solid circles) results. The vertical lines denote the calculated  transition pressures. From~\textcite{Cohen1997}.}
\label{fig:Cohen}
\end{figure}

Using Local Density Approximation (LDA) and Generalized Gradient Approximation (GGA) band structure approaches, \textcite{Cohen1997,Cohen1998} have performed systematic calculations  of the magnetic moment of the series of early transition metal monoxides (MnO, FeO, CoO, and NiO) which are prototypes of strongly correlated materials and the simplest transition metal oxide systems. At ambient conditions, the monoxides are insulators of mostly charge-transfer character~\cite{Hufner1994,Bocquet1992a} ($\Delta<U$) with large $U$ of about 5--10 eV. The authors argue that at high pressure LDA still holds as the system becomes metallic as $U/W<\sqrt{N}$ (with $N$ the $d$ orbital degeneracy) and  the magnetic stability was checked using a refined form of the Stoner criterion of Eq.~(\ref{eq:Stoner}) including the spin-polarization of the density of states. Fig.~\ref{fig:Cohen} shows the variation of the calculated magnetic moment with pressure. In MnO, CoO, and FeO, the calculations yield a LS ground state at high pressure. 
Note that in the latter, the HS-LS transition is pushed to very high pressures when using LDA+U compared to LDA or GGA~\cite{Gramsch2003}.
No spin transition occurs in NiO as expected from a $3d^8$ configuration in $O_h$ symmetry. The transition pressures, indicated by vertical bars in Fig.~\ref{fig:Cohen}, roughly coincide with the experimental metal-insulator transition (or associated structural transition) pressures in these materials. 

Magnetic collapse has been investigated in MnO, FeO, CoO, and NiO in the megabar range~\cite{Mattila2007} by XES. The experimental conditions are given in~\textcite{Rueff2005}.
\begin{figure}[htbp]
\includegraphics[width=0.80\linewidth]{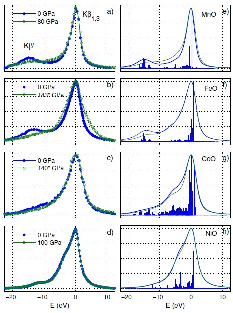}
\caption[K$\beta$-XES spectra in MnO, FeO, CoO and NiO]{(Color online) (a--d) K$\beta$-XES spectra measured in MnO, FeO, CoO and NiO in both low (close circles) and high pressure (open circles) phases. (*) indicates spectra obtained after offline laser-heating. (e--h) Calculated spectra at ambient pressures (thick) and high pressures (thin lines). Ticks represent the multiplet states, before broadening. From~\textcite{Mattila2007}.}
\label{fig:kbeta4}
\end{figure}
Fig.~\ref{fig:kbeta4}(a--d) summarizes the K$\beta$ emission spectra measured in the transition-oxides series at low and high pressures. All the spectra but NiO show significant modifications in the lineshape, essentially observed in the satellite region, through the transition\footnote{The asymmetric broadening of the main line in FeO at 140 GPa is an artifact}. Following the description put forward by \textcite{Peng1994a} within the atomic multiplet formalism, the satellite is expected to shrink with decreasing $3d$ magnetic moment, and move closer to the main peak. This agrees well with the observed spectral changes when going from MnO ($3d^5$, $S=5/2$) to NiO ($3d^8$, $S=1$). It also qualitatively accounts for the collapse of the satellite at high-pressure observed in MnO, FeO and CoO, viewed as the signature of the HS to LS transition on the given metal ion. The unchanged spectra in NiO, where no such transition is expected, confirm the rule. 

The atomic description however omits the crucial role played by the O($2p$)-M($3d$) charge-transfer effects and finite ligand bandwidth. To take these into account full multiplet calculations within the Anderson impurity model~\cite{Kotani2001,Groot2001} can be used. In contrast to band-like treatment of $d$ electrons, crystal-field, ligand bandwidth, and charge transfer are here explicitly introduced as parameters. The model, derived from the configuration interaction approach, was first put forward to explain the core-photoemission spectra of transition metals~\cite{Zaanen1986,Mizokawa1994}. It was later applied to the K$\beta$ emission line in Ni-compounds~\cite{Groot1994} and more recently in transition metal oxides~\cite{Tyson1999,Glatzel2001,Glatzel2004}. The multiplet calculation scheme yields an accurate model of the emission lineshape. More interestingly, it allows a direct estimate of the fundamental parameters, which are adjusted in the calculations with respect to the experimental data. 
\begin{table}[htbp]
\caption[Calculations parameters in monoxides]{Parameters used in the calculations (in eV): $\Delta$ is the charge transfer energy; $U$ the $d$-$d$ correlation; $V_{e_{g}}$ the hybridization strength; the crystal-field splitting is given by $10Dq$, and the core-hole Coulomb interaction by $U_{dc}$; $W_{2p}$ denotes the ligand bandwidth.}
\begin{tabular}{lcccccc}
\hline
\hline
P (GPa) & $\Delta$ &  U & $V_{e_{g}}$ & $10Dq$ & $U_{dc}$	& $W_{2p}$ \\
\hline
\multicolumn{6}{l}{MnO}\\
0 (HS) & 5 & - & 2.2 &  1 & 10 & 3 \\
80 (LS) & 6 & - & 3.06 &  1.6 & 10 & 4 \\
100 (LS) & 6 & - & 3.7 & 2.3 & 10 & 6 \\
\hline
\multicolumn{6}{l}{FeO}\\
0 (HS) & 5  & -  & 2.4  & 0.5 & 7 &3  \\
140 (LS) & 5 & -  & 3.2  & 0.8 & 7 & 9  \\
\hline
\multicolumn{6}{l}{CoO}\\
0 (HS) & 6.5 & 6 &  2.5 & 0.7 & 7 & 4  \\
140 (LS) & 6.5 &  6 &  4.2 & 1.2 & 7 & 9 \\
\hline
\multicolumn{6}{l}{NiO}\\
0  & 3.5 & 8.2   & 2.4  & 0.3  & 9 & 5  \\
140 & 4.5  & 9.2  & 3  & 0.65 & 9 & 7.5  \\
\hline
\hline
\end{tabular}
\label{tab:MO_param}
\end{table}

Inclusion of the charge-transfer in the multiplet calculations substantially improves the simulated main-peak to satellite intensity ratio, via a transfer of spectral weight to the main peak. In the cluster model charge-transfer enters the calculations through a configuration interaction scheme. Details on the computational method is given in \textcite{Mattila2007}. 
%
The model parameters were first chosen to reproduce the emission spectra at ambient pressure and subsequently fitted to the high pressure data. The parameters, charge-transfer energy $\Delta$, hybridization strength in the ground state ($V_{e_{g}}$ and $V_{t_{2g}}$), the ligand bandwidth $W$(O-$2p$), the on-site Coulomb interaction $U$ and the crystal field splitting $10Dq$ are summarized in table  \ref{tab:MO_param}. $U$ is a second order perturbation of the spectral lineshape and was dropped in the calculations when not necessary. 
\begin{figure}[htb] 
\includegraphics[width=0.90\linewidth]{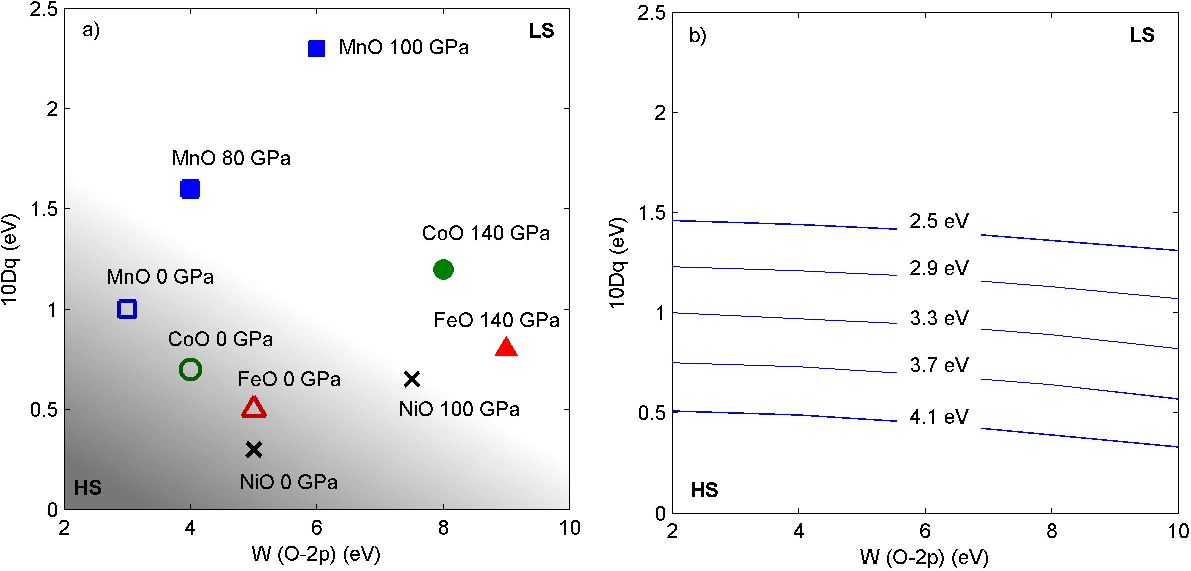}
\caption[Phase diagram in M-oxides]{(Color online) Phase diagram of the magnetic collapse in the transition-metal oxides. a) The point coordinates refer to calculated values of the crystal-field splitting $10D_q$ and ligand bandwidth $W(O-2p)$, both in the HS (open symbols) and LS (closed symbols) states, as obtained from comparison with the experimental spectra. b) Solid lines mark the calculated HS-LS transition boundary for CoO for different values of $V_{e_{g}}$. From~\textcite{Mattila2007}.}
\label{fig:phase_diag} 
\end{figure}

Fig.~\ref{fig:kbeta4}(e-h) shows the calculated spectra for both the ambient and high pressure phases. For MnO, CoO and FeO the calculations in the highest pressure phases yield a LS ground state, even though the high spin state multiplet stays energetically close. An intermediate regime is therefore expected where both HS and LS states coexist on the same ion, which indeed has been observed in several transition metals (cf.\ \ref{sec:Fe_HP}) and oxides (cf.\ \ref{sec:MgFeO}). The HS-LS transition is then understood as resulting from the conjugated effects of increase of the crystal-field parameter $10Dq$ \emph{and} a broadening of the O-($2p$) bandwidth $W_{2p}$ together with an increasing covalent contribution from the hybridization to the ligand field at high pressures. The increase of $10Dq$ is seen as the driving force toward a LS state. It traces back to the atomic description of the magnetic collapse. More notable is the interplay of the ligand bandwidth together with the increased hybridization. The parallel evolution of these parameters across the magnetic collapse transition is represented in a phase diagram, Fig.~\ref{fig:phase_diag}. The lines mark the calculated HS-LS transition boundary for CoO for different values of $V_{e_{g}}$. The results illustrate the dual behavior---both localized and delocalized---of the correlated $d$ electrons at extreme conditions.

\subsubsection{Fe$_2$O$_3$: Magnetic metastable states}
\label{sec:Fe2O3_HP}
\begin{figure*}[htbp]
\begin{tabular}{cc}
{\small a)}\includegraphics[width=0.45\linewidth]{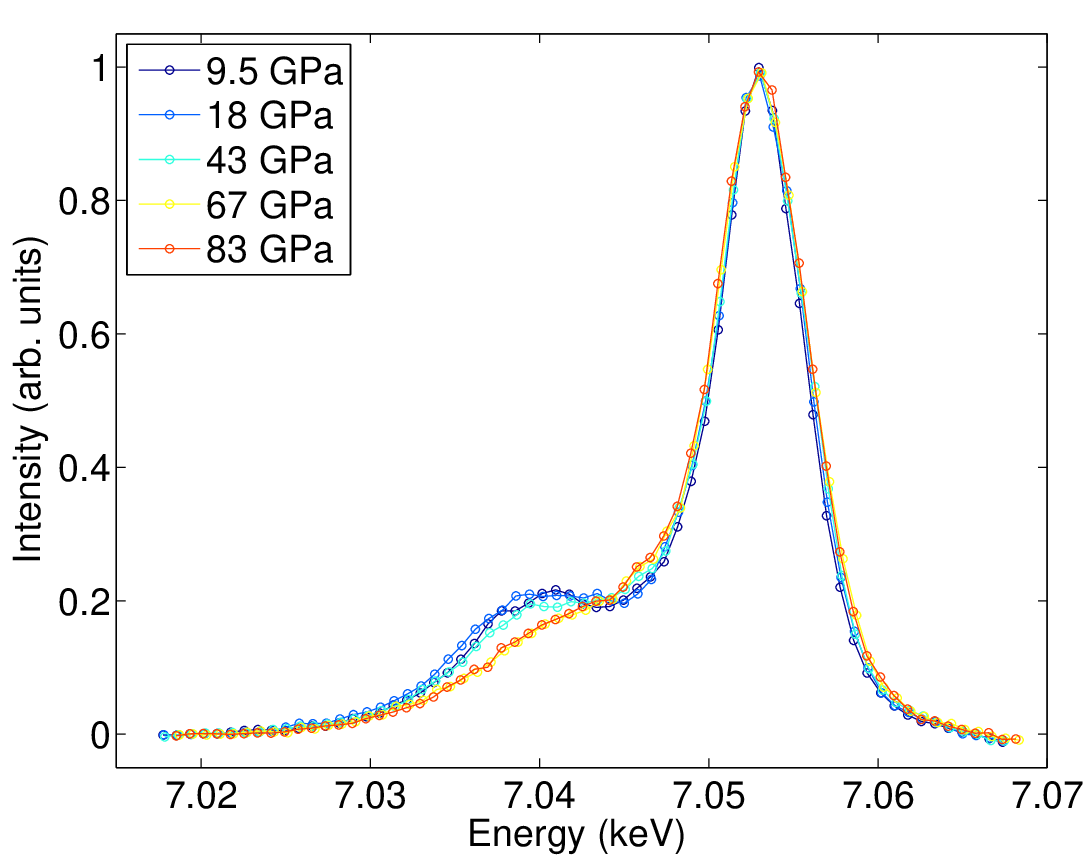} &
{\small b)}\includegraphics[width=0.45\linewidth]{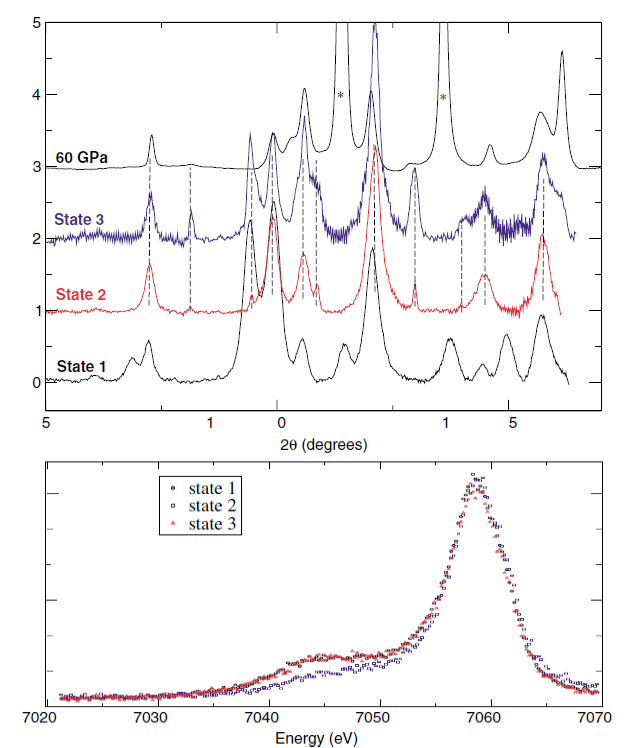} \\
\end{tabular}
\caption[Fe$_2$O$_3$ XES]{(Color online) a) Fe K$\beta$ emission in Fe$_2$O$_3$ as a function of pressure; b) Diffraction patterns through the metastable phase transition and corresponding XES spectra. From~\textcite{Badro2002}.}
\label{fig:Fe2O3}
\end{figure*}

Hematite is another wide-gap AF insulator archetypical of Mott localization but presents a distinct behavior from the monoxides as it involves reportedly metastable spin states. The stable $\alpha$-phase at ambient pressure crystallizes in the corundum structure. At a pressure of about 50 GPa, Fe$_2$O$_3$ transforms into a low-volume phase of still debated nature. A recent M\"ossbauer study has demonstrated that the structural change is accompanied by an insulator to metal transition, understood by the closure of the correlation gap and the emergence of a non-magnetic phase~\cite{Pasternak1999}. The abrupt magnetic collapse at 50 GPa was confirmed by XES experiment at the Fe K$\beta$ line (cf.\ Fig.~\ref{fig:Fe2O3}(a)) again showing  the relationship between the Mott transition and magnetic collapse as in the monoxides. More interestingly, in a similar XES experiment at the Advanced Photon Source the evolution of the spectral lineshape was measured simultaneously with the x-ray diffraction pattern~\cite{Badro2002}. Hematite was compressed to 46 GPa without noticeable change of spin or structure (state 1 in Fig.~\ref{fig:Fe2O3}(b)). At that point, the sample was laser heated using an offline Nd:YAG laser and immediately quenched in temperature, leaving it in a metastable state (state 2) characterized by a LS state and a structure typical of the high pressure phase. After relaxation, the electronic spin state reverts to the initial HS state while the structure stays unchanged showing that the LS magnetic state is not required to stabilize the high pressure structural phase.

\subsubsection{Measuring the insulating gap}
In correlated materials, low lying excited states involve charge excitations across the correlation or charge transfer gaps which are clearly relevant to the physics of metal-insulator transitions. Laser-excited optical reflectivity yields such information~\cite{Syassen1982a} with high pressure compatibility and excellent resolution.  Alternatively, RIXS can be used to measure these low energy excitations, as explained in section~\ref{sec:RIXS}. In contrast to the optical response, the x-ray measurements are carried out at finite $\mathbf{q}$ which means that potentially this method could be used to study dispersion. We comment on results obtained in the transition metal monoxides under pressure similar to the  studies of section~\ref{sec:mono_MO_HP}, but here using RIXS.

\paragraph{NiO}
The first example of such measurements concerns NiO which however shows no  magnetic collapse. NiO nevertheless is a prototype charge-transfer insulator with a band gap of about 4 eV~\cite{Hufner1994}. Several calculations of the electronic structure of NiO exist as this has proved to be a good testing ground for theories due to the influence of strong correlations. Pressure-induced electronic changes may provide complementary information on the correlated state because it involves high electrons density and modify their motion through the lattice. 

\begin{figure*}[htbp]
\begin{tabular}{cc}
{\small a)}\includegraphics[width=0.45\linewidth]{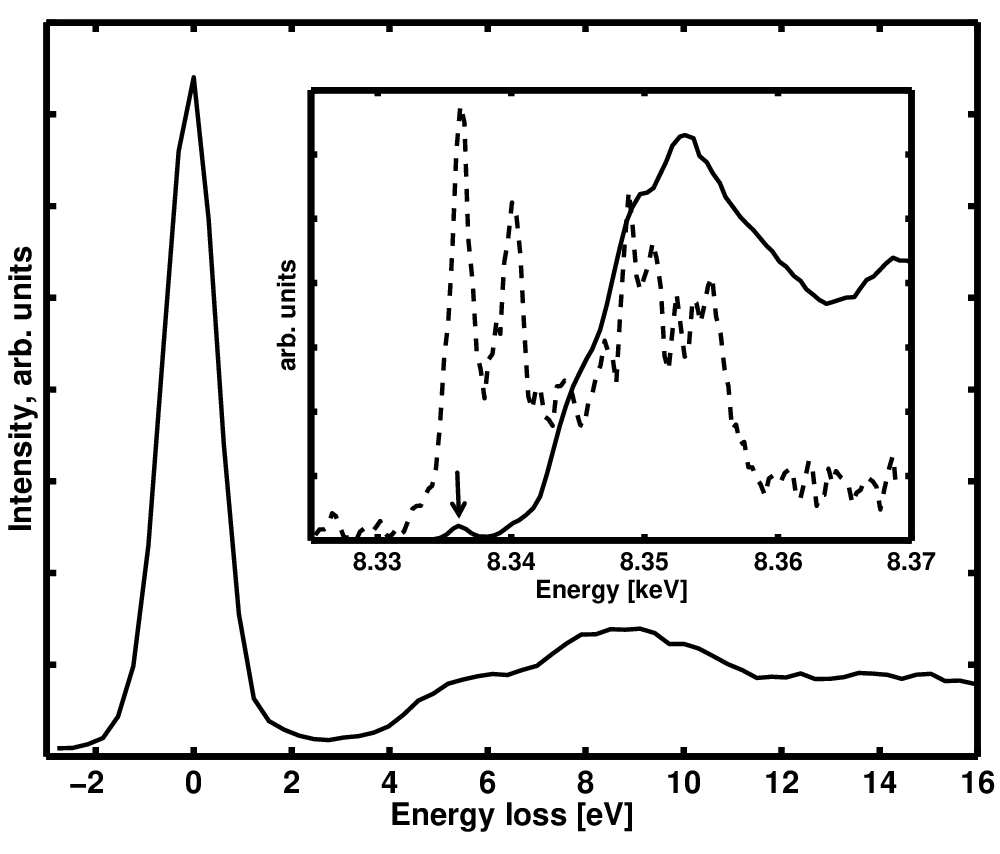} &
{\small b)}\includegraphics[width=0.45\linewidth]{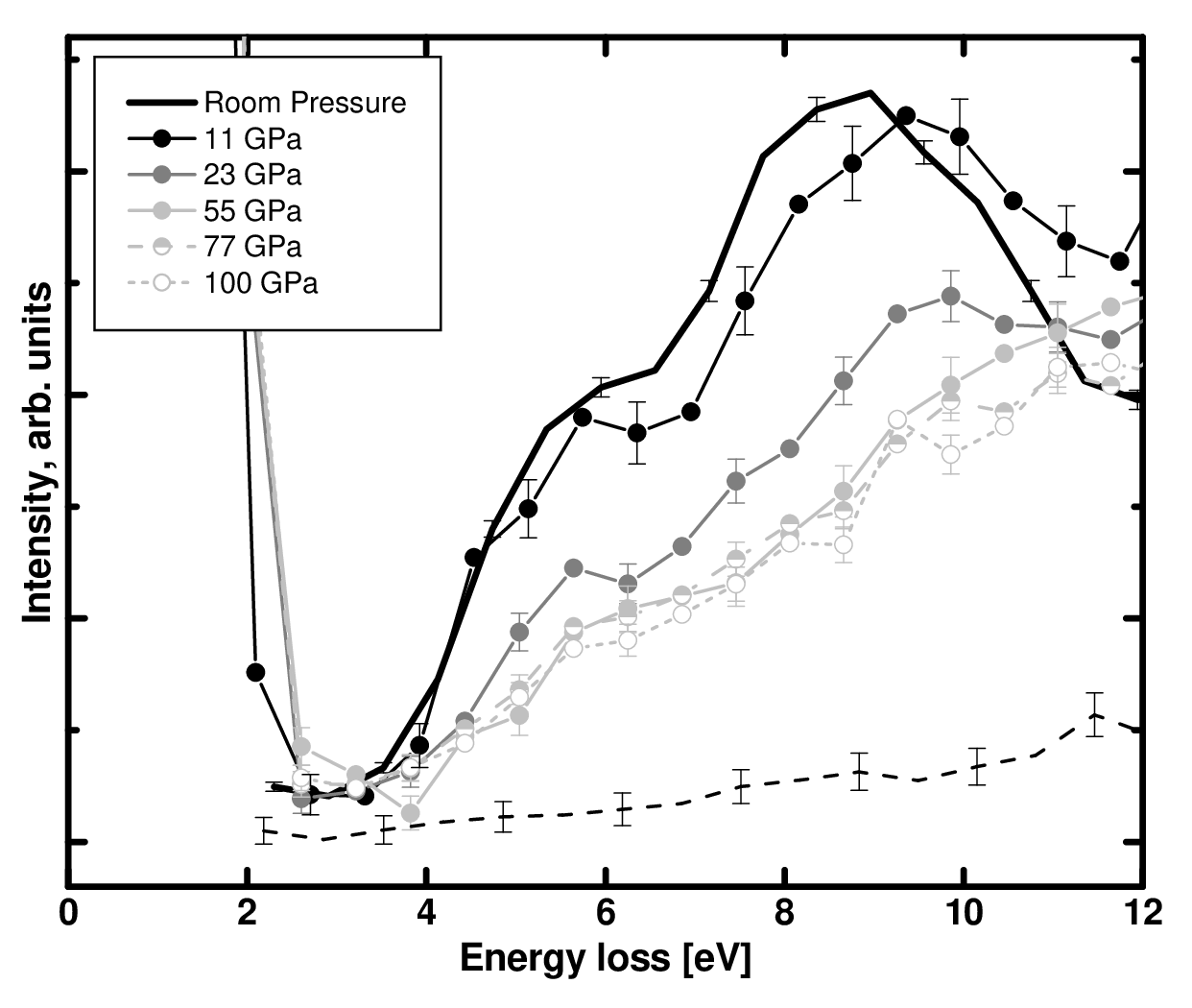} \\
\end{tabular}
\caption[RIXS in NiO]{a) RIXS in NiO at ambient pressure; the incident energy is tuned to the quadrupolar pre-peak in the absorption spectrum (arrow in inset); Inset: PFY-XAS spectrum of the Ni K-edge (solid line) with an arrow showing the choice of incident energy, and constant-final-state scan (dashed line) at an energy loss of 5 eV. b) Dependence of the RIXS spectra as a function of pressure; the dashed line is the non-resonant background. From~\textcite{Shukla2003}.}
\label{fig:NiO}
\end{figure*}

RIXS in NiO was carried out under very high pressure conditions~\cite{Shukla2003}. In Fig.~\ref{fig:NiO}, the RIXS spectra was obtained by tuning the incident energy to the pre-peak resonance with the following two benefits: quadrupolar transitions (favored at large scattering angles) are associated to the pre-peak and the lowest-energy excited states relevant to the electronic properties of NiO are the $3d^{n+1}$ configurations.

The RIXS spectrum consists of two peaks, centered around 5.3 and 8.5 eV above the elastic line. Following the interpretation of \textcite{Kao1996}, the first feature is associated to the charge transfer excited state $d^{n+1}\underline{L}$ where $\underline{L}$ denotes a ligand hole; the energy loss to the edge of this shoulder corresponds to the charge-transfer gap in NiO. The nature of the second peak at 8.5 eV is less clear. It can be tentatively ascribed to the metal-metal transitions leading to $d^{n+1}d^{n-1}$ excited states and thus to the correlation energy $U$. 
As pressure is increased, the RIXS features progressively decreases. Secondly the double structure (shoulder and peak) clearly resolved at ambient and lower pressures, smears at pressures above 50~GPa into a poorly-defined line-shape. This tendency primarily reflects the increasing band dispersion at high pressure. In particular overlap with the ligand states increases since the lattice parameter changes by about 10\% at 100 GPa. Calculations suggest that the shape of the electronic density of states does not change much with pressure but the density of states decreases uniformly and band width increases \cite{Cohen1997}. This is compatible with the behavior of the 5.3 eV peak which increases in width without appreciable change in position. It is the increase in width, or the increased dispersion which reduces the value of the charge transfer gap. The behavior of the 8.5 eV peak would suggest an initial growth of the $d$-$d$ Coulomb interaction with pressure. Though this is unexpected since screening increases with pressure, it seems to be limited to the lower pressure regime. Finally the observed trends suggest that a metal-insulator transition would happen mainly due to the closing of the charge-transfer gap as predicted by theory in NiO~\cite{Feng2004} (in addition to band-broadening and crystal-field effects) and in other charge-transfer insulators at lower pressures~\cite{Dufek1995a}.

\paragraph{CoO} 
\begin{figure}[htbp]
\includegraphics[width=0.90\linewidth]{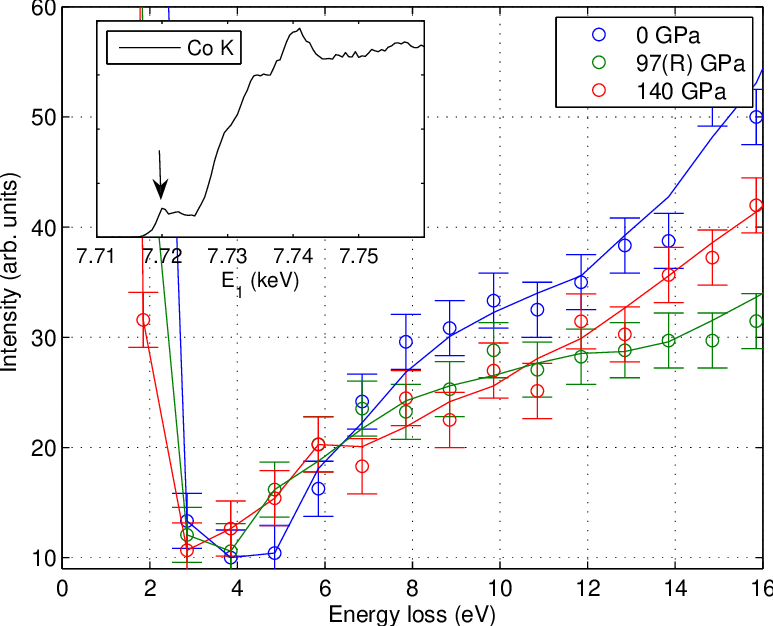}
\caption[RIXS in CoO]{(Color online) RIXS in CoO as a function of pressure (open circles); lines are 3-point average of the data; the incident energy is tuned to the quadrupolar pre-peak in the absorption spectrum (arrow in inset). (R) indicates pressure release.}
\label{fig:CoO}
\end{figure}
The insulating gap in CoO is supposedly intermediate between Mott-Hubbard and charge-transfer type. RIXS spectra obtained in CoO under pressure are illustrated in Fig.~\ref{fig:CoO}. The incident energy was tuned to the pre-edge region in the absorption spectra (shown in inset). In the ambient pressure spectrum, a sharp increase is observed around 6~eV energy loss, characteristic of the insulating gap. No other RIXS features show up at higher energy contrary to NiO. Following the interpretation of the RIXS spectra in the latter, this could indicate that $U$ and $\Delta$ are of similar magnitude in CoO, in agreement with~\textcite{Shen1990}. The spectral features are smeared out upon pressure increase, while the gap region is filled. The tendency points to a metal-insulator transition which could occur in the megabar range. This is consistent with the magnetic collapse pressure deduced from XES. 

Note that, in these experiments, NiO and CoO powder samples were loaded in the pressure cell without a transmitting medium. Though the powder to some extent preserves hydrostaticity, a pressure-gradient of about 10\% is expected in the megabar range; an estimate of the pressure gradient in FeO yields 10 GPa at 135 GPa~\cite{Badro1999}.  Future, better, setups would consist of single-crystal samples loaded with gaseous He which is hydrostatic up to several 100 GPa.

\subsection{Magnetovolumic effects}
\begin{figure}[htbp]
\includegraphics[width=0.90\linewidth]{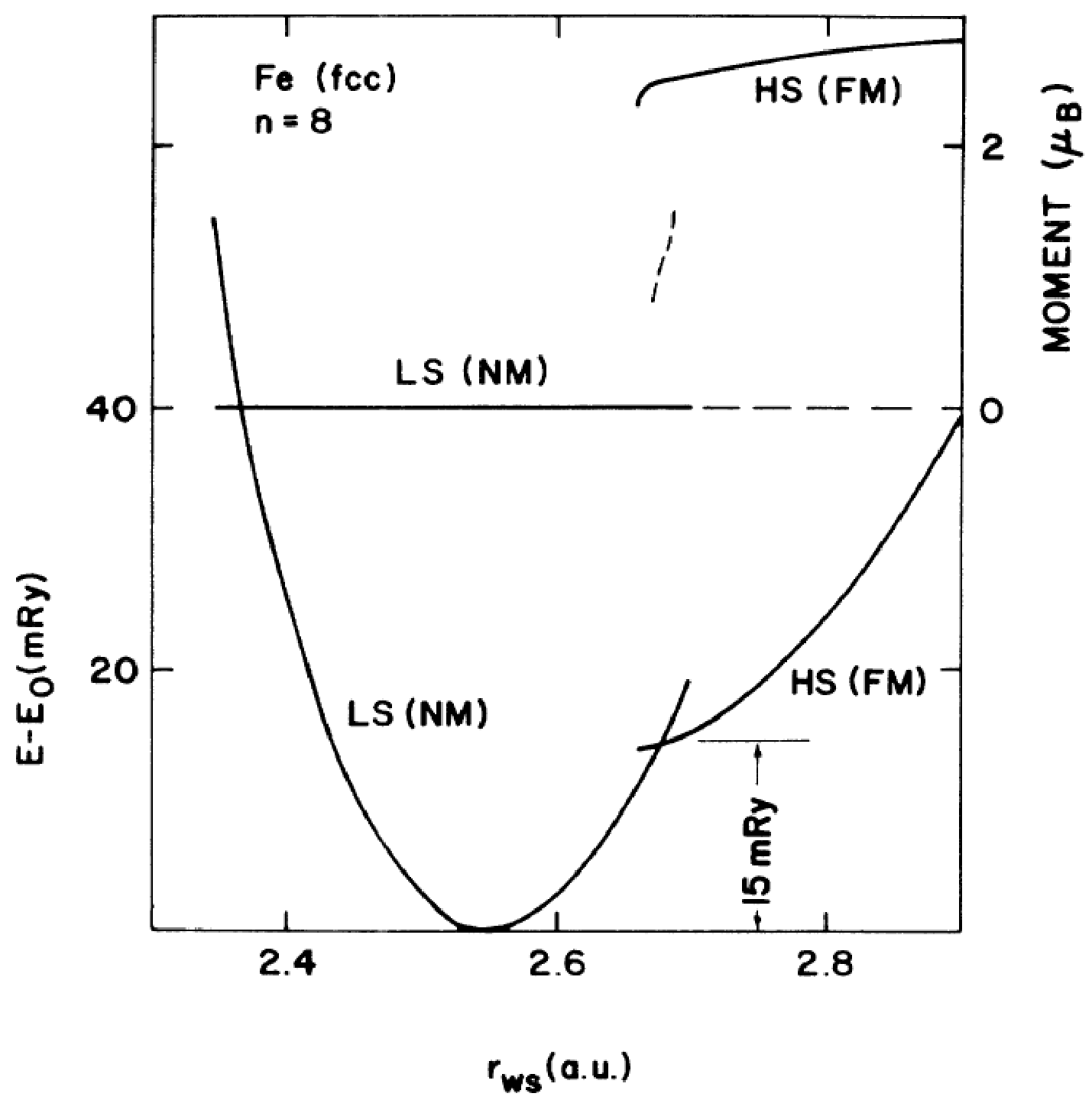}
\caption[Fixed Spin Moment]{Total energy (left scale) and magnetic moment (right scale) of Fe in the fixed spin moment method. From~\textcite{Moruzzi1990}.}
\label{fig:FSM}
\end{figure}

High spin to low spin transitions are often associated with structural changes. These magnetovolumic effects of prime importance for the structural stability of solids are related to the electron occupation of the crystal-field states. Intuitively one expects the $d$ orbital extension, and thus the atomic volume, to be smaller in the low spin state than in the high spin state. Theoretically magnetovolumic instabilities have been investigated by the fixed spin moment method \cite{Moruzzi1990}. Fig.~\ref{fig:FSM} shows the total energy and magnetic moment of Fe calculated in this framework with varying Wigner size radius $r_{WS}$. The total energy of the HS configuration forms a parabolic branch shifted towards higher volumes with respect to the LS one. This model confirms that the system preferentially adopts a high volume (high spin) state at low pressure, and inversely a low spin (low volume) state at high pressure. The transition between the two is of first order and entails a sudden decrease of the local magnetic moment at the pressure (volume) where the two branches cross.

Two of the best known examples of magnetovolumic effects under pressure are found in Fe-based Invar alloys and pure Fe. 

\subsubsection{Fe Invar}
The Invar effect is the anomalously low thermal expansion of certain metallic alloys over a wide range of temperature. One of the most commonly accepted models of the Invar anomaly is the so-called 2$\gamma$-state model proposed by Weiss~\cite{Weiss1963}. According to this model, iron can occupy two different states: a high volume state and a slightly less energetically favorable low volume state. With increasing temperature, the low volume state is thermally populated thus compensating lattice expansion. This model is supported by fixed-spin moment calculations in Invar which show that, as a function of temperature, the Fe magnetic state switches from a high spin to a low spin state of high and low atomic volume respectively~\cite{Moruzzi1990}. The same effect is also expected under applied pressure at ambient temperature: Pressure tends to energetically favor a low volume state, eventually leading to a HS to LS transition. 
\begin{figure}[htbp]
\includegraphics[width=0.90\linewidth]{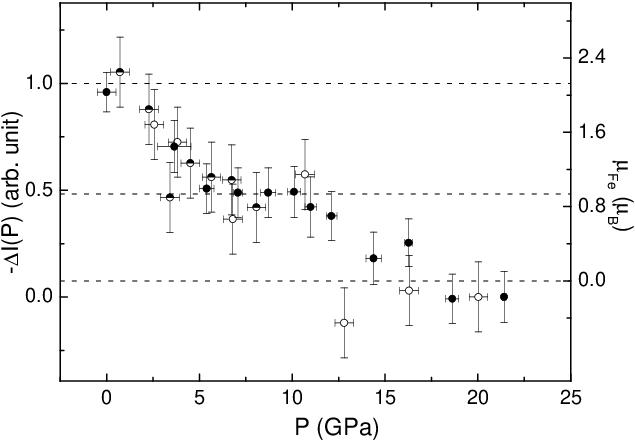}
\caption[Integrated absolute difference: Invar]{
Integrated absolute difference calculated from the XES spectra in the Fe-Ni Invar. The open, half filled, and solid circles represent IAD values for the consecutive series of measurements. The right scale is deduced from the pure Fe XES
data. Horizontal lines emphasize the three magnetic states (high spin, low spin, non-magnetic) of the Fe atom. From~\textcite{Rueff2001}.}
\label{fig:Invar_IAD}
\end{figure}

To verify this experimentally with IXS, the spin state of Fe in Fe$_{64}$Ni$_{36}$ was monitored by XES at the Fe K$\beta$ line up to 20 GPa in a diamond-anvil cell \cite{Rueff2001}. The pressure dependence in the form of IAD($P$) is shown in Fig.~\ref{fig:Invar_IAD}. In the low pressure region below 5 GPa, the curve presents a linear decrease followed by a plateau in the intermediate pressure region which extends up to about 12 GPa. At higher pressures, the intensity drops to zero around 15 GPa and remains unchanged up to the highest measured pressure point around 20 GPa. The existence of two plateaus supports the interpretation of two magnetic transitions taking place in the 2--5 GPa and in the 12--15 GPa ranges, respectively. This demonstrates the existence of three distinct magnetic states that are successively reached as pressure is increased: HS ($S=5/2$), LS ($S=1/2$) and finally diamagnetic ($S=0$). Further information on the magnetism in Invar can be obtained comparing the Invar XES spectra with those previously measured in pure iron under pressure. This confirms that: i) the Fe atom in the Invar alloy is in a high-spin state at zero pressure as it is in iron and, ii) at high pressure 20 GPa, the Fe atom in the Invar alloy is in a nonmagnetic state as it is in $\epsilon$ iron, iii) therefore the plateau in the intermediate pressure region can be associated with the existence of a low spin magnetic state. 

This interpretation supports the 2$\gamma$-state model and is also in qualitative agreement with XMCD~\cite{Odin1999} and M\"{o}ssbauer measurements carried out on both Fe-Pt Invar and Fe-Ni Invar under pressure. Note that these two techniques probe the long range magnetism contrary to XES which is a local probe on the atomic scale. The deduced magnetic moment (here in the sense of magnetization) is thus sensitive to the reported decrease of the Curie temperature with pressure in Invar alloys. In the recently discovered Invar Fe$_3$C for instance, a HS-LS transition was found at 10 GPa~\cite{Duman2005} by XMCD whereas the magnetic instability manifests itself in the XES spectra at 25 GPa~\cite{Lin2004a}. The difference is related to the occurrence of a paramagnetic phase (at room temperature) above 10 GPa. 

\subsubsection{Fe}
\label{sec:Fe_HP}
The interplay between the structural properties and magnetism is best exemplified in elemental Fe as a model system for $d$ electronic properties. Under pressure, Fe is known to undergo a phase transition from the ferromagnetic $\alpha$-phase (bcc) to the non-magnetic $\epsilon$-phase (hcp) around 13 GPa at room temperature while at high temperature, Fe is stabilized in the paramagnetic $\gamma$-phase (fcc). 
\begin{figure}[htbp]
\includegraphics[width=0.90\linewidth]{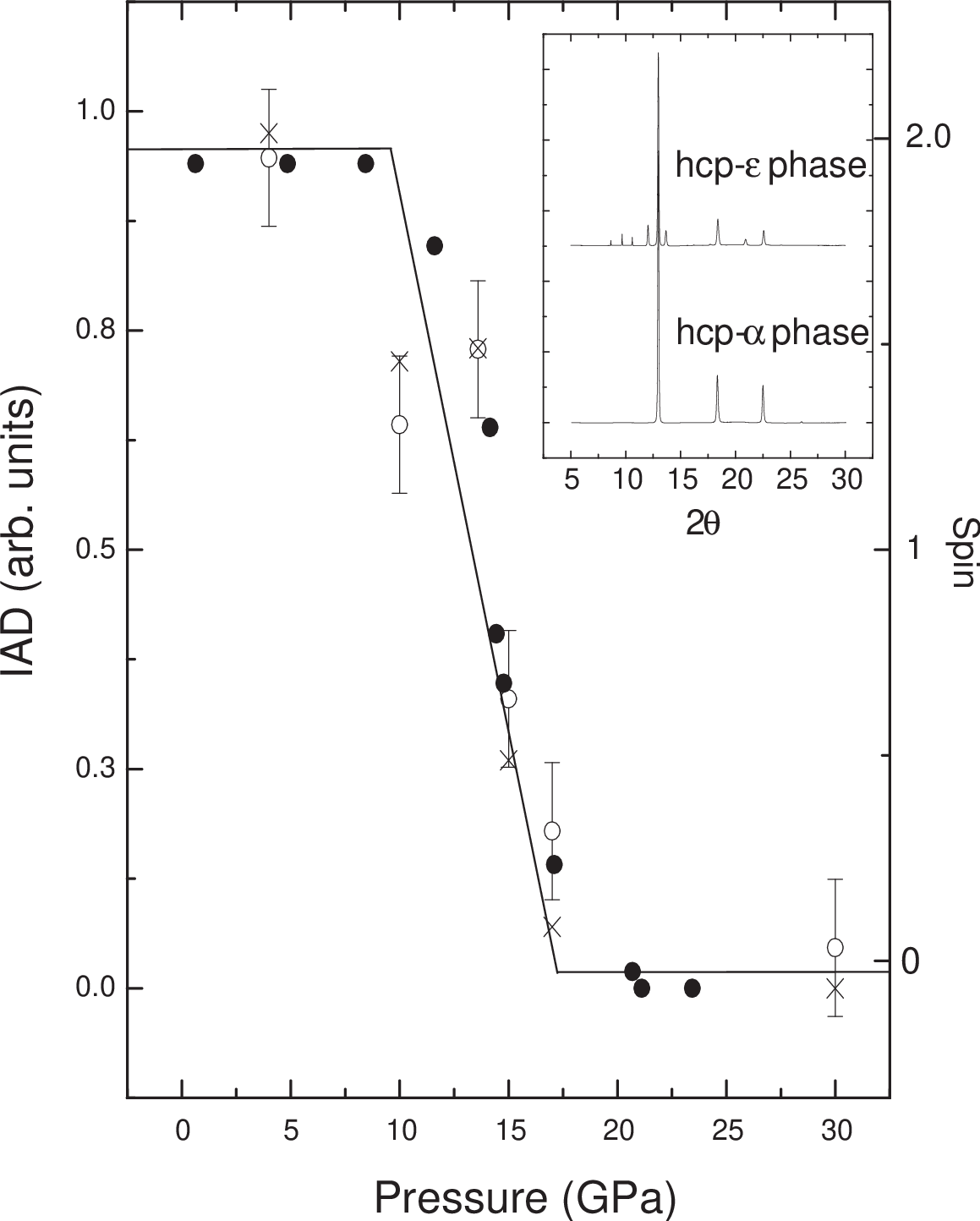}
\caption[Integrated absolute difference: $\alpha$-$\epsilon$ Fe]{Integrated absolute difference calculated from the XES spectra (cross) in pure Fe compared to the $\alpha$-phase fraction determined by M\"{o}ssbauer spectroscopy \cite{Taylor1991} (solid circles) across the transition. The diffraction pattern measured in both phases, shown in the inset, confirm the structural change. Spin state is indicated. From~\textcite{Rueff1999a}.}
\label{fig:Fe_IAD}
\end{figure}

Experimentally, the $\alpha$-$\epsilon$ transition is well documented by M\"{o}ssbauer spectroscopy~\cite{Taylor1991}. Lately, the Fe spin state was studied by XES~\cite{Rueff1999a}. The IAD analysis applied to Fe (cf.\ Fig.~\ref{fig:Fe_IAD}) shows a decreased spin state around 10 GPa before it reaches a full low spin state at 15~GPa in good agreement with the M\"{o}ssbauer data. A magnetic transition of comparable amplitude was reported recently in the $\gamma$-phase within the paramagnetic domain at 1400 K \cite{Rueff2008}. A puzzling aspect is the width of the  $\alpha$-$\epsilon$ magnetic transition which has been observed by several other techniques as well, including M\"{o}ssbauer. In addition to the pressure gradient which could account for a fraction of the width, we suggest that the HS and LS magnetic states are mixed at finite temperature close to the magnetic collapse pressure as evoked in section~\ref{sec:thermal_broadening}, which in turn broadens the phase transition. 

Whether the magnetic collapse precedes the structural change or not is still a matter of debate. Recent XMCD experiments in pure Fe combined with EXAFS seem to suggest that magnetic collapse precedes structural changes \cite{Mathon2004}, 
though recent calculations allowing for non collinear magnetic structure show clearly that Fe would remain antiferromagnetic in the hcp structure up to 50 GPa~\cite{Mukherjee2001,Cohen2002,Cohen2004,Steinle-Neumann2004,Steinle-Neumann2004a}. In addition, XMCD can only detect ferromagnetic states and is not suited to such a study. The pressure dependence of the magnetic state of Fe was reassessed in a recent work by XES in Fe both at room temperature through the $\alpha$-$\epsilon$ transition and in the $\gamma$-fcc phase at high temperature~\cite{Rueff2008}. The results are in good agreement with non collinear magnetic structures both in $\epsilon$- and $\gamma$-Fe.
Finally, an investigation of the spin state in hematite (Fe$_2$O$_3$) are indicative of structural changes that precedes the electronic transition \cite{Badro2002}.

\subsection{Geophysical Implications} 
Not surprisingly, the magnetic properties of Fe under extreme conditions have been of large interest to the geophysics community since it is the most abundant element of the Earth's interior. A widely accepted picture is that the solid inner core is made of pure Fe whereas  the liquid outer core is composed of Fe mixed with light elements such as S or O. In the mantle, Fe is present as an impurity in rocks, mostly silicates. These different forms of Fe have been investigated by XES.   

\subsubsection{FeS}
FeS is an anti-ferromagnetic insulator ($T_N$=598~K) and crystallized in the NiAs-related (troilite) structure. FeS falls at the boundary between charge-transfer and Mott-Hubbard insulators in the ZSA phase diagram ($\Delta<U$ with $U$ relatively small)~\cite{Bocquet1992a}. Under pressure and at ambient temperature, FeS undergoes two structural phase transitions, from the NiAs-related to a MnP-related structure at 3.5 GPa, and then to a monoclinic phase at 6.5 GPa. The last transition is further accompanied by an abrupt shortening of the $c$ parameter from 5.70 to 5.54 Å. Pressure-induced structural phase transitions in FeS have been extensively studied because the material is considered to be a major component of the cores of terrestrial planets~\cite{Taylor1970,Fei1995,Sherman1995}.
\begin{figure}[htbp]
\includegraphics[width=0.90\linewidth]{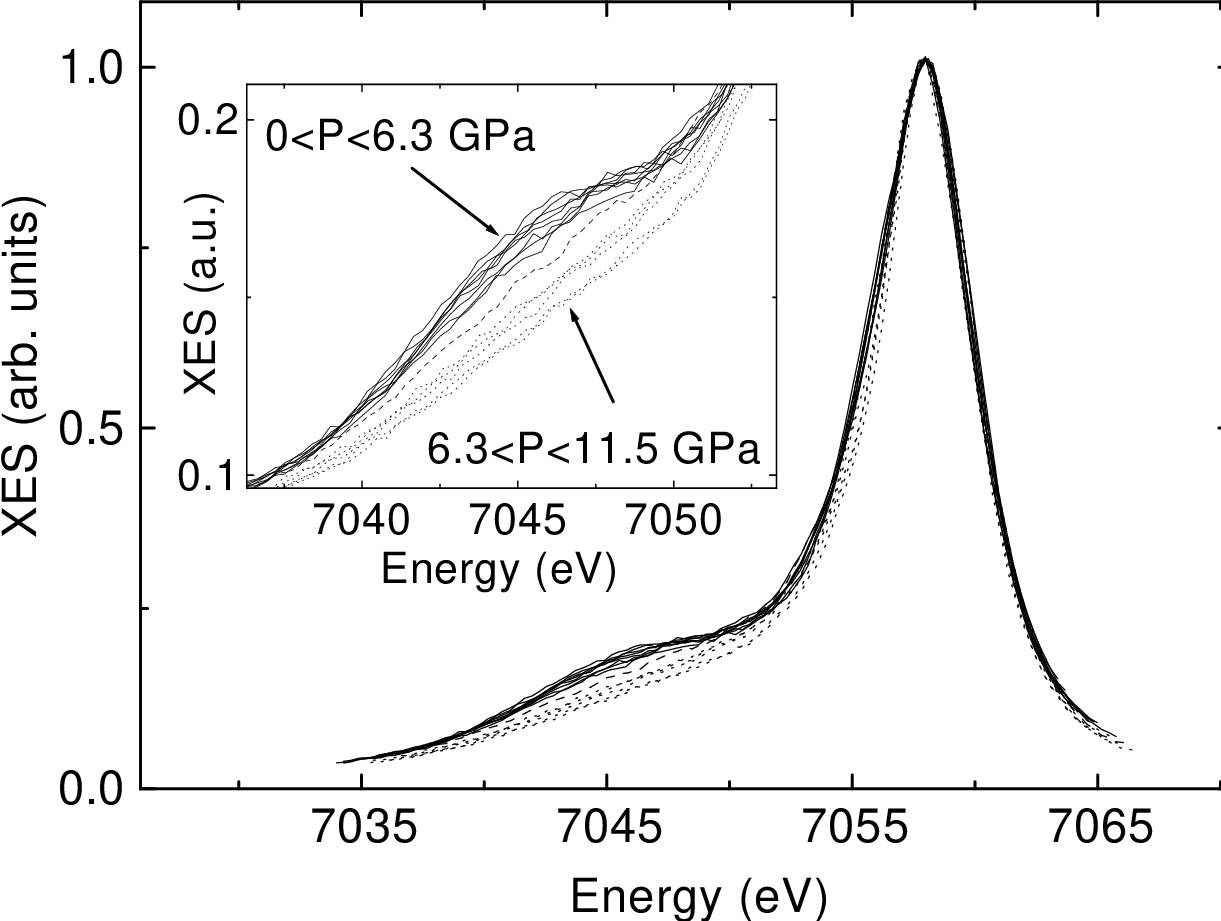}
\caption[Fe K$\beta$ emission in FeS]{Fe K$\beta$ emission line in FeS as a function of pressure. At low pressure, the satellite at 7045.5 eV is characteristic of the HS state. The decrease in the satellite  at high pressure denotes the transition to the LS state (dotted lines). From~\textcite{Rueff1999}.}
\label{fig:FeS_XES}
\end{figure}
Fig.~\ref{fig:FeS_XES} shows the changes in XES spectra measured as pressure increases from 0 to 11.5 GPa~\cite{Rueff1999}. The well defined satellite at low pressure is indicative of the local magnetic state of the Fe$^{2+}$ ion (despite the anti-ferromagnetic long-range order). The satellite intensity disappears for pressures ranging between 6.3 and 11.5 GPa. The width of the main line also shows significant narrowing in this pressure range, as expected in the low-spin state. 

\subsubsection{Fe solid-solutions}
In solid solutions, the transition metal ions form an assembly of isolated magnetic impurities bearing a local moment. At low pressure, electron correlations are important because of the narrow $d$-band. Under pressure, electron itinerancy sets in while the $d$-band broadens and solid solutions then provide an interesting counterpart to compounds or alloys. As it turns out, Fe solid solutions are often found in minerals relevant for geophysics. This is the case of Fe-perovskite and magnesiow\"{u}stite which are discussed below. Being able to describe the electronic properties of these materials under high pressure (and temperature) is crucial for the description of their properties (elasticity, thermodynamics, transport) under realistic conditions for planetary studies.

\begin{figure*}[htbp]
\begin{tabular}{cc}
{\small a)}\includegraphics[width=0.45\linewidth]{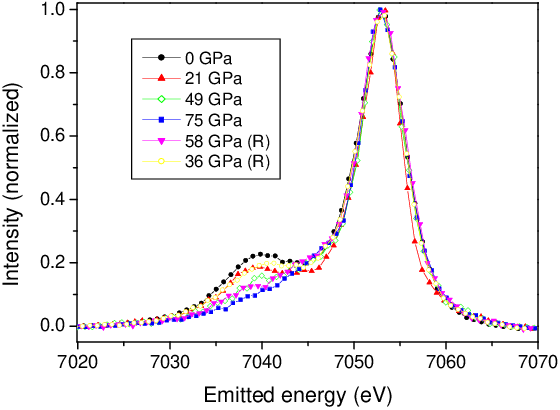}	
{\small b)}\includegraphics[width=0.45\linewidth]{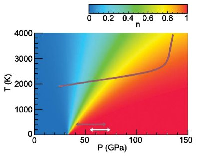}	
\end{tabular}
\caption[Fe K$\beta$ emission in magnesiow\"{u}stite]{(Color online) a) Fe X-ray emission spectra measured in magnesiow\"{u}stite solid solution (Mg$_{0.83}$,Fe$_{0.17}$)O. From~\textcite{Badro2003}; b) computed LS fraction $n$ in the ($P$,$T$) space ; The arrows correspond to the transition pressure range as obtained from XES measurements, the full line is a lower mantle geotherm. From~\textcite{Tsuchiya2006}.}
\label{fig:MgFeO_XES}
\end{figure*}

\paragraph{(Mg,Fe)O}
\label{sec:MgFeO}
Magnesiow\"{u}stite is considered as the dominant phase of the Earth's lower mantle. At ambient conditions of pressure and temperature, (Mg,Fe)O is a paramagnetic insulator of moderate charge transfer character with Fe$^{2+}$ in the high spin state. AF correlation builds up when temperature is decreased with a N\'{e}el temperature $T_N$ of about 25~K (with 20\% of iron). \textcite{Badro2003} have measured the spin state of iron in (Mg$_{0.83}$,Fe$_{0.17}$)O by XES. The (cf.\ Fig.~\ref{fig:MgFeO_XES}) change with pressure indicates a broad transition from HS to LS. The transition starts around 20 GPa and full conversion to LS state is completed in the 60--70 GPa region. The magnetic transition was confirmed recently by M\"{o}ssbauer spectroscopy~\cite{Speziale2005,Lin2006}.

~\citet{Tsuchiya2006} have used an LDA+U approach to investigate the electronic properties and magnetic transition in magnesiow\"{u}stite under pressure. Both $U$ and the $d$-bandwidth $W$ are found to increase with pressure but the latter at a faster rate so that $U/W$ eventually decreases, an indication of less correlation in the high pressure phase. Because the LS and HS states have comparable energies, the average spin state at finite temperature is a Boltzmann average of the two spins as shown in Fig.~\ref{fig:MgFeO_XES}(b). The experimental temperature dependence of the spin state in (Mg,Fe)O under pressure was  recently investigated by XES~\cite{Lin2007} up to 95 GPa and 2000 K. At high temperature, the spin transition broadens and becomes more gradual as expected due to Boltzmann averaging. Other factors could contribute to broadening of the transition (such as pressure gradient) as seen at room temperature. 

The magnetic collapse also affects the compressional behavior of the Fe mineral: The high-pressure LS state exhibits a much higher bulk modulus and bulk sound velocity than the HS phase at low pressure~\cite{Lin2005,Lin2007} which can be traced back to the lower atomic volume of the LS state.

\paragraph{(Mg,Fe)SiO$_3$}
Iron perovskite is another important component of the Earth's lower mantle. In the perovskite, iron is present as ferrous (Fe$^{2+}$) or ferric (Fe$^{3+}$) species. It is admitted that the ferrous iron occupies the large dodecahedral $A$ site, whereas the smaller octahedral $B$ site is the host of ferric iron together with lesser amounts of the ferrous species. 

\begin{figure}[htbp]
\includegraphics[width=0.90\linewidth]{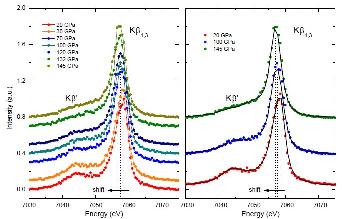}
\caption[Fe K$\beta$ emission in Mg-perovskite]{(Color online) a) Fe X-ray emission spectra measured in (Mg$_{0.9}$,Fe$_{0.1}$)SiO$_3$ between 20 and 145 GPa; b) XES spectra in model compounds (solid lines) are superimposed on to three spectra representative of the three
different spin states. From~\textcite{Badro2004}.}
\label{fig:Perov_XES}
\end{figure}
\citet{Badro2004}) have reported two successive magnetic transitions around 70 GPa and 120 GPa under high pressure conditions by XES at the Fe K$\beta$ line (cf.\ Fig.~\ref{fig:Perov_XES}(a)). The transitions are characterized by a sudden decrease of K$\beta'$ peak intensity and a shift of the K$\beta_{1,3}$ feature to lower energy.
In the measured sample, perovskite is supposed to contain 75\% of Fe$^{2+}$ and 25\% Fe$^{3+}$, with 75\% of the Fe$^{2+}$ being in the A site and 25\% in the B site and all Fe$^{3+}$ being in the B site. This leads to 56\% ferrous dodecahedral, 25\% ferric octahedral and 19\% ferrous octahedral sites; i.e., 56\% of total iron in the A site and 44\% in the B site. To understand the nature of the two transitions, the XES spectra in (Mg,Fe)SiO$_3$ are compared in Fig.~\ref{fig:Perov_XES}(b) to spectra obtained in Fe model compounds containing Fe$^{2+}$ or Fe$^{3+}$ iron in pure HS and LS spin states. Composite spectra were built from the model compounds, starting from the known abundance of Fe$^{2+}$ and Fe$^{3+}$ ions. Best agreement is obtained by combining model spectra of ferrous and ferric iron both in the HS states in the low pressure region, while the high pressure regime is well described by having both ions in the low spin states.  In the intermediate regime, a mixed state is seemingly realized with relative amounts of the HS and LS iron species of $\sim$55\% and 45\%, respectively. This could indicate that the transition is site-specific; the first and second transitions could correspond to electron pairing in the A then B site, respectively.

This interpretation however contrasts with the more recent XES results of \citet{Lin2008} in (Mg$_{0.6}$,Fe$_{0.4}$)SiO$_3$ at pressure–temperature conditions of the lowermost mantle. M\"{o}ssbauer analysis shows no evidence of Fe$^{3+}$ ions in this sample while XES strongly supports a $S=1$ intermediate state in both perovskite (300 K) and post-perovskite (2500 K) structure.\\

To conclude this section, we notice that the modifications of the electronic properties in both Fe magnesiow\"{u}stite and Fe perovskite which accompanies the magnetic collapse are expected to affect the heat transport in the Earth's interior \cite{Badro2004,Tsuchiya2006,Lin2008}.

\subsection{Coupling to thermal excitation} 
\label{sec:thermal_excitation}
As just discussed, the proximity of the first excited spin state above the ground state may provoke a mixing of different spin states upon thermal excitation. In Fe compounds, this usually happens at elevated temperatures but in the cobaltates, three possible spin states of Co$^{3+}$ have been identified with competing occupancies at room temperature: low-spin ($S=0$), high-spin ($S=2$) and an extra intermediate spin (IS) state ($S=1$). Although the existence of the IS is still debated, temperature effects should be markedly enlarged in cobaltates. Pressure may further lead to ground state inversion, yielding peculiar behavior of the local Co magnetic moment.

\subsubsection{Co compounds}
\label{XES_Co}
\paragraph{LaCoO$_3$}
The rhombohedral perovskite LaCoO$_3$ is an unusual case of a non-magnetic semi-conducting ground state. Because of the large crystal field splitting, Co is trivalent ($e_g^6$ configuration) at 0 K with a low-spin state ($S=0$). As a function of temperature, two broad transitions have been observed in the magnetization measurements at around 90 K and 500 K. The first transition is conventionally interpreted by the occurrence of a $t_{2g}^4e_g^2$ ($S=2$) high spin state while the metallization which goes with the second transition at high temperature remains of unclear origin. More recent interpretations based on LDA+U calculations have proposed the formation of an intermediate spin state (IS) ($S=1$) above 90 K, characterized by a doubly degenerate $t_{2g}^5e_g^1$ configuration \cite{Korotin1996}. 
In this model, the non-metallic state in the low temperature range is attributed to the intermediate spin: The degeneracy of the IS state is lifted by the Jahn-Teller effect, which supposedly leads to orbital ordering on the Co sites and the opening of the semi-conducting gap. The high temperature insulator/metal transition is then accounted for by the ``melting'' of the orbital-ordered state. Hints of a LS to IS transition have been identified in the T-dependence of the magnetic susceptibility~\cite{Zobel2002}.

\begin{figure*}[htbp]
\begin{tabular}{cc}
{\small a)}\includegraphics[width=0.45\linewidth]{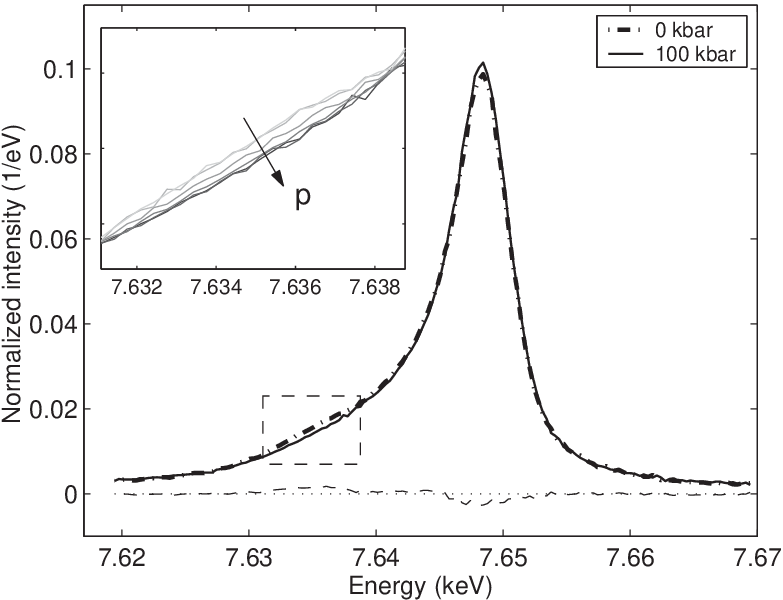} &
{\small b)}\includegraphics[width=0.45\linewidth]{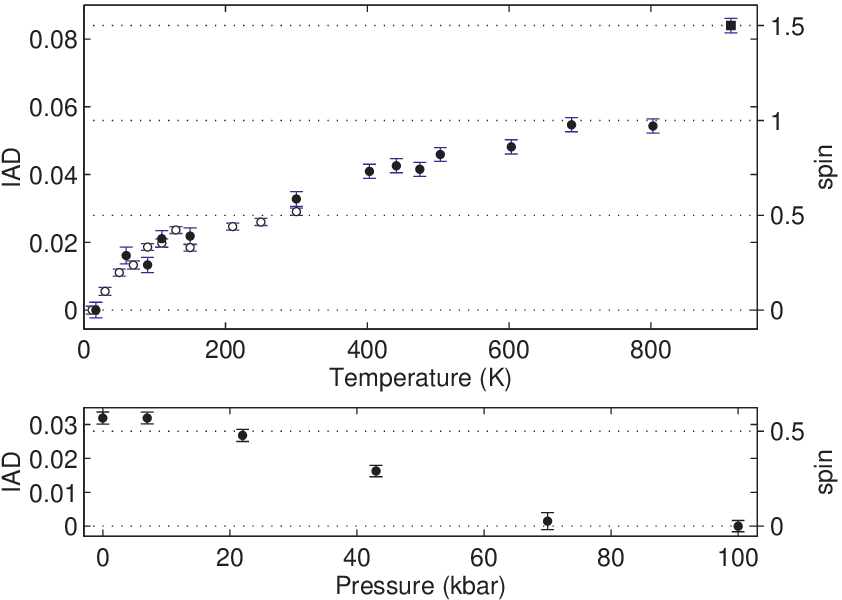} \\
\end{tabular}
\caption[LaCoO$_3$ XES]{a) Evolution of the K$\beta$ emission line in LaCoO$_3$ as a function of pressure; b) Pressure ($T=300$ K) and temperature (at ambient pressure) dependence of IAD values derived from XES (left scale) and estimated Co spin-state (right scale). From~\textcite{Vanko2006}.}
\label{fig:LaCoO3}
\end{figure*}
Pressure can also cause spin-state transitions in LaCoO$_3$, as the crystal field splitting sharply increases when bond lengths shrink. Since the transitions of LaCoO$_3$ are associated to anomalous volume expansions relatively low pressure can have considerable effect on the spin state. \textcite{Asai1998} showed that the energy gap between the LS and the higher spin state increases with pressure. More recently, \textcite{Vogt2003} using x-ray powder diffraction interpreted the pressure-induced changes as a continuous transition from IS to LS state. Chemical pressure, introduced by a partial substitution of La$^{3+}$ with the smaller Eu$^{3+}$, leads to a similar stabilization of the LS state.

The pressure (and temperature) dependence of the Co spin state was investigated by XES at the Co K$\beta$ line~\cite{Vanko2006}. Fig.~\ref{fig:LaCoO3}(a) shows the evolution of the emission spectra as a function of pressure. A gradual variation of the K$\beta$ lineshape is observed up to 70 kbar. 
For extracting the spin moment the IAD values (cf.\ Eq.~(\ref{eq:IAD})) were scaled to spin moment by comparison with model Co-compounds with a well characterized spin-state: $S=2$ in CoF$_3$; $S=1.5$ in LaCoO$_{2.5}$; $S=1$ in Co$^{2+}$-molecular compounds; $S=0$ in LiCoO$_2$ (right scale, Fig.~\ref{fig:LaCoO3}(b)). The pressure dependence can be analyzed in terms of excited spin-states: At ambient temperature and low pressure, both $S=0$ and $S=1$ states are populated leading to an average of $S\sim 0.5$. Upon pressure increase, the $S=0$ state is increasingly favored with respect to the $S=1$ state. Full LS state is reached around 100 kbar. In contrast, the spin state increases with temperature. Starting from the LS state at low $T$, $S$ ramps up progressively to $S=1$ at 800 K. In this $T$-region, the spin variation is well described by Boltzmann statistics~\cite{Vanko2006} involving both LS and IS states but not the HS state. Above 800 K, the sample is no longer stoichiometric. Oxygen vacancies form and $S$ jumps to $3/2$, a value characteristic of LaCoO$_{2.5}$. 

The XES analysis therefore points to a $S=1$ ground state. Notice however that XES cannot distinguish between a true IS state and a superposition of HS and LS states. Recently, \textcite{Haverkort2006} have ruled out the existence of the IS in LaCoO$_3$ based on experimental data at the Co L$_{2,3}$ edges and multiplet calculations. 


\paragraph{La$_{1-x}$Sr$_x$CoO$_3$}
\begin{figure*}[htbp]
\begin{tabular}{cc}
{\footnotesize a)}\includegraphics[clip=on,width=0.45\linewidth]{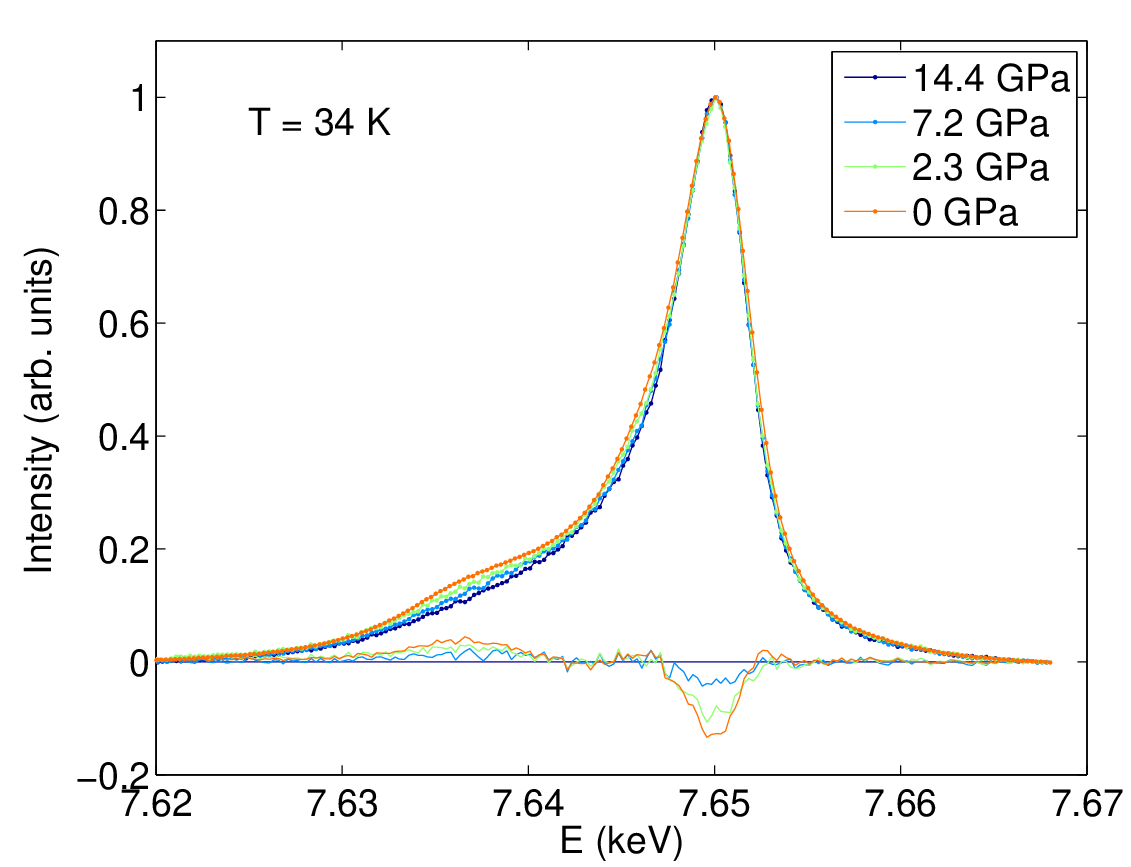} &
{\footnotesize b)}\includegraphics[width=0.45\linewidth]{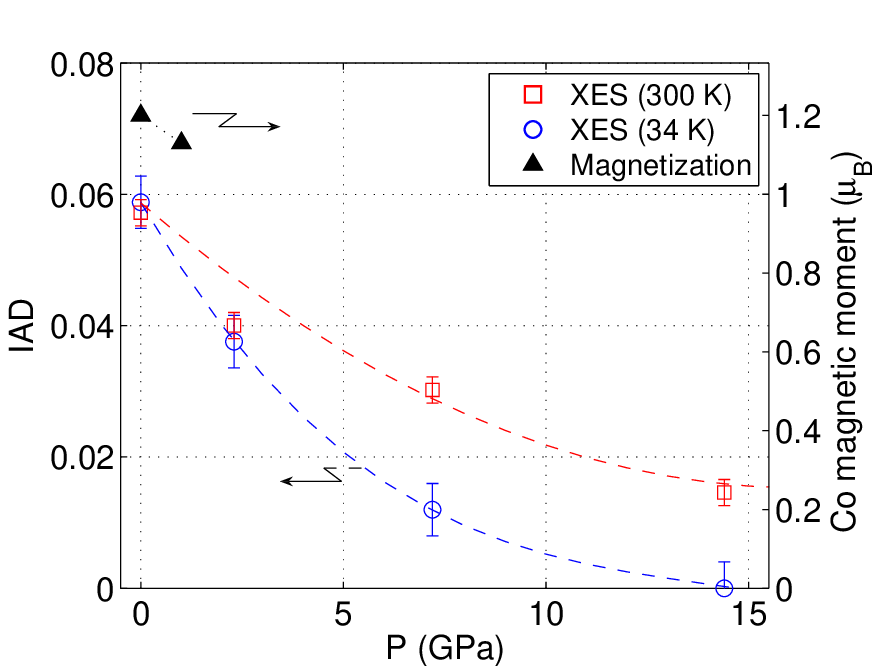} \\
\end{tabular}
\caption[LaSrCoO$_3$ XES]{(Color online) a) Evolution of the Co K$\beta$ emission line in La$_{1-x}$Sr$_x$CoO$_3$ ($x=0.18$) as a function of pressure at $T=34$ K; b) Pressure dependence at 34 and 300 K of spin state (IAD values) derived from XES. From~\textcite{Lengsdorf2007}; Dashed lines are guides to the eyes.}
\label{fig:LaSrCoO3}
\end{figure*}
When substituting La$^{3+}$ by Sr$^{2+}$ ions, hole-type carriers are introduced in LaCoO$_3$. The hole-doped compounds are particularly useful to study the interplay between the spin degrees of freedom and electronic and magnetic properties close to a metal-insulator transition. Depending on the doping level, La$_{1-x}$Sr$_x$CoO$_3$ changes from a spin glass $x>0.05$ to a ferromagnetic metal at $x\geq 0.18$. At ambient pressures, the $x=0.18$ compound is reasonably conducting and metallic in the FM regime. The behavior differs drastically under pressure ~\cite{Lengsdorf2004}: above 2 GPa the compound departs from metallicity and turns into an insulator over the whole temperature range. More remarkably, the resistivity increases continuously under pressure until it saturates around 5.7 GPa. This contrasts for example with La$_{1-x}$Sr$_x$MnO$_3$ which is metallic under pressure, a behavior more in line with the expected band widening in the compressed unit cell. 

Compared to the manganites, the metal ion in La$_{1-x}$Sr$_x$CoO$_3$ has an additional degree of freedom related to the spin state. The spin state was investigated by K$\beta$ x-ray emission spectroscopy under pressure and temperature~\cite{Lengsdorf2007} (cf.\ Fig.~\ref{fig:LaSrCoO3}(a)). Fig.~\ref{fig:LaSrCoO3}(b) illustrates the variation of the spin state (in arbitrary units) with pressure at 300 K and 34 K. As in LaCoO$_3$, $S$ diminishes with pressure tending to a LS configuration. At 300 K, the final spin value is higher than at low temperature probably due to thermally excited spin states. 
\begin{figure}[htbp]
\includegraphics[width=0.90\linewidth]{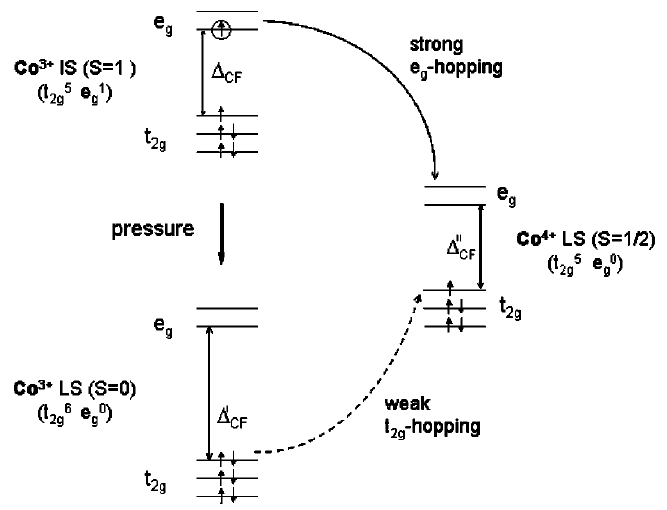}
\caption[Co hopping]{Electron hopping mechanism in the Co$^{3+}$/Co$^{4+}$ lattice in La$_{1-x}$Sr$_x$CoO$_3$. From~\textcite{Lengsdorf2004}.}
\label{fig:Co_hopping}
\end{figure}
The conversion of the Co$^{3+}$ ions into LS species under high pressure reflects on the electron transport properties through the lattice. As illustrated in Fig.~\ref{fig:Co_hopping}, the $e_g$-type electron hopping which takes place between the Co$^{3+}$ and Co$^{4+}$ sites is strongly suppressed when the trivalent ion converts into a LS state and only the weak $t_{2g}$ hopping remains. This blocking mechanism turns out to be efficient enough to provoke a metal-insulator transition at high pressure despite the bond length shortening (and increase of the Co-O-Co bond angle) which acts an opposite way by favoring double exchange.


%

\section{Hybridized $f$ states}
\label{sec:RE_HP}
\begin{figure}[htbp]
\centering
\includegraphics[width=0.90\linewidth]{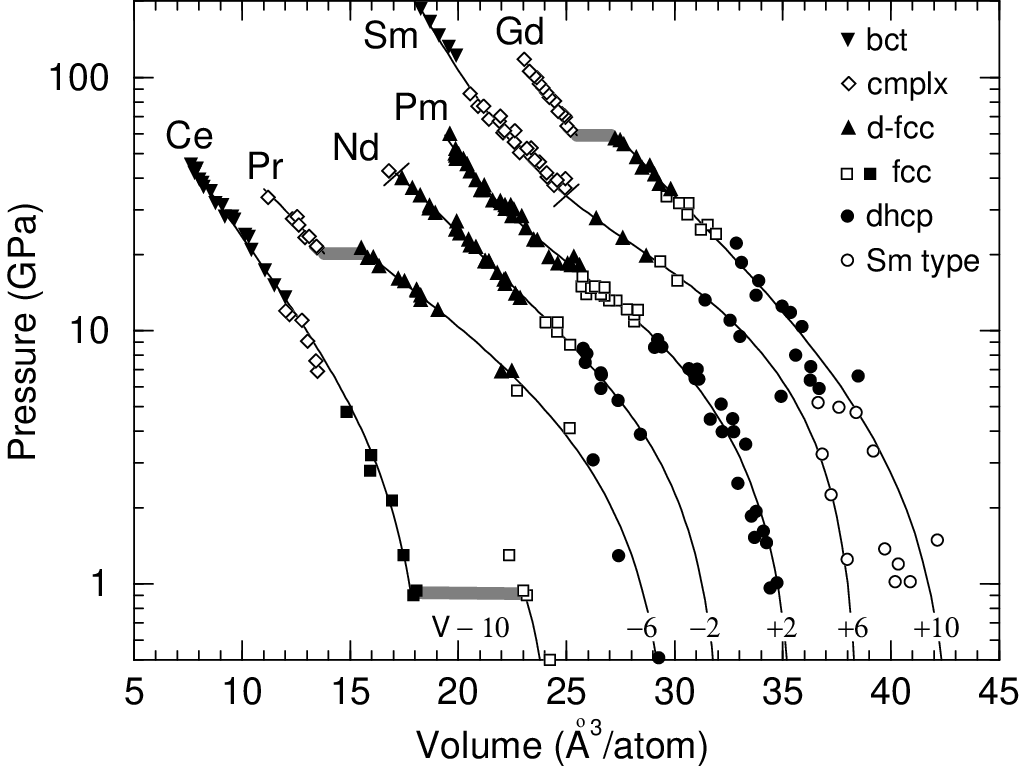}
\caption[Equation of state of $4f$ elements]{Equation of state of several $4f$ elements. From \textcite{McMahan1998}.}
\label{fig:EOS_f_P}
\end{figure}

As opposed to the $d$ levels, the $4f$ electrons in solids are considered localized and as such unaffected by the proximity of the conduction band. The almost constant molar volume dependence reported in Fig.~\ref{fig:molar_vol} as a function of  band filling---with the exception of the divalent Eu and Yb---shows that the $4f$ electrons do not contribute significantly to the cohesive energy. Under pressure however, the conduction bandwidth and Fermi energy will change, eventually modifying the $f$-electronic behavior and with a lowering of symmetry for the crystal structure (cf.\ Fig.~\ref{fig:EOS_f_P}); in rare-earths and actinides, the structural changes are often correlated to a sudden contraction of the lattice leading to volume collapse transitions (VCT)~\cite{McMahan1998}; the magnetic susceptibility evolves from Curie-Weiss behavior to a Pauli-like paramagnetism, yielding a loss of magnetism such as in Ce; pressure may induce metallization of the $f$ electrons like the black to golden phase transition in SmS. Even so, because the Hund coupling energy is much larger than the $f$ bandwidth, the $f$ electrons are expected to retain their localized character to a large extent, hybridization being considered as a second-order perturbation. The structure changes  reported in Fig.~\ref{fig:EOS_f_P} are indeed mostly determined by the pressure dependence of the $5d$ band with little influence from the $f$ states, with the notable exception of Ce and Gd. In a similar fashion, while the VCT in mixed valence rare-earths is mostly caused by the pressure-induced change of occupation of the $f$ and $d$ bands, the general consensus is that only a minute fraction of the $f$ electrons delocalizes to hybridize with the conduction band, the other part being considered still well localized. This situation is at odds with $d$ electron behavior where the bandwidth is one or two orders of magnitude larger than in hybridized $f$ electrons.
%
%

In the 70's, \textcite{Johansson1976} modeled the electronic changes of rare-earths under pressure in the more general context of the intermediate valency. The physical picture is that the $4f$ energy ($\varepsilon_f$) comes closer to the Fermi energy ($E_F$) as $P$ increases until $\varepsilon_f\approx E_F$. At this stage, the $4f$ level progressively empties into the conduction band. At still higher pressure, the empty $4f$ band (now located above the Fermi level) marks the transition to a new valent state. Considering a linear variation $\Delta E(P)$ of the energy distance between $\varepsilon_f$ and $E_F$, 
\begin{equation}
\lim_{P\rightarrow 0} \frac{d\Delta E(P)}{dP}=\mathcal{V}_{n+1}-\mathcal{V}_{n}
\label{eq:limEG}
\end{equation}
with $\mathcal{V}_{n}$ the atomic volume of the $|4f^n\rangle$ configuration, Johansson was able to predict the valence transition pressures in rare earth ions. Thus, the stability of one valency over the other is seen to result from the balance between the gain in cohesive energy and the energy price to promote an $f$ electron into the conduction band. However, the latter process is poorly taken into account in the model. The description of a hybridized $f$-state is in fact a formidable task, which lies at the core of modern treatment of $f$ electrons, involving the formation of a Kondo singlet state or the heavy fermion (HF) behavior, two most challenging aspects of the $f$ mixed valent state (cf.~\textcite{Flouquet2005a} for a review). 
Both Kondo and HF phenomena deal with low temperature physics on an energy scale far different from that envisaged in this review. Nevertheless, since the competition between the localization and delocalization can be tuned effectively by applying pressure, we can also access to and from transitions between Kondo screening and Fermi liquid behavior.

\subsection{Interaction with the conduction states}
\subsubsection{Mixed valency}\label{mixval}
It is convenient to treat the hybridized $f$ state as a mixed (or fluctuating) valent state, the $f$ electron acquiring partial conduction electron character. Hybridization or mixed-valency may be defined formally in terms of the configuration interaction. Then, the ground state is written as a linear combination of degenerate states $|g\rangle =  c_{n-1}\left|f^{n-1}v^{m+1}\right\rangle+c_{n}\left|f^{n}v^m\right\rangle+\cdots$ where $v$ represents the valence electrons and $\left|c_i\right|^2$  the weight of the $f^i$ configuration~\cite{Gunnarsson1983}. 
Large fluctuations are excluded due to the strong Coulomb repulsion, but the $f$ valency may vary around the ground state value on a characteristic timescale determined by the $f$ bandwidth~\cite{Varma1976,Lawrence1981}. In the case of the $f$ systems with a narrow bandwidth, fluctuations are slower than typical core-hole lifetimes in x-ray spectroscopy. Hence, mixing of configuration can be in principle resolved by such techniques since the ground degeneracy is lifted in presence of a core-hole. Indeed, $f$ electron states can be clearly identified in spectroscopic data such as obtained by x-ray photoemission (XPS) or x-ray absorption (XAS)~\cite{Fuggle1983,Fuggle1983a}. These are the two probes that have contributed most to unraveling the electronic properties of $f$-electron materials in the past. Resonant IXS turns out to be a powerful alternative. Here, the ground state degeneracy is lifted in both intermediate and final state which allows a detailed investigation of the mixed valent state as initially shown in the soft x-rays~\cite{Kotani2000}. Sections~\ref{sec:2p3dRXES} and following are mainly devoted to exploring this aspect but applied to hard x-rays and high pressure conditions. 

\subsubsection{Anderson Impurity Model}
A good starting point for describing the hybridization of a single $f$ level with band states is the Anderson impurity model (AIM). The AIM Hamiltonian applies to the case of a single magnetic impurity of energy $\varepsilon_f$ weakly interacting with conduction electrons, described by the dispersion $\varepsilon(k)$. The AIM is related to the Hubbard Hamiltonian in Eq.~(\ref{eq:MH}) in that it also accounts for correlations between the $f$ electrons with an on-site Coulomb parameter $U$. 
\begin{eqnarray}
\mathcal{H}_{AIM}&=&\sum_{km}\varepsilon(k)c^+_{km}c_{km}+\varepsilon_f\sum_m \hat{f}_m^+\hat{f}_m \nonumber \\
&+&\sum_{km}V(k)(\hat{f}_m^+c_{km}+c^+_{km}\hat{f}_m)+\frac{U}{2}\sum_{m\neq m'}n_m^f n_{m'}^f \nonumber\\
&+&\mathcal{H}_0
\label{eq:AIM}
\end{eqnarray}
$\mathcal{H}_0$ represents the conduction electrons term which do not couple to the impurity; $\hat{f}_m^+$ creates an $f$ electron with a magnetic quantum number $m$ from a previously empty site while $\hat{f}_m$  annihilates it. The hybridization strength $V$ when finite drives the ground state towards a singlet Kondo state ($S=0$). $U$ is often considered in the limit $U\rightarrow\infty$, which implies that double occupancy configuration is not allowed.

\begin{figure}[htbp]
\centering
\includegraphics[width=0.90\linewidth]{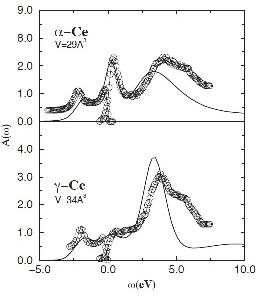}
\caption[Ce Kondo resonance]{(Color online) Kondo resonance in Ce, near the Fermi energy (set at $\omega=0$ eV). The photoemission and inverse photoemission data borrowed from \textcite{Liu1992} are compared to DFMT calculations. From \textcite{McMahan2003}.}
\label{fig:Ce_Kondo_resonance}
\end{figure}

The AIM grasps the most important concepts of hybridized $f$ electrons such as the stabilization of singlet ground state and especially the building-up of the Abrikosov-Suhl resonance near the Fermi energy which governs the low energy excitations and is seen as a fundamental signature of the Kondo excited states. A remarkable experimental results was the observation of this feature in Ce by XPS (Fig.~\ref{fig:Ce_Kondo_resonance}). 
A more accurate treatment would include $f$-electron correlation and their hybridization with an electron bath using a self-consistent band structure approach~\cite{Held2001}. The influence of pressure can be clearly seen when one expresses the energy gain $\Delta\varepsilon$ due to the formation of the singlet state~\cite{Fulde1995}:
\begin{equation}
\Delta\varepsilon=-D\mathrm{e}^{-|\varepsilon_f|/(\nu_f N(0)V^2)}
\label{eq:singlet}
\end{equation}
$D$ is the half bandwidth of the conduction band, $\nu_f$ the $f$-orbital degeneracy and N(0) the conduction electron density of states (per spin). This energy is usually associated to a characteristic temperature $T_K$. The hybridization strength $V$ is known to be strongly pressure dependent because the band overlap is reinforced when the unit cell is reduced. As in the $\gamma$-$\alpha$ transition in Ce, the Kondo state is favored at high pressure. Similarly, $T_K$ is expected to increase with pressure as hybridization becomes stronger. This is indeed the case in Ce but not in Yb which shows the opposite behavior. We will come back to this issue in section~\ref{sec:Kondo_behavior}. 

Another important parameter which is derived from the generalized AIM Hamiltonian is the $f$-occupancy ($n_f$) and the double occupancy. In the case of strong coupling $V$, i.e.\ at high $T_K$, $n_f$ significantly deviates from unity, while the double occupancy is expected to increase~\cite{McMahan2003}. In the Kondo regime, a proportion $1-n_f$ of the $f$ electrons are delocalized. Having an experimental access to the $f$-occupancy and following its evolution when pressure is applied is therefore crucial for a proper description of the $f$ hybridized state. This is the main object of the next sections. 

\subsubsection{Actinides}
Due to the hierarchy between crystal field and spin-orbit interactions, the $5f$ states in actinides are considered as intermediate between $4f$ and $3d$ electrons in terms of electron localization~\cite{Johansson1975,Lander1991}. Thus, in the early actinides (Th--Np), the decrease of the atomic volume (cf.\ Fig.~\ref{fig:molar_vol}) is well described by normal band structure calculations pointing to an itinerant behavior of the $f$ electrons. From Am onwards, the $5f$ states start showing localized characteristics at ambient conditions, while Pu seems to lie at the borderline of localization-delocalizaton: the $f$ states show an itinerant character in the numerous phases of Pu, while in $\delta$-Pu they are close to localization. 

In the U-compounds, which will be more specifically discussed in section~\ref{sec:U_comp}, the $5f$ electron bandwidth is of the same order of magnitude as the spin-orbit energy and the on-site Coulomb interaction, and all these parameters must be taken into account on the same footing. The $5f$ electronic behavior in U is then expected to be highly sensitive to modification of these interactions when pressure increases, and simplistic treatment of the $f$ electron as band states is no longer valid. Similarly, in Am under pressure the $5f$ electrons were shown to delocalize in a Mott sense, and a proper treatment must be included~\cite{Griveau2005}.

\begin{table}[htbp]
\caption[$f$-sample properties]{Summary of main properties of the studied transition rare earth and actinide samples under pressure. $\bar{v}$ is the mean valence of the $f$ ion determined from RXES.}
\begin{tabular}{lcccc}
\hline\hline
Sample & P (GPa) & Structure & Properties  & $\bar{v}$\\
\hline
Ce & 0 & fcc & (PM)M & 3.03 \\
& 20 & fcc & (NM)M & 3.2\footnote{\textcite{Rueff2006a}}\\
Gd & 0 & hcp & (PM)M & 7  \\
& 110 & hcp & (NM)M & 7+$\delta$\footnote{\textcite{Maddox2006}}\\
SmS & 0 & NaCl & (NM)SC & 2 \\
& 20 & NaCl & (AF)$^\dag$M & 3\footnote{\textcite{Annese2006}}\\
TmTe & 0 & NaCl & (AF)SC & 2 \\
& 4 & NaCl & (FM)M & 2.5 \\
& 10 & Tetragonal & (AF)I & 2.8\footnote{\textcite{Jarrige2008}}\\
Yb & 0 & fcc & (NM)M & 2 \\
& 20 & bcc & (PM)M & 2.55 \\
& 60 & hcp & (PM)M & 2.7\footnote{\textcite{Dallera2006}}\\
YbAl$_2$ & 0 & MgCu$_2$ & (NM)M & 2.3 \\
& 40 & MgCu$_2$ & (NM)M & 2.9\footnote{\textcite{Dallera2003}}\\
YbS & 0 & NaCl & (NM)SC & 2.35\\
& 40 & NaCl & (NM)M & 2.6\footnote{\textcite{Annese2004}}\\
\hline
UPd$_2$Al$_3$ & 0 & P6/mmm & (AF)M & 4-$\delta$\\
& 40 & Cmmm & (NM)M & 4\footnote{\textcite{Rueff2007}}\\
UPd$_3$ & 0 & dhcp & (AF)M & 4\\
& 40 & dhcp & ( - ) M & 4\footnotemark[8]\\
\hline\hline
\end{tabular}
\label{table:f-sample}
\end{table}

\subsection{2p3d-RXES}
\label{sec:2p3dRXES}
\begin{figure*}[htbp]
\begin{tabular}{cc}
\includegraphics[width=0.45\linewidth]{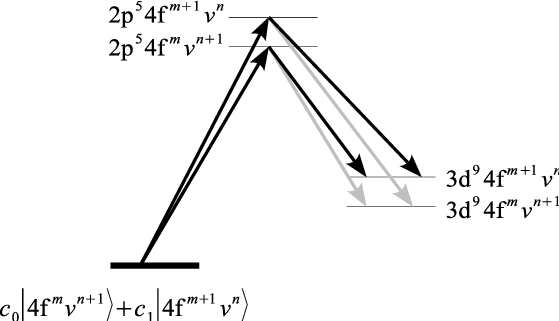}&
\includegraphics[width=0.45\linewidth]{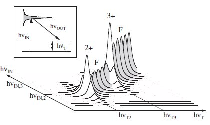} \\
{\small a)} & {\small b)}\\
\end{tabular}
\caption[$2p3d$ RXES process]{(a) $2p3d$-RXES process in a model mixed-valent rare-earth represented by the superposition of two valent states; $v$ stands for the valence electrons. Gray arrows indicate less probable transitions. (b) Illustration of the $2p3d$-RXES in a mixed-valent ions in the incident energy ($\hbar\nu_{IN}$) vs.\ transfer energy $(\hbar\nu_{T}$) plane; $F$ indicates the fluorescence contribution. From~\textcite{Dallera2003}.}
\label{fig:2p3d-RXES}
\end{figure*}

As pointed out in section~\ref{mixval}, hybridization can be introduced by a superposition in the ground state of degenerated $|f^n\rangle$ configurations. The degeneracy is lifted in the XPS or XAS final state, as the core-hole is screened differently by the various $f$ states. Well separated features, each assigned to a different $f$ valency then indicate that there is more than just a single component. This allows an estimation of the various $f$-electron weights, and therefore helps to characterize the degree of hybridization of the $f$-electron. Resonant inelastic x-ray scattering with a degeneracy lifting final state core hole turns out to be a powerful complementary probe of these systems (cf.\ Table~\ref{table:f-sample}). Of particular interest is the $2p3d$-RXES process as explained in Fig.~\ref{fig:2p3d-RXES} in the case of a mixed valent $\left|4f^{m}\right\rangle+\left|4f^{m+1}\right\rangle$ ion. It consists of measuring the L$\alpha_{1,2}$ emission ($3d\rightarrow 2p$) in resonant conditions at the L$_{2,3}$ edges ($2p\rightarrow 5d$). As in first order spectroscopies, the different $f$-states, supposedly mixed in the ground state, are split in the RXES final states. But by tuning the incident energy to particular intermediate states (here in presence of $2p$ core hole), one of the multiple $f$-states can be specifically enhanced through the resonance. The gain in resolution and intensity of selective spectral features is notable and markedly helps the comparison with theoretical calculations. 
In particular, the average valent state ($\bar{v}$) can be simply derived by computing the ratio of different $f$ components in the $2p3d$-RXES spectra: 
\begin{equation}
\bar{v}=n+\frac{I(n+1)}{I(n+1)+I(n)}
\label{eq:val}
\end{equation}
where $I(n)$ ($I(n+1)$) represents the integrated intensity of the $f^n$ ($f^{n+1}$) spectral features. These intensities can be estimated either from the PFY spectra or the resonant emission spectra combined with standard fitting routines. \textcite{Dallera2002} first proposed $2p3d$-RXES technique to study YbAgCu$_4$ and YbInCu$_4$ compounds. A similar procedure was later applied to other mixed valent systems under pressure. Details about the data analysis can be found in the cited works.

\subsection{Kondo behavior}
\label{sec:Kondo_behavior}
\subsubsection{Double occupancy in Ce}
\begin{figure}[htbp]
\includegraphics[width=0.90\linewidth]{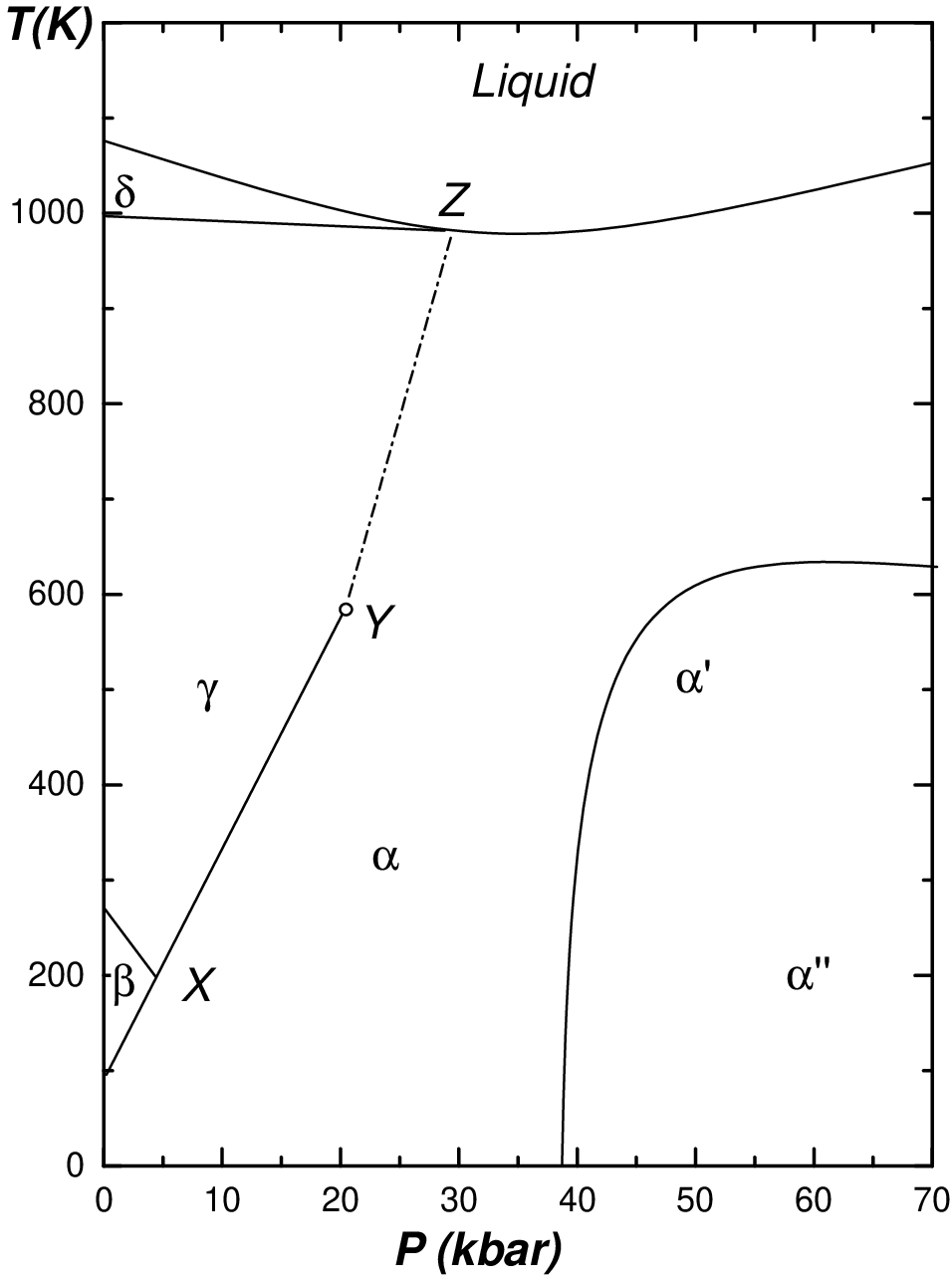}
\caption[Ce phase diagram]{Phase diagram of elemental Ce. Adapted from \textcite{Eliashberg1998}.}
\label{fig:Ce_phase_diag}
\end{figure}
The $\gamma$-$\alpha$ transition in Ce is archetypical of Kondo phenomenon encountered in $f$-electron systems, and one of the best known examples of volume collapse transition. In Ce, the VCT is accompanied by $\sim$15\% volume contraction and ends in a tricritical point as shown in Fig.~\ref{fig:Ce_phase_diag}. It is instructive to recall briefly the various theories put forward to explain the $\gamma$-$\alpha$ transition in Ce. The promotional model~\cite{Coqblin1968} first considered an integer valence change with the transition of one $4f$ electron into the conduction band. This model was soon ruled out by melting-point and cohesive energy arguments by Johansson~\cite{Johansson1974}, predicting that the $4f$ electrons undergo a Mott transition from localized in the $\gamma$ phase to weakly itinerant in the $\alpha$ phase ($f$-band model). In the following years, the Kondo lattice model~\cite{Lavagna1982,Allen1982} (Kondo volume collapse, KVC) has envisaged the disappearance of Ce magnetism in the $\alpha$ phase by an extremely high Kondo coupling. The KVC model differs from Johansson's scenario essentially by the active role played by the conduction electrons which hybridize with the $f$ states.\\

Important results were accumulated over the years in Ce by XPS, among which the observation of the Abrikosov-Sulh resonance which builds up at the $\gamma$-$\alpha$ transition (Fig.~\ref{fig:Ce_Kondo_resonance}). RXES is not sensitive to these low energy excitations but can provide complementary information with the advantage of bulk sensitivity and the added parameter of pressure. We discuss in the following applications of $2p3d$-RXES to Ce solid-solutions (chemical pressure) \cite{Rueff2004a,Dallera2004a} and elemental Ce under pressure across the transition. We next consider the case of Yb, a hole-type Kondo system which shows similarities with Ce, before discussing the possibility of multi-channel Kondo screening in TmTe.

\paragraph{Ce(Sc,Th)}
The $\gamma$-$\alpha$ transition is normally triggered by applying external pressure, but it may also be tracked as a function of temperature by using chemical pressure (cf.\ Fig.~\ref{fig:Ce_phase_diag}). In this case, the formation of the parasitic $\beta$-phase is normally avoided and sample handling is simplified although alloying effects with the doping element cannot be excluded. To this end, $2p3d$-RXES was applied to Ce solid solutions (Ce$_{0.93}$Sc$_{0.07}$, Ce$_{0.90}$Th$_{0.10}$, and Ce$_{0.80}$Th$_{0.20}$). Though no external pressure was applied here, the results of RXES in the solid solutions are useful as they highlight the main spectroscopic changes at the $\gamma$-$\alpha$ transition in Ce.
\begin{figure*}[htbp]
\begin{tabular}{cc}
{\small a)}\includegraphics[width=0.45\linewidth]{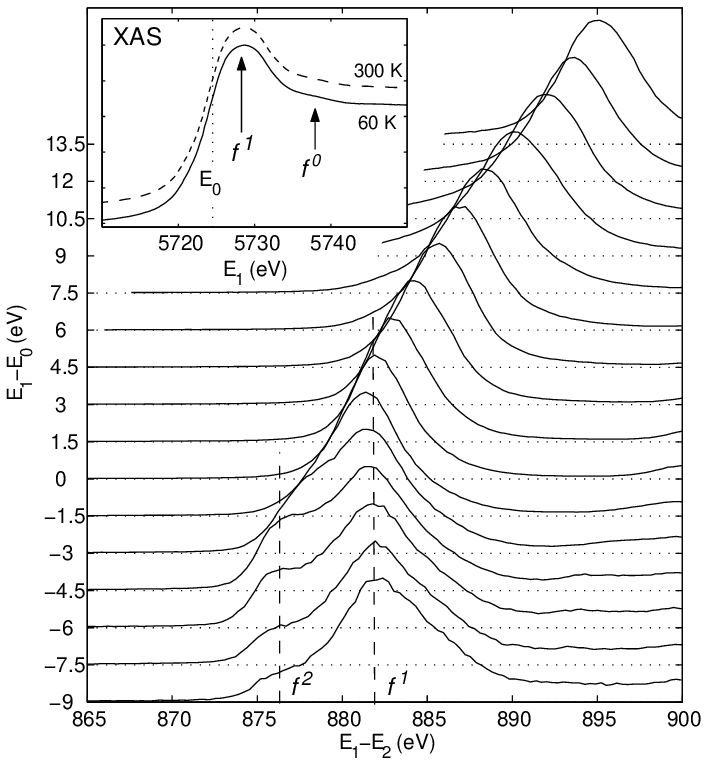} &
{\small b)}\includegraphics[width=0.45\linewidth, trim=0 -60 0 0]{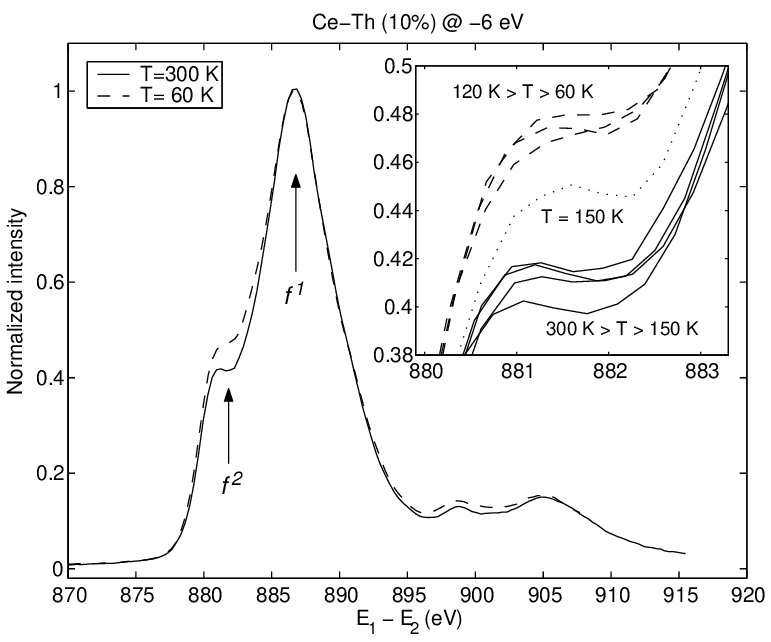} \\
\end{tabular}
\caption[Ce-Th RXES]{(a) Ce-$2p3d$ RXES spectra for Ce$_{0.90}$Th$_{0.10}$ at 60 K as a
function of the transfer energy; inset shows the XAS spectra at 60 K and 300 K. (b) Variation of the spectrum measured at the $f^2$ resonance while changing the temperature through the transition. From~\textcite{Rueff2004a}.}
\label{fig:CeSc_RXES}
\end{figure*}
The  Ce L$_3$ XAS spectrum measured in the total fluorescence-yield mode in one of the compounds (Ce$_{0.90}$Th$_{0.10}$) is shown in the inset to Fig.~\ref{fig:CeSc_RXES}(a)). The intense whiteline at 5728.8~eV and the very  weak feature at $\approx 5736$~eV are respectively ascribed to the mainly  $\underline{2p}4f^1$ and $\underline{2p}4f^0$ components. The $\underline{2p}4f^2$ configuration which is expected below the white line is not visible in the Ce L$_3$ XAS spectra but was observed by PFY-XAS \cite{Dallera2004a} in Sc-doped Ce. The changes across the $\gamma$-$\alpha$ transition are barely visible in the XAS spectra but come out clearly in the $2p3d$-RXES measurements.

The L$\alpha_1$-RXES spectra for Ce$_{0.90}$Th$_{0.10}$ measured at 60~K are shown in Fig.~\ref{fig:CeSc_RXES}(a) on a transfer energy scale. In the Raman regime ($E_1-E_2<$~$E_{edge}$), the spectra consist of a well-resolved double structure peaking at 876.1 eV  and 881.9~eV transfer energies---each component corresponding to an identifiable final state---while a single feature dominates in the fluorescence regime ($E_1-E_2>$~$E_{edge}$). The 881.9~eV peak resonates in the whiteline region and is assigned to the $\underline{3d}4f^15d^{n+1}$ final state. The peak at 876.1~eV  has its maximum intensity for excitations well below the whiteline and corresponds to the well screened $\underline{3d}4f^25d^{n}$ final state. The extra stability compared to the $f^2$ configuration results from the strong intrashell Coulomb interaction $U_{ff}$.

The resonant enhancement due to the RIXS process allows one to derive the variation of the $f^1/f^2$ ratio with temperature as it is cycled through the transition. Fig.~\ref{fig:CeSc_RXES}(b) illustrates the temperature dependence of the normalized RIXS spectra in Ce$_{0.90}$Th$_{0.10}$ measured at fixed incident energy. The $f^2$ shoulder shows a marked relative increase in intensity when the temperature is lowered below the transition. \textcite{McMahan2003} predict that the weight of doubly occupied states increases at the expense of single occupancy when the system goes from $\gamma$ to $\alpha$: At the $\gamma$-$\alpha$\ transition, the Ce--Ce interatomic distance dramatically shrinks, which strengthens the $f$ itinerant character through hybridization. The RIXS data confirms this tendency. However, electron interactions with the doping element (Sc or Th) can perturb the Ce-$4f$ electronic properties. Such a perturbation of the $f$-states has been observed, for instance, in the Sc-doped Kondo system YbAl$_2$ by XPS~\cite{Vesoco1991}.

\paragraph{Ce}
Thanks to the combination of perforated diamonds and the bulk sensitivity of RIXS, it has been possible to investigate directly the $\gamma$-$\alpha$\ transition in elemental Ce under pressure~\cite{Rueff2006a}. Without the alloying effects inherent to chemical substitution, the Anderson impurity model can be correctly applied. From there, one can derive the ground state $f$-counts in both $\gamma$ and $\alpha$ phases and more particularly the variation of $n_f$ and double occupancy across the transition. 
\begin{figure}[htb]
\includegraphics[width=0.90\linewidth]{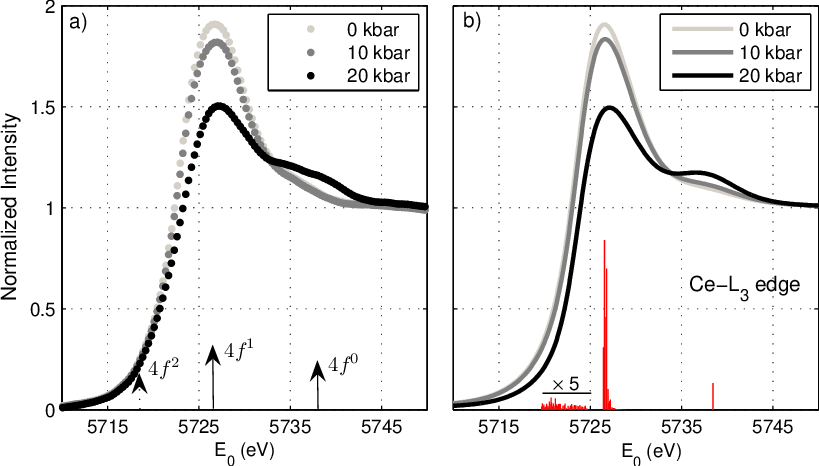}
\caption[XAS in Ce at high pressure]{(Color online) Experimental (a) and calculated (b) $L_3$ XAS spectra in elemental Ce as a function of pressure. Ticks in panel b) are the multiplet states (shown at 0 kbar). From~\textcite{Rueff2006a}.}
\label{fig:Ce_XAS}
\end{figure}
Figure~\ref{fig:Ce_XAS}(a) shows the experimental $L_3$ XAS spectra as a function of pressure. The white line exhibits a marked decrease in intensity as Ce is driven through the $\gamma$-$\alpha$ transition, while the feature denoted $4f^0$ progressively builds up at higher energy, the difference with the doped compounds being attributed to the sample purity. The overall spectral shape and the spectral changes at the transition are consistent with early results by Lengeler \emph{et al.}~\cite{Lengeler1983}. The $4f^2$ component is masked by the $2p_{3/2}$ core-hole lifetime. Fig.~\ref{fig:Ce_RXES}(a) illustrates the evolution of the $2p3d$-RXES spectra measured on resonance (at $E_0=5718.3$ eV) as pressure is increased.  The spectrum at 1.5~kbar barely shows a difference with the ambient pressure data.  However, a striking increase ($\approx 40$\%) in the $4f^2/4f^1$ intensity ratio is observed as the systems passes the $\gamma$-$\alpha$ transition pressure.

The data were analyzed by carrying out full multiplet calculations within the Anderson impurity model and a $f^0$,$f^1$ and $f^2$ configuration mixing. Details of the calculations in Ce can be found in \textcite{Rueff2006a}. The model calculations and Hamiltonian have been described in previous works~\cite{Kotani2001,Ogasawara2000a}. The XAS spectra (Fig.~\ref{fig:Ce_XAS}(b)) are well reproduced throughout the transition. The overall agreement for RIXS is equally good (Fig.~\ref{fig:Ce_RXES}(a)), except on the high energy-transfer side.  The discrepancy likely results from a fluorescence-like contribution to the spectra, which is not taken into account in the calculations. 
\begin{figure*}[htb]
\begin{tabular}{cc}
{\small a)}\includegraphics[width=0.50\linewidth]{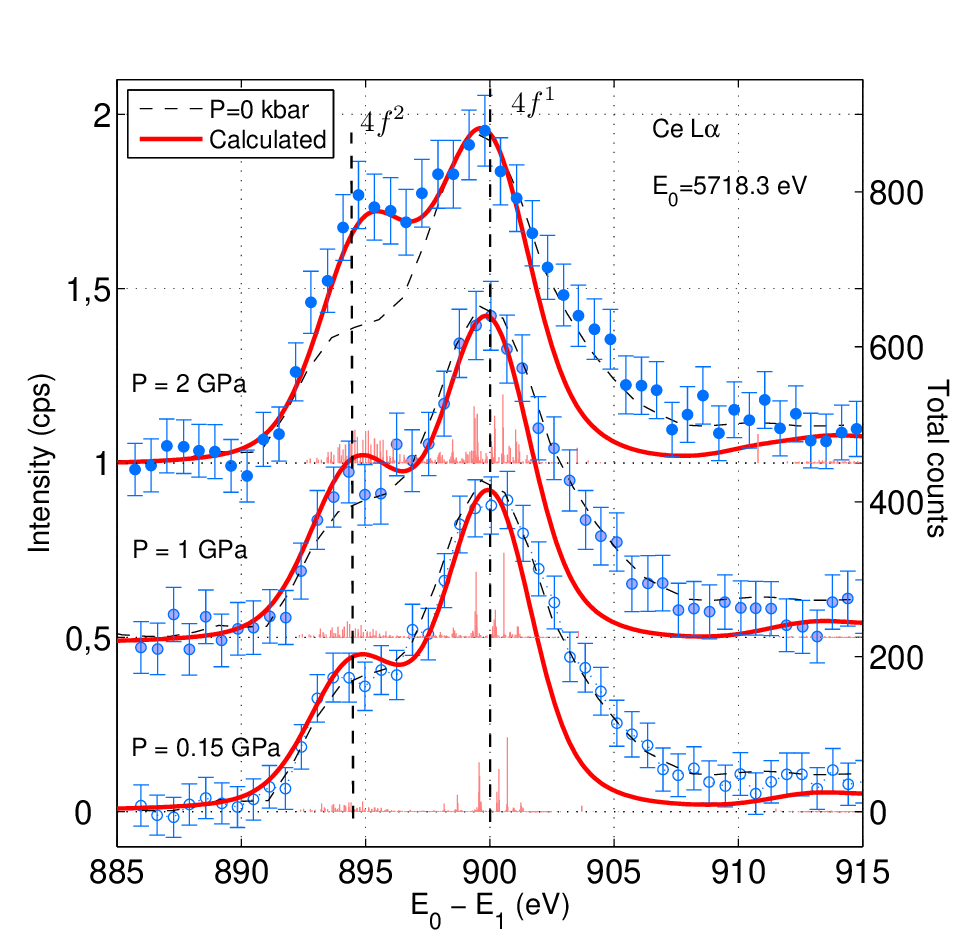} &
{\small b)}\includegraphics[width=0.40\linewidth]{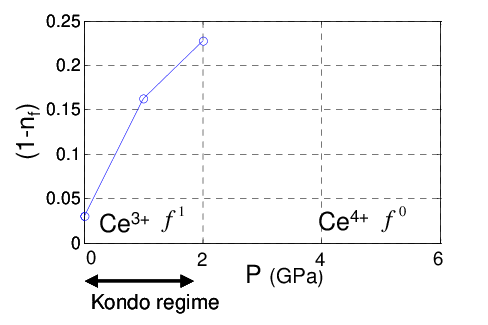}\\
\end{tabular}
\caption[RIXS in Ce at high pressure]{(Color online) (a) $2p3d$-RXES spectra in elemental Ce as a function of pressure ; thick lines are calculated spectra. From~\textcite{Rueff2006a}; (b) Change in $f$-occupation number $1-n_f$.}
\label{fig:Ce_RXES}
\end{figure*}
The main effect, according to the calculation, is the sharp decrease in the $4f^1$ component with respect to the $4f^0$-related feature, which gains intensity as Ce becomes more $\alpha$-like.  Such a trend is consistent with the spectral changes in the XAS spectra.  Formally, the transfer of spectral weight from the $4f^1 (5d^1)$ configuration toward a more $4f^0 (5d^2)$ configuration in the $\alpha$-phase can be understood as a partial delocalization of the $4f$ electrons. Interestingly enough, the highly hybridized $4f^2$ state also shows a sizable ($\sim$40\%) increase with pressure. The expanding contribution of the doubly-occupied state at high pressure stresses the reinforcement of the interaction between the $4f$ and the conduction electrons in the $\alpha$-phase, a characteristic feature of Kondo-like behavior. This growth of the double occupancy at low volume has another important consequence: it points to less correlation in the $\alpha$-phase as electron hopping is favored. Therefore, the picture that arises from the RIXS analysis at the $\gamma$-$\alpha$  transition is that of the coexistence of competing effects: partial delocalization of the $4f$ electrons through band formation with the conduction states on the one hand, and reduced electron-electron correlations on the other hand that allows the system to accommodate stronger on-site repulsion.  

The change in $n_f$ can be obtained from the calculated weights of the $4f$-components (cf.\ Fig.~\ref{fig:Ce_RXES}(b)). The results are consistent with earlier estimations obtained by photoemission~\cite{Wuilloud1983,Liu1992} for the $\gamma$-phase, but not for the $\alpha$-phase where the RXES values differ substantially; $n_f$-value is found to be 10--15\% lower. These new values of the $f$-occupation can be compared to recent {\textit{ab initio}} calculations using dynamical mean-field theory (DMFT)~\cite{Georges1996,Kotliar2006}. The discontinuous dependence of $n_f$ at the transition is well accounted for by DMFT~\cite{McMahan2003,Zolfl2001} in the low temperature limit.  On the other hand, the drop in $n_f$ at the transition is largely underestimated (4--10\% in the DMFT calculations while $\approx 20$\% according to the RIXS results).

$n_f$ deviates from unity in Ce as a direct consequence of non-zero hybridization.  As explained earlier, a remarkable manifestation of this Kondo behavior is the occurrence of the quasiparticle resonance at $E_F$ in the single-particle spectral function $\rho_f(\omega)$. The sharp decrease of $n_f$ in the $\alpha$-phase can be related to the enhancement of the quasiparticle peak and that of the renormalization of the bare particle which scales as $(1-n_f)$. The former effect is partly smeared out at temperatures comparable to the Kondo temperature $T_{\mathrm{K}}$~\cite{McMahan2003}.  $T_{\mathrm{K}}$ is here the key quantity to characterize the $4f$-electron coupling with the Fermi sea.  It can be evaluated thanks to the Friedel sum rule and given the approximate relationship $(1-n_f)/{n_f}\sim(\pi k_B T_K)/(N_f\Delta)$ \cite{Gunnarsson1983} in the limit of large $N_f$. The derived values of T$_K$ were 70~K in the $\gamma$-phase and 1700 K in the $\alpha$-phase assuming $\Delta\sim110$~meV.  The temperatures show a fair agreement with neutron scattering data~\cite{Murani1993} obtained in Ce-Sc alloys but differ very significantly from the generally accepted XPS-derived values~\cite{Liu1992}. They are consistently smaller by a factor $\sim 2$ in the $\alpha$-phase. The RIXS results demonstrates that the full characterization of the hybridized $f$-state necessitates an access to the bulk properties.

\subsubsection{A hole type Kondo system: Yb}
Like Ce, metallic Yb is characterized by a tendency to form an intermediate valence ground state. At ambient condition, Yb is divalent with an almost filled $4f$ shell, a configuration reminiscent of the quasi empty $f$ states in Ce at the opposite end of the rare earth series. Thus, $f$ holes in Yb are expected to play a role similar to that of the $f$ electrons in Ce. In contrast to Ce though, Yb undergoes two consecutive structural transitions: at 4 GPa from fcc to bcc phases and at 30 GPa where it transforms to the hcp phase. Furthermore, while pressure induces $f$ electron delocalization in Ce, it suppresses valency fluctuation in Yb and leads the Yb ions towards a localized trivalent state.

\begin{figure*}[htb]
\begin{tabular}{cc}
{\small a)}\includegraphics[width=0.50\linewidth]{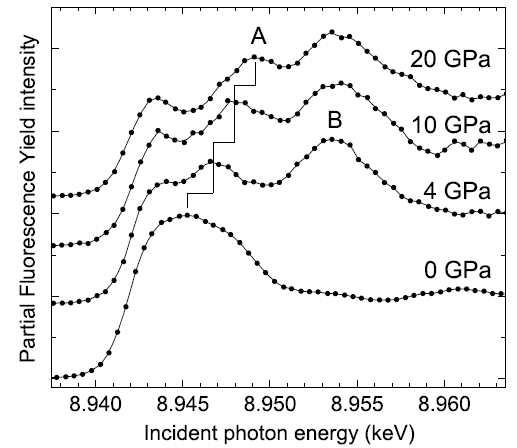} &
{\small b)}\includegraphics[width=0.40\linewidth, clip=true]{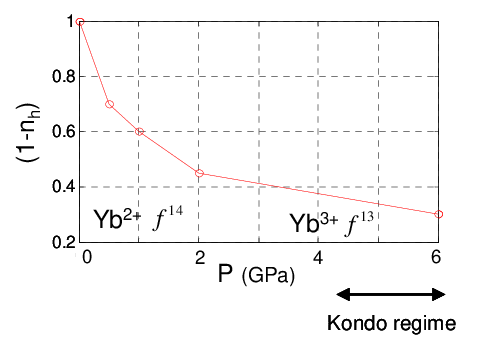}\\
\end{tabular}
\caption[PFY in Yb at high pressure]{(Color online) (a) PFY absorption spectra in elemental Yb as a function of pressure. From~\textcite{Dallera2006}; (b) Change of the Yb hole number $1-n_h$.}
\label{fig:Yb_PFY}
\end{figure*}

The Yb L$_3$ XAS spectra were measured as a function of pressure in the PFY mode at the Yb L$\alpha_1$ line (Fig.~\ref{fig:Yb_PFY}(a))~\cite{Dallera2006}. With increasing pressure, spectral weight is transferred from the edge region to a new peak $B$ at $\sim$10 eV higher energy. An additional peak $A$ is also observed in the mid energy region, which progressively shifts to higher energy. 
The spectral line shape is understood as the superposition of two replicas of the $d$ density of states, shifted in energy and weighted by the proportion of Yb$^{2+}$ and Yb$^{3+}$ in the ground state. The two extreme features correspond to the $\underline{2p}4f^{14}v^2$ and $\underline{2p}4f^{13}v^3$ final states split by the Coulomb interaction admixed with $5d$ character. Feature A however, cannot be associated to any $f$ states. It is in fact well accounted for by ab-initio calculations including dynamical screening of the core-hole \cite{Colarieti-Tosti2004,Dallera2006}. The transfer of spectral weight as $P$ increases reflects the enhancement of the Yb$^{3+}$ contribution at high pressure, in accordance with early XAS measurements~\cite{Syassen1982}. More precisely, Fig.~\ref{fig:Yb_PFY}(b) shows the evolution of the Yb mean valence $\bar{v}$ derived from Eq.~(\ref{eq:val}) as a function of pressure. We use the quantity $1-n_h$ where $n_h$ is the number of $f$ holes. Similar results are obtained by decomposition of the $2p3d$-RXES spectra (not shown) measured in the pre-edge region. The steep decrease at low pressure is indicative of the structural transition at 4 GPa. At higher pressure, $\bar{v}$ progressively increases until it reaches $\sim$2.55 at 20 GPa, the maximum pressure obtained during the experiment. An extra pressure point was simulated at 60 GPa, yielding a valency of 2.72 ($n_h=0.28$). This value is significantly lower than the previous estimation of near trivalency at 30 GPa of \textcite{Syassen1982}. 

The pressure dependence of the hole occupation number reported in Fig.~\ref{fig:Yb_PFY}(b) is a mirror image of the electron occupation number in Ce. But Yb differs from Ce in the proximity of the VCT to the Kondo regime. The latter settles when $1-n_h$ ($1-n_f$) is close to zero which corresponds to the low pressure region in Ce and the high pressure region in Yb. While the $\gamma$-$\alpha$ transition in Ce falls well within the Kondo regime, Yb only enters it at high pressure after the $f$-$d$ electron system is already significantly altered by band broadening effect. A striking point is that the dependence $T_K$ for the Ce$^{3+}$ configuration is expected to increase continuously as $n_f$ decreases. On the contrary due to the interplay of the $5d$ electron in Yb$^{3+}$, $T_K$ is expected to go through a maximum before decreasing \cite{Flouquet2005}. The difference between electron-type (Ce) and hole-type (Yb) Kondo temperature has been more precisely explained by the influence of two competing and contradictory effects under pressure: increase of hybridization and suppression of valency fluctuation~\cite{Goltsev2005}.

\subsubsection{Multi-Kondo channel: TmTe}
Could other electronic channels participate in Kondo screening? This question, known as the n-channel Kondo (NCK) problem has been invoked to explain the exotic behavior of materials such as magnetic nanodots and heavy fermion compounds. In their review article on exotic Kondo phenomena, \textcite{Cox1998} conjecture that pressure may induce an NCK effect in intermediate valent $f$-electron systems, eventually leading to non Fermi liquid behavior, as pressure can fine tune the hybridization between the impurity and the conduction bands.
A particularly intriguing case of NCK effect is foreseen in intermediate valent Tm compounds where the valence fluctuation of the Tm ion occurs between two magnetic states ($J=6$ and $J=7/2$). This is in contrast with the more usual Kondo ions where at least
one of the two fluctuating configurations is non magnetic: Ce$^{4+}$ ($f^0$), Yb$^{2+}$ ($f^{14}$) are all characterized by a zero angular momentum.
%
%
\begin{figure*}[htb]
\begin{tabular}{cc}
{\small a)}\includegraphics[width=0.35\linewidth]{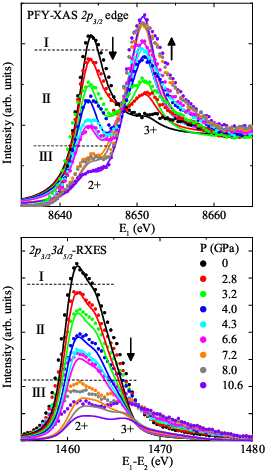} &
{\small b)}\includegraphics[width=0.50\linewidth]{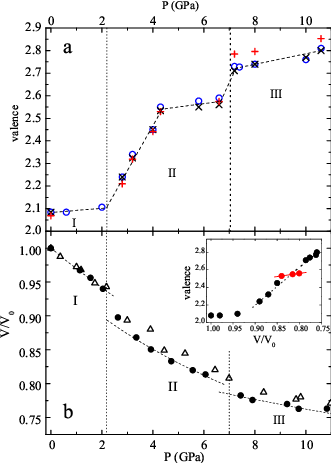}\\
\end{tabular}
\caption[Summary of TmX at high pressure]{(Color online) (a) (top) PFY-XAS spectra measured for TmTe at the Tm L$_{3}$-edge at pressures up to 10.6 GPa; (bottom) RXES spectra measured at the 2+ resonance (solid circles); solid lines are multiplet calculations; (b) (top) Pressure dependence of the Tm valency in TmTe as obtained by PFY-XAS (circles) and RXES (crosses) and calculations (full squares); (bottom) Relative volume change in TmTe from x-ray diffraction. From~\textcite{Jarrige2008}.}
\label{fig:TmX}
\end{figure*}

\textcite{Jarrige2008} have studied the Tm valence in TmTe by $2p3d$-RXES and x-ray diffraction under pressure. Figure~\ref{fig:TmX}(a) summarizes the spectroscopic results in TmTe under high-pressure. The pressure-dependence of the Tm valence was estimated by fitting independently the PFY and RXES spectra using a phenomenological approach similar to that described by~\textcite{Dallera2003} (cf.\ Fig.~\ref{fig:TmX}(b)). The valence $v$, initial found around 2 at the pressure (region I), increases abruptly above 2 GPa to reach 2.5 at 4.3 GPa (region II). The jump in $v$ coincides with the transition to the metallic regime \cite{Matsumura1997} and collapse of the unit cell volume. Above 4.3 GPa, the valence levels off as the volume recovers a normal compressibility behavior. The valence anomaly persists up to the structural transition near 7 GPa (region II to III) where $v$ suddenly increases from 2.58 to 2.72, and is expected to reach trivalency near 25 GPa.

The valence plateau in the intermediate pressure range above 4 GPa is regarded as a signature of NCK effects. In Tm ions, $T_K$ and hence the NCK effects are supposed to reach a maximum near $v=2.4$ \cite{Saso1989}, a value that matches the measured TmTe valence in the 4.3-–6.5 GPa pressure range. It is argued that, when more than one screening channels are involved, the contribution of the Kondo screening to the localization is sufficient to counterbalance the pressure-induced delocalization through band widening. Also, the variation of the Tm valence with pressure is clearly different form the continuous change usually observed in other compressed $f$-electron systems that are associated with a single-channel Kondo picture.

\subsection{Delocalization and mixed valent behavior}
\label{sec:Delocalization}
\begin{figure}[htb]
\includegraphics[width=0.90\linewidth]{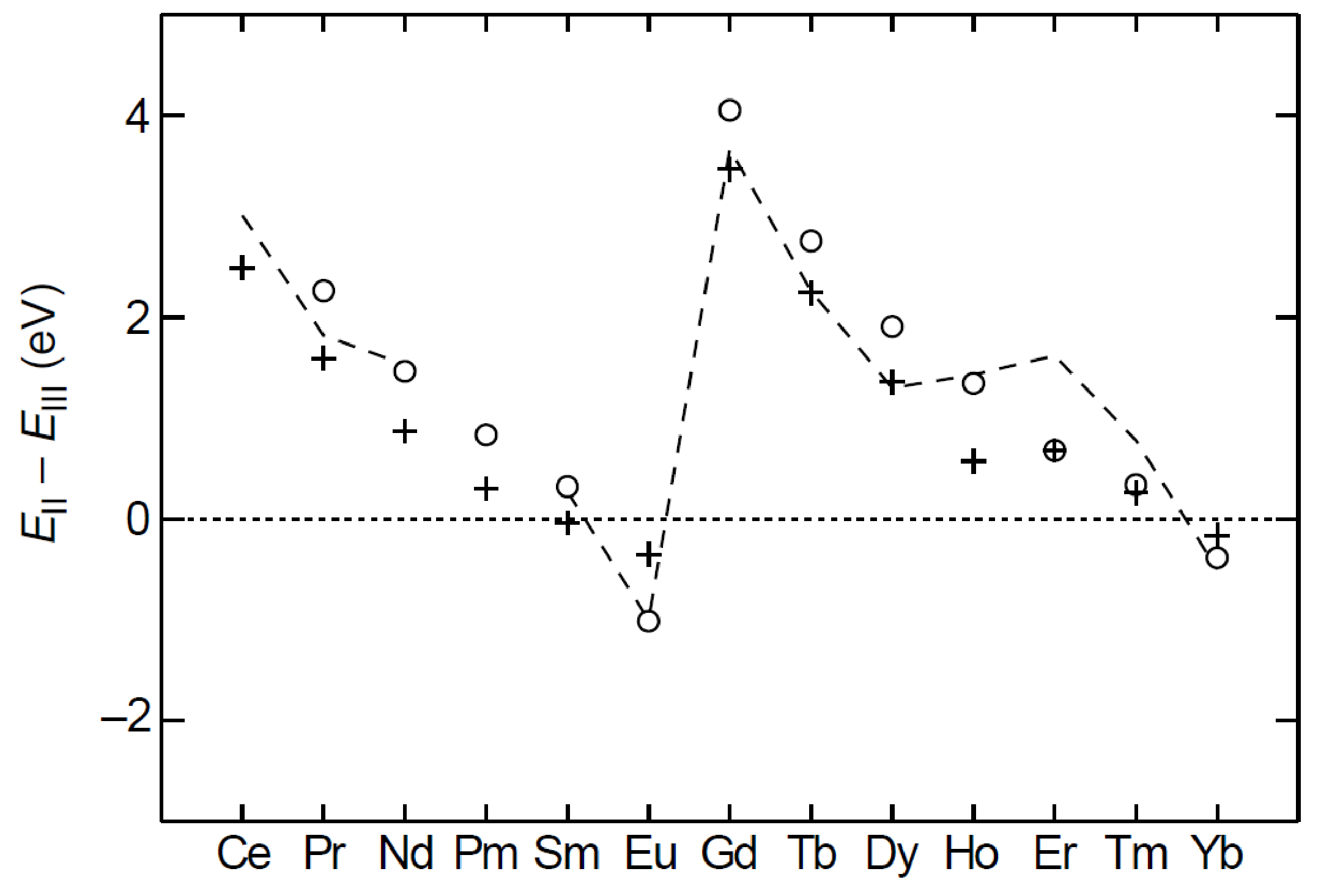}
\caption[Rare earth Valency]{Energy difference between divalent and trivalent $4f$ ions  calculated for rare earth metal (open circles) and sulfides (crosses). From~\textcite{Strange1999}.}
\label{fig:Strange}
\end{figure}

Besides intervening in Kondo effects, the $4f$ electrons play an important role in magnetic and structural properties of rare earth. \textcite{Strange1999} have estimated the rare-earth valency in the metallic phase and in sulfides from first principle local spin density (LSD) calculations including self interaction correction (SIC). In the SIC-LSD approach, $f$ electrons can be treated both as localized (where they experience a potential corrected from self-interaction) and band electrons (moving in a mean field potential) that are found only in trivalent systems. The merit of this approach is to allow for non-integer $f$ occupancy in contrast to other theoretical frameworks~\cite{Temmerman1999}. In the SIC-LSD picture, the effective valency $n_\text{eff}$ (i.e.\ the number of non-$f$ valence electrons) results from the hybridization of the $f$ band-state and the broad conduction band. The stability of either nominal divalent or trivalent configurations results from a trade-off between the localization energy and the energy gained by hybridization (cf.\ Fig.~\ref{fig:Strange}). YbS and SmS for instance are predicted to be divalent, on the verge of valence instability, while the trivalent Gd state is found to be highly stable. These opposite tendencies are expected to be at the root of the $f$ electron properties under pressure from delocalization at moderate or high pressures to metal insulator transitions.

\subsubsection{YbS and YbAl$_2$}
YbAl$_2$ and YbS both exhibit signatures of non-integer valence at ambient conditions. In YbAl$_2$ especially, a strong valence fluctuation and a correspondingly large Kondo temperature $T_K$=2000~K is inferred from inelastic neutron scattering~\cite{Gunnarsson1985a}. The $T$-dependence of the Yb valence in YbAl$_2$ has been estimated by a variety of bulk techniques such as thermodynamic measurements, magnetic susceptibility, and thermal expansion, as well as by spectroscopic probes including PES and inverse photoemission. Yet, these results are far from being consistent. Furthermore, pressure-dependent studies are limited, although valence change could be probed on a larger scale: A 0.2 valence increase was deduced from standard XAS in Yb solid solutions, using chemical pressure via Ca and Sc substitution~\cite{Eggenhoeffner1990}. 

\begin{figure*}[htb]
\begin{tabular}{ccc}
\includegraphics[width=0.30\linewidth]{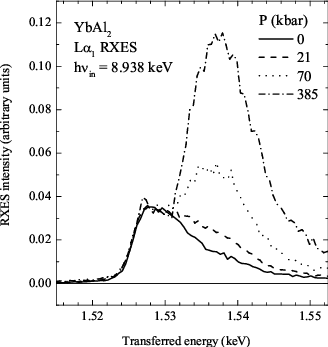} &  
\includegraphics[width=0.30\linewidth]{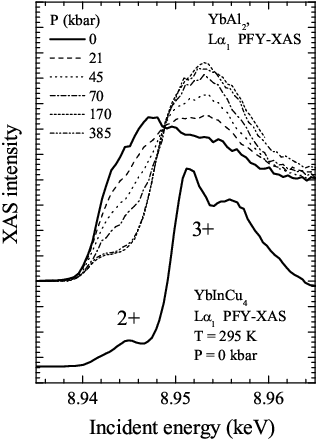} &
\includegraphics[width=0.30\linewidth]{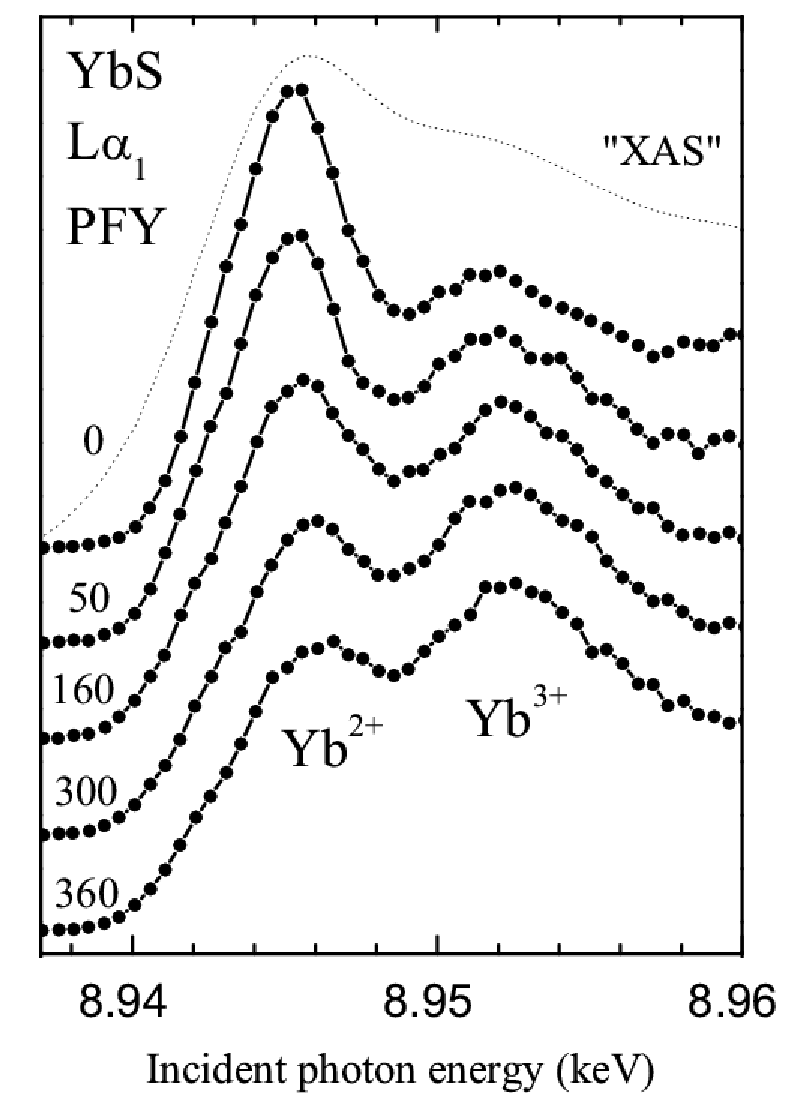} \\
{\small a)} & {\small b)} & {\small c)}\\
\end{tabular}
\caption[RXES in YbAl$_2$ and YbS at high pressure]{(a) $2p3d$-RXES in YbAl$_2$ measured at fixed incident energy set in the pre-edge region; (b) and (c) PFY-XAS spectra in YbAl$_2$ and YbS as a function of pressure. From~\textcite{Dallera2003} and \textcite{Annese2004}.}
\label{fig:YbX_PFY}
\end{figure*}

\begin{figure}[htb]
\includegraphics[width=0.90\linewidth]{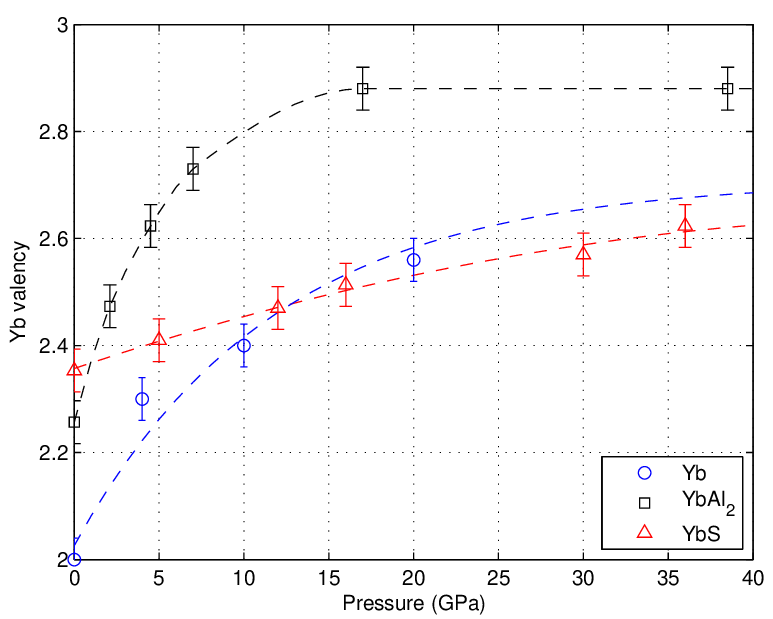}
\caption[Summary of Yb valency]{(Color online) Summary of the Yb valency as a function of pressure. From \textcite{Dallera2003,Annese2004,Dallera2006}; dashed lines are guide to the eyes.}
\label{fig:YbX_v}
\end{figure}

The RIXS spectra were obtained in YbAl$_2$ and YbS under high pressure~\cite{Dallera2003,Annese2004}. The pressure dependence of the L$\alpha_1$ PFY-XAS spectra in YbAl$_2$ is shown in Fig.~\ref{fig:YbX_PFY}(b). The ambient pressure spectrum is characterized by two well-separated features that are assigned to Yb$^{2+}$ and Yb$^{3+}$ components in the final state. The latter gains in intensity as pressure is increased, indicating the valence increase. Fig.~\ref{fig:YbX_v} illustrates this evolution in YbAl$_2$. Results obtained in Yb and YbS are shown for the purpose of comparison. The simple lineshape of the PFY-XAS spectra allows a direct estimation of the Yb valence from regular fitting procedure. The Yb valency is found to grow from $\sim$2.2 to 2.9 over 40 GPa pressure increase, as deduced from the PFY data. Consistent results are obtained using the RXES spectra (cf.\ Fig.~\ref{fig:YbX_PFY}(a)).

Fig.~\ref{fig:YbX_PFY}(c) displays the PFY spectra measured in YbS as a function of pressure. In contrast to YbAl$_2$ where two similar lineshapes for the 2+ and 3+ PFY component were used, the fitting procedure in YbS takes into account quadrupolar excited states observed in the pre-edge region (see the kink around 8.942 keV in Fig.~\ref{fig:YbX_PFY}(c)). This excitation involves $2p\rightarrow4f$ transitions in the intermediate states, which is only realized for Yb$^{3+}$ since Yb$^{2+}$ has a full $4f$ shell. The nature and position in energy of the quadrupolar peak was confirmed by multiplet calculations. Notice that quadrupolar features are visible neither in Yb nor in YbAl$_2$, even in the high pressure regime where the weight of trivalent Yb supposedly dominates the mixed-valent states. As discussed below, this difference reflects the various degree of $f$ electron localization in Yb compounds which also shows up in the variations of the Yb valence with pressure in Fig.~\ref{fig:YbX_v}: $\bar{v}$ in YbS slowly increases with pressure from 2.3 at 0 GPa up to 2.6 at 38 GPa, contrasting with the steeper increase in YbAl$_2$ and also Yb. 

SIC-LSD calculations in Yb compounds~\cite{Svane2000,Svane2001} predict that YbS is strongly divalent ($n_\text{eff}=2$ at $T$=0~K), as the trivalent excited state is located far above in energy. On the other hand, YbAl$_2$ is supposed to be weakly trivalent ($n_\text{eff}=2.46$). The RIXS-extracted valency (at $P=0$) is coherent with this picture in YbAl$_2$ though the experimental $n_\text{eff}$ is slightly underestimated compared to theory. This discrepancy may be ascribed to temperature effects as the divalent state is expected to contribute more at finite $T$. Calculations in YbS are more difficult to reconcile with the RIXS experimental value which shows a stark departure from divalency (note that indirect estimate of the Yb valency in YbS by diffraction~\cite{Syassen1985} did conclude on a divalent state at low pressure). The predicted stability of the divalent state in the sulfide presumably rules out temperature effects. On the other hand, YbS is a semiconductor, contrasting with the metallic character of YbAl$_2$ and also Yb, and is less accurately described by the SIC-LSD approach. 
YbS differs also by the sluggish variation of the valence state as a function of pressure. In divalent YbS, the two electrons provided by the rare earth ion fill the S-$3p$ band whereas they occupy the $s$-$d$ band in Yb. When going to the more trivalent state, the $f$ hybridized band state is pulled closer to the Fermi energy while the $s$-$d$ electronic structure is more or less unchanged contrary to the metallic materials. This indicates that $f$ electrons in Yb-sulfide are less affected by bonding, and significantly less sensitive to the lattice compression at high pressure compared to Yb. No structural change has indeed been reported in YbS even though the compressibility shows a small anomaly around 15--20 GPa. In the light of the RXES results, the anomaly cannot be attributed to a 2+ to 3+ valence transition as first proposed~\cite{Jayaraman1974} or the onset of valence instability~\cite{Syassen1985} but has to be related to gap closure and metallization. 
%

\subsubsection{A trivalent $4f$ ion: Gd}
\begin{figure*}[htb]
\begin{tabular}{cc}
{\small a)}\includegraphics[width=0.55\linewidth]{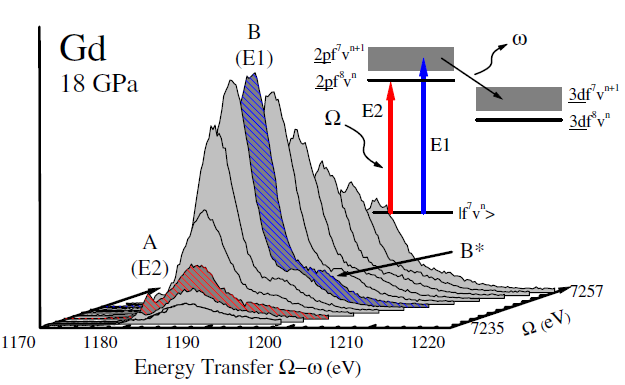}&
{\small b)}\includegraphics[width=0.35\linewidth]{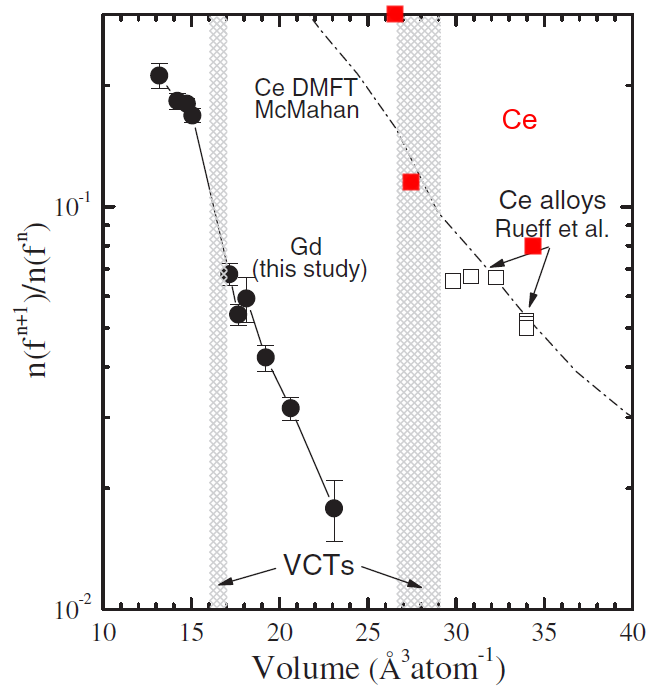}\\
\end{tabular}
\caption[Gd at high pressure]{(Color online) $2p3d$-RXES in Gd as at high pressure; b) experimental and calculated $f^{n+1}/f^n$ ratio as a function a pressure. Results in pure Ce (solid squares) from \textcite{Rueff2006a} have been added for comparison purpose. Adapted from~\textcite{Maddox2006}.}
\label{fig:Gd_RXES}
\end{figure*}
Similarly to Ce, Gd undergoes a volume collapse ($\Delta V/V\sim 5$\%) transition. But the latter occurs at much high pressure around 59 GPa with respect to other rare earth due to the exceptional stability of the Gd trivalent state. The $f$-delocalization in Gd under high pressure has been studied by \textcite{Maddox2006} using $2p3d$-RXES. Fig.~\ref{fig:Gd_RXES}(a) shows resonant spectra taken at 18 GPa. The spectra consist of dipolar ($B$ and $B^*$) and quadrupolar ($A$) excited states. At high pressure, an additional feature $C$ grows on the low energy side of $A$ which is interpreted as the signature of the increased valency. The data was interpreted assuming a $c_1\left|4f^6v^4\right\rangle+c_2\left|4f^7v^3\right\rangle+c_3\left|4f^8v^2\right\rangle$ mixed ground state. Feature $B$ and $C$ are attributed to $\underline{2p}4f^7v^4$ and $\underline{2p}4f^8v^3$ respectively. The progress of $4f$ delocalization was estimated
by $f^8/f^7$ spectral ratio. The results are shown in Fig.~\ref{fig:Gd_RXES}(b) which also display results from Ce-Sc and Ce under pressure. 

The smaller $f^8/f^7$ ratio compared to Ce at the same volume is consistent with the more localized $f$ electron in Gd since the $f$ shell is more tightly bound in the heavier
lanthanides due to the ever-increasing but incompletely screened nuclear charge. The continuous decay in Gd as a function of pressure suggests a Kondo-like aspect of the delocalization of the $f$ electron in fair agreement with DMFT predictions (in the Ce case). Interestingly enough, the volume instability in Gd falls approximately within the same $f^{n+1}/f^n$ region as in Ce where a low volume Kondo-state is favored.  

\subsubsection{Connection to Metal-insulator transition: SmS}
\label{sec:SmS}
\begin{figure*}[htb]
\begin{tabular}{cc}
{\small a)}\includegraphics[width=0.35\linewidth,trim=0 0 370 0,clip=true]{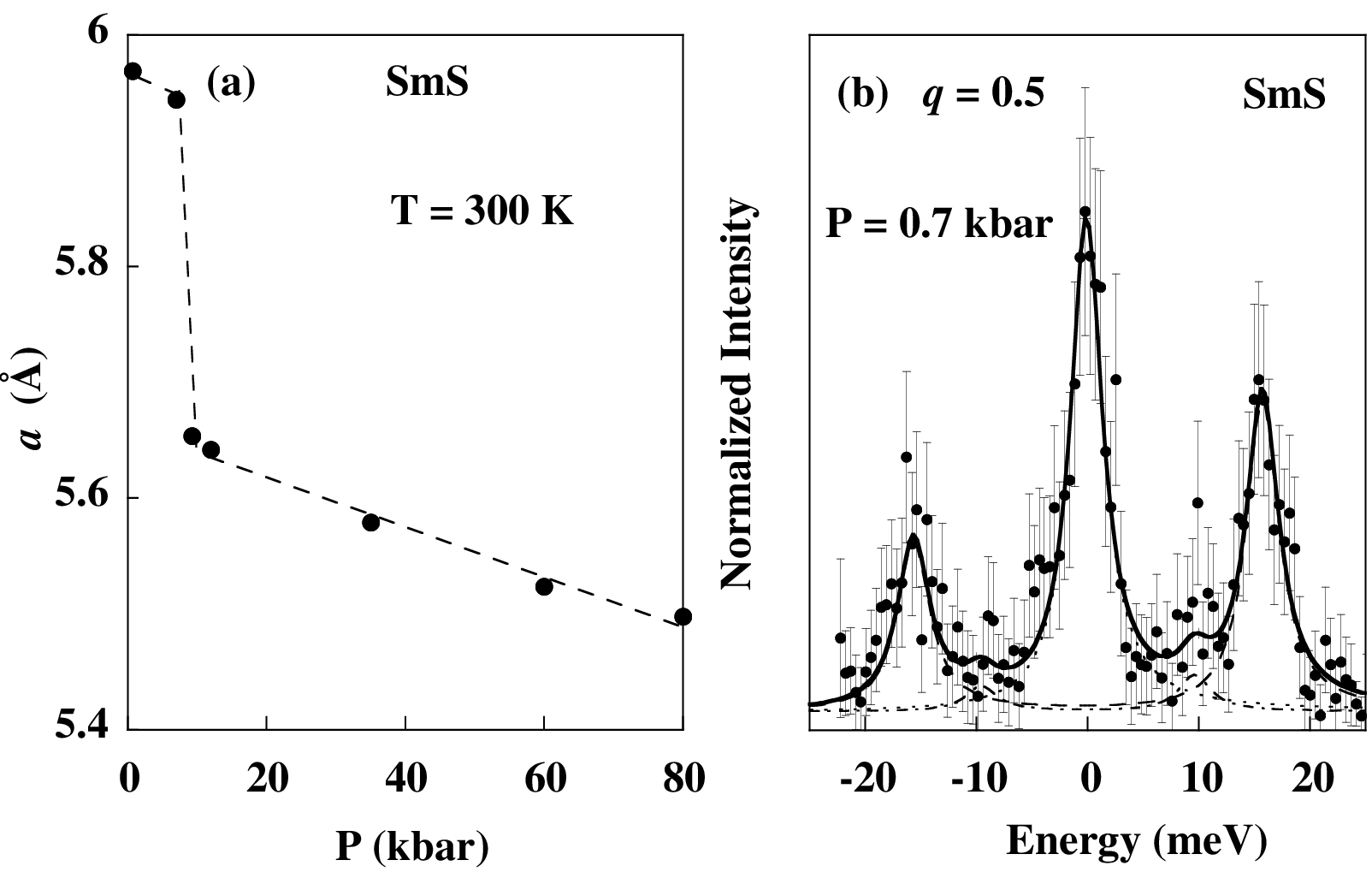} &
{\small b)}\includegraphics[width=0.55\linewidth]{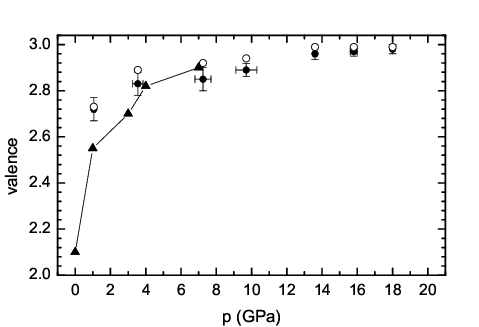}\\
\end{tabular}
\caption[SmS at high pressure]{(a) Pressure variation of the lattice parameter $a$ of SmS
at $T$=300~K. From~\textcite{Raymond2002}; (b) Variation of the Sm valence in SmS obtained by RIXS (circles)  and standard x-ray absorption (triangles). From~\textcite{Annese2006}.}
\label{fig:SmS_RXES}
\end{figure*}

SmS is considered as a model system for $f$ electron delocalization as the interplay between charge, lattice, and magnetic degrees of freedom is at its strongest among the rare earth series. At ambient pressure, SmS is a semiconductor which crystallizes in the NaCl structure (black phase) with a divalent non-magnetic configuration ($4f^6$). At 0.65 GPa and room temperature, it undergoes a first-order isostructural phase transition to a metallic state (gold phase), marked by a significant contraction of the unit cell (cf.\ Fig.~\ref{fig:SmS_RXES}(a)). In the high pressure phase, the Sm ion is supposedly in an intermediate valence state. In contrast to the room temperature behavior, the semi-conducting state persists at T=0 K up to $P_\Delta$=2~GPa, where the sample ultimately becomes metallic. The transition towards a magnetic ground state at 2 GPa was confirmed by nuclear forward scattering (NFS) experiments performed at low temperature~\cite{Barla2004}. A value of 0.5~$\mu_B$ was estimated for the Sm magnetic moment, which points to a trivalent state. Magnetism was found stable up to 19 GPa
with an ordering temperature continuously increasing with pressure. This makes unlikely the presence of a QCP in SmS as initially thought.

The pressure dependence of the Sm valency was studied by~\textcite{Annese2006} using $2p3d$-RXES from 3 to 18 GPa at room temperature. The results (Fig.~\ref{fig:SmS_RXES}(b)) compare well with early estimations obtained by XAS at the Sm L$_3$ edge~\cite{Roehler1982} that are extended here to much higher pressure. RXES indicates that the onset of long-range magnetic order at $P_\Delta$ is not correlated with transition to the full trivalent state ($P_c\sim 13$ GPa), but occurs beforehand at a valence $\bar{v}$=2.8. On the other hand, NFS data from \textcite{Barla2004} do not show any anomaly around $P_c$. In fact, barely any change in the magnetic properties is observed from 2 to 19 GPa. It seems therefore that the Sm $4f$ electrons behave magnetically like a completely trivalent ion well before the pressure $P_c$ for the transition to the pure trivalent state is reached.

\subsection{$f$ band states: Actinides}
\subsubsection{U heavy fermions}
\label{sec:U_comp}
In the U-compounds, U normally exists in two valencies with nominal U$^{3+}$ (5$f^3$) and U$^{4+}$ (5$f^2$) ionic configurations. Yet its exact valency is poorly characterized since bulk magnetic measurements cannot distinguish between two valent states having a similar paramagnetic moment. In fact, the tendency of the $5f$ states to hybridize with the conduction electrons is likely to lead to a non-integer $5f$ occupancy. That the U structure in the $\alpha$ phase is reminiscent of the Ce structure further supports this idea as first suggested by \textcite{Ellinger1974}.

The extension of $2p3d$-RXES to actinides provides an alternative way for determining the U valency and following its evolution with pressure. But it is only recently that the delocalization of $5f$ states under high pressure has been investigated by RXES, following the surge of activity related to superconducting U heavy fermion compounds. 

\paragraph{UPd$_3$,UPd$_2A$l$_3$}
UPd$_{3}$ is a clear-cut example of a well defined $5f^{2}$ state. Divalency is confirmed by neutron spectroscopy via the measurement of crystal field excitations~\cite{Buyers1987} and photoemission~\cite{Ito2002}. Pressure-induced delocalization toward a $5f^1$ state at 25~GPa was predicted by~\textcite{Petit2002} using self-interaction corrected local spin density (SIC-LSD). However no corresponding effect of volume collapse was observed experimentally up to 53 GPa by x-ray diffraction under pressure~\cite{Heathman2003}. UPd$_{2}$Al$_{3}$ is an antiferromagnetic superconductor ($T_{N}$=14 K, $T_{c}$= 2K), characteristic of the interplay between magnetism and superconductivity and of a moderate heavy fermion character. Contrary to UPd$_3$, UPd$_{2}$Al$_{3}$ undergoes a structural phase transition around 23.5 GPa. The doubling of the compressibility was interpreted by a valence change~\cite{Krimmel2000} induced by the partial delocalization of the $f$ electrons~\cite{Zwicknagl2003}. The dual nature of the $5f$ electrons is supposedly illustrated by the behavior at ambient conditions of another U-compound, UPt$_{3}$, a heavy fermion superconductor ($T_c$=0.5 K). The unconventional character of UPt$_3$ is considered to be partly related to the coupling between the localized $f^2$ state and delocalized $f$ electrons~\cite{Zwicknagl2002} which are found at the Fermi energy~\cite{Allen1992}. 

\begin{figure}[htbp]
\includegraphics[width=0.90\linewidth]{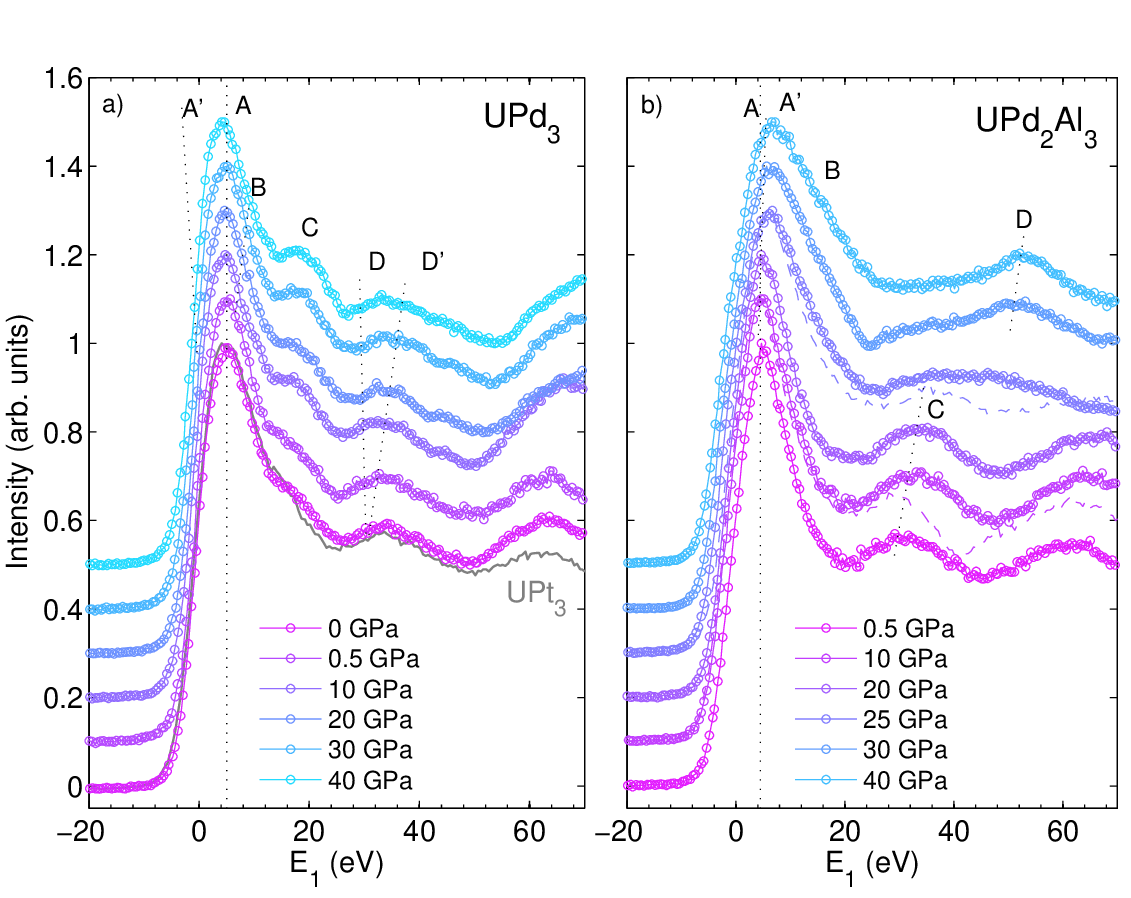}
\caption[PFY-XAS in UPd$_3$ and UPd$_2$Al$_3$ at high pressure]{(Color online) PFY-XAS spectra in UPd$_3$ (a) and UPd$_2$Al$_3$ (b) as a function of pressure (open circles). Ambient PFY-XAS spectrum of UPt$_3$ is shown (black line) for comparison purpose. From~\textcite{Rueff2007}.}
\label{fig:UPdX_PFY}
\end{figure}

Figure~\ref{fig:UPdX_PFY} illustrates the PFY x-ray absorption spectra at the U-L$_3$ edge in UPd$_3$ and UPd$_2$Al$_3$ as a function of pressure~\cite{Rueff2007}. 
The UPd$_3$ spectra show well defined peaks ($A$, $C$ and $D$ in the figure) in the edge region but no $f$-related features in the pre-edge region contrary to the rare earth compounds. As pressure increases, the structures $C$ gains in intensity while the high energy features $D$ and $D'$ seemingly split in energy. Simultaneously, a shoulder $A'$ appears on the low energy side of the white line that gradually becomes asymmetric.
The evolution of the U-L$_3$ edge  in UPd$_2$Al$_3$ as a function of pressure (Fig.~\ref{fig:UPdX_PFY}(b)) strongly differs from UPd$_3$. The spectra barely vary 
up to 20 GPa except for the progressive increase of a second feature which appears as a shoulder to the white line ($A'$) along with a slight energy-shift of the white line itself ($A$) and of the high energy oscillations ($C$). The white line suddenly broadens above the structural transition while the high energy oscillating pattern reduces to a single peak ($D$).


The data is compared to \textit{ab-initio} calculations of the U L$_3$ edges with the linear muffin tin orbital (LMTO) method in the LDA approximation. In UPd$_3$, this method was shown to give a pertinent solution with two localized $f$ electrons~\cite{Yaresko2003}. The $5f$ levels in UPt$_3$ are found to be partly delocalized in agreement with the XPS results. 
\begin{figure}[htb]
\includegraphics[width=0.90\linewidth]{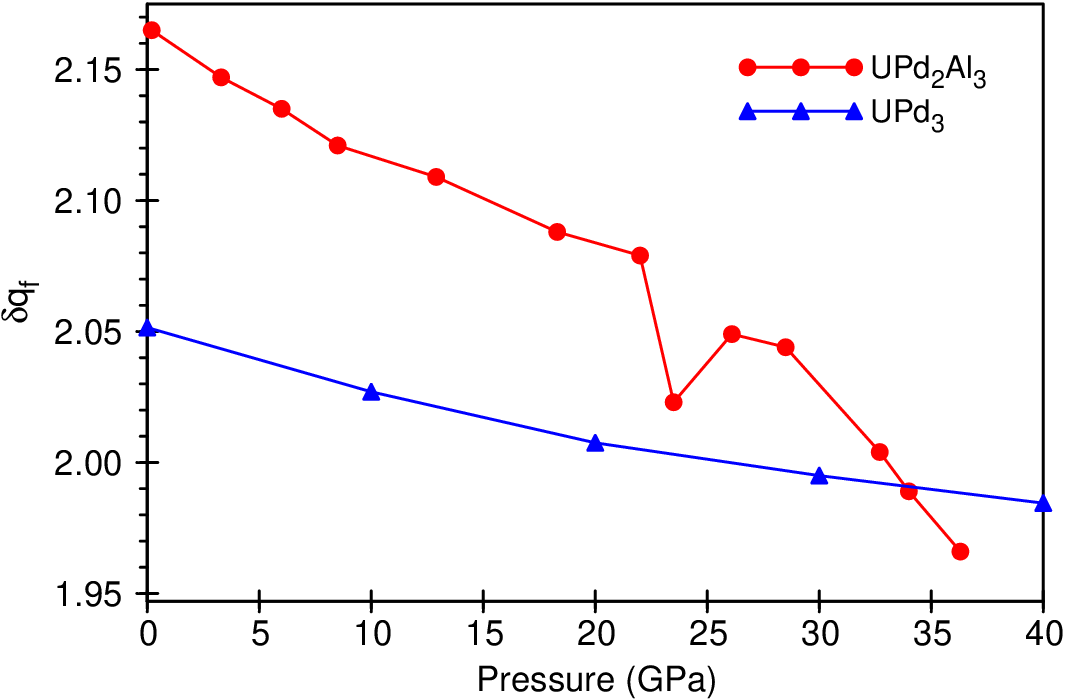}
\caption[Occupation number in UPd$_2$Al$_3$ and UPd$_3$ with pressure]{(Color online) Calculated occupation number in UPd$_2$Al$_3$ and UPd$_3$ as a function of pressure. From~\textcite{Rueff2007}.}
\label{fig:UPdX_qf}
\end{figure}
The LMTO calculation cannot reproduce the exact lineshape of the absorption spectra but yields a reasonable estimation of the $5f$ occupancy number (cf.\ Fig.~\ref{fig:UPdX_qf}). 
When discussing the U 5$f$ occupation, two problems arise: the first is the unknown structure of UPd$_2$Al$_3$ at high pressure; the second is due to the tails of the states centered on neighboring sites that have the $f$ character inside the uranium atomic sphere. 
The structure of ThPd$_2$Al$_3$ was used with the atomic radius of uranium. To solve the second issue, the difference $\delta q$ between the $f$ electron density inside the U atomic sphere ($q^{U}_{f}$) and that inside a Th sphere ($q^{Th}_{f}$) which substitutes a U atom is calculated. The correction serves to exclude the contribution from the Pd and Al states with $f$ symmetry inside the U atomic sphere. 
This approach is valid as long as the $f$ electrons are not too strongly hybridized.

In UPd$_3$ the occupation of the U 5$f$ shell averaged over two non equivalent U sites monotonously decreases from 2.05 at ambient pressure to 1.98 at 40 GPa. In the whole pressure range $\delta q_f$ remains very close to 2 which suggests that the valence state of U ions does not change under lattice compression. Thus, the picture that emerges is that of a localized $f^2$ configuration, consistent with the diffraction data of \textcite{Heathman2003} and former band calculations by Ito et al.\ \cite{Ito2002}. It definitely rules out the prediction of a $f^2$ to $f^1$ transition under pressure reported in \textcite{Petit2002}. 

In UPd$_2$Al$_3$,  $\delta q_f$=2.17 at ambient pressure indicates that the U ion is in an intermediate valence state. $\delta q_f$ then gradually decreases with pressure with a somewhat higher rate above the structural transition. Comparing the UPd$_2$Al$_3$ data to the $\delta q_f(0)$=2.05 for UPd$_3$, for which the U 5$f^2$ configuration is well established, one can suppose that the structural transition at 23.5 GPa is related to the change of the valence state of a U ion from an intermediate U$^{(4-\delta)+}$ valency to U$^{4+}$. This semi-qualitative analysis does not allow to answer the question whether the U valency in the high pressure phase remains integer or becomes $U^{(4+\delta)+}$. It nevertheless agrees with the current understanding of the U valence in UPd$_2$Al$_3$ which is described by a coexistence of localized and delocalized $f$ electrons. The mixed valent state already formed at ambient pressure deviates from the preceding assumption of 2 localized and 1 delocalized $f$ electrons~\cite{Petit2003}. 

\section{Bonding changes in light elements}
\label{sec:light_element}
The discussion so far was limited to resonant spectroscopy where absorption and emission are combined in a single process defined as RIXS. X-ray absorption, the first step of the RIXS process, is itself a very widely used tool and can be used as a probe of the electronic structure when the near edge structure is measured. XAS is notably advantageous in the soft x-ray range because of applications to elements of wide interest but also because the low photon energy can result in a high resolving power. In this last section we look at a particular aspect of IXS where absorption edges of light elements are measured via the inelastic scattering of hard x-rays.

\subsection{Soft x-ray XAS vs.\ XRS}
In this section, we address the particular case of the K-edges of light elements which has been one of the major domains of application of XAS for its sensitivity to chemical bonding,  coordination, or molecular level~\cite{Stohr1992}. The methods of detection may vary from fluorescence yield, to Auger or electron yield or sample photocurrent. Despite differences in probing depth related to different methods of detection, soft x-ray XAS is highly sensitive to the sample surface. In the soft x-ray range, the penetration depth of x-rays is typically of the order of 50 \AA. This is not an issue as long as the surface is clean and the sample environment transparent. However, when it comes to high pressure experiments in a pressure cell, soft x-ray XAS is not applicable.  

\begin{figure}[htb]
\includegraphics[width=0.90\linewidth]{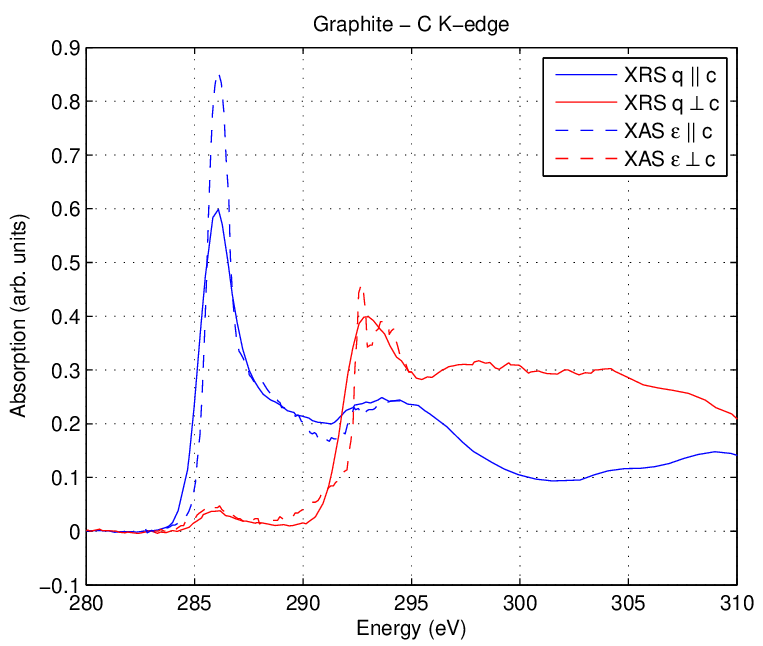}
\caption[C K-edge in graphite]{(Color online) C K-edge in highly oriented pyrolitic graphite (HOPG) as obtained by x-ray Raman scattering (XRS) from~\textcite{Rueff2002} and soft x-ray absorption spectroscopy (XAS) from~\textcite{Bruhwiler1995}. The energy scale refers to the transfer energy (resp.\ incident energy) for XRS (resp.\ XAS). The momentum transfer ($q$) and polarization vector ($\epsilon$) are set either parallel ($\|$) or normal ($\perp$) to the graphite $c$ axis.}
\label{fig:graphite}
\end{figure}

In contrast, X-ray Raman scattering offers the possibility to access core electronic level through a high energy scattering process. As explained in section~\ref{sec:XRS}, the method is equivalent to soft x-ray XAS providing the momentum transfer $q$ is chosen small enough compared to the core wave function spatial extension (forward scattering geometry). Second, $\mathbf{q}$ acts as the polarization vector $\boldsymbol\epsilon$ in XAS and can be used to project the final states onto directions of high symmetry. This is illustrated in Fig.~\ref{fig:graphite} in the case of highly oriented pyrolytic graphite (HOPG). The XRS spectra were measured at a scattering angle of 10$^\circ$ with $\mathbf{q}$ set parallel or normal to the $c$ axis~\cite{Rueff2002}. The XRS spectra (solid lines) compare well, though less resolved, with polarization dependent soft x-ray XAS at the C K-edge (dashed lines)~\cite{Bruhwiler1995}. 

Evidently, the price to pay for using non-resonant scattering is the low cross section with respect to resonant spectroscopy. The poor efficiency of XRS explains the limited number of experiments, summarized in Table~\ref{table:light-elements}, which have been performed under high pressure. This difficulty nevertheless can be overcome in a straightforward manner by enlarging the collecting solid angle of the spectrometer. Combining several crystal analyzers in an array~\cite{Bergmann1998} or  diminishing the crystal bending radius~\cite{Gelebart2007} allows a significant gain in intensity. 
\begin{table}[htbp]
\caption[Light Elements]{Changes in light elements under pressure investigated by x-ray Raman scattering. $P_{max}$ stands for the maximal pressure reported in the experimental work.}
\begin{tabular}{lcccc}
\hline\hline
Sample & K-edge & $P_{max}$ (GPa) & changes\\
\hline
v-B$_2$O$_3$ & B,O & 22.5 & tri$\rightarrow$tetra-coordinated\footnote{\textcite{Lee2005a}}\\
BN & B,N & 18 & $sp^2\rightarrow sp^3$\footnote{\textcite{Meng2004}}\\
C (graphite) & C & 23 & $\pi\rightarrow\sigma$\footnote{\textcite{Mao2003}}\\
C$_{60}$ & C & 20 & $sp^2\rightarrow sp^3$\footnote{\textcite{Kumar2007}}\\
C$_6$H$_6$ & C & 20 & hybridization change\footnote{\textcite{Pravica2007}}\\
H$_2$O & O & 0.25 & ordering of O-H bonds\footnote{\textcite{Cai2005}}\\
H$_2$O & O & 0.03 & supercritical water\footnote{\textcite{Wernet2005}}\\
H$_2$O & O & 12.5 & H$_2$,O$_2$ dissociation\footnote{\textcite{Mao2006a}}\\
H$_2$O & O & 0.6 & H-bonding increase\footnote{\textcite{Fukui2007}}\\
\hline\hline
\end{tabular}
\label{table:light-elements}
\end{table}

\subsection{Coordination chemistry under pressure}
In the low $q$ limit, the XRS cross section is dominated by the dipolar term (cf.\ section~\ref{sec:XRS}). At the K-edges therefore the XRS final states are orbitals of $p$-symmetry whose nature ($\pi$, $\sigma$) or bonding character can be probed, similarly to soft x-rays XAS or electron energy loss spectroscopy (EELS). But because high energy photons are involved in the XRS process, it is now possible to investigate how these evolve in-situ under extreme conditions

\begin{figure}[htb]
\includegraphics[width=0.90\linewidth]{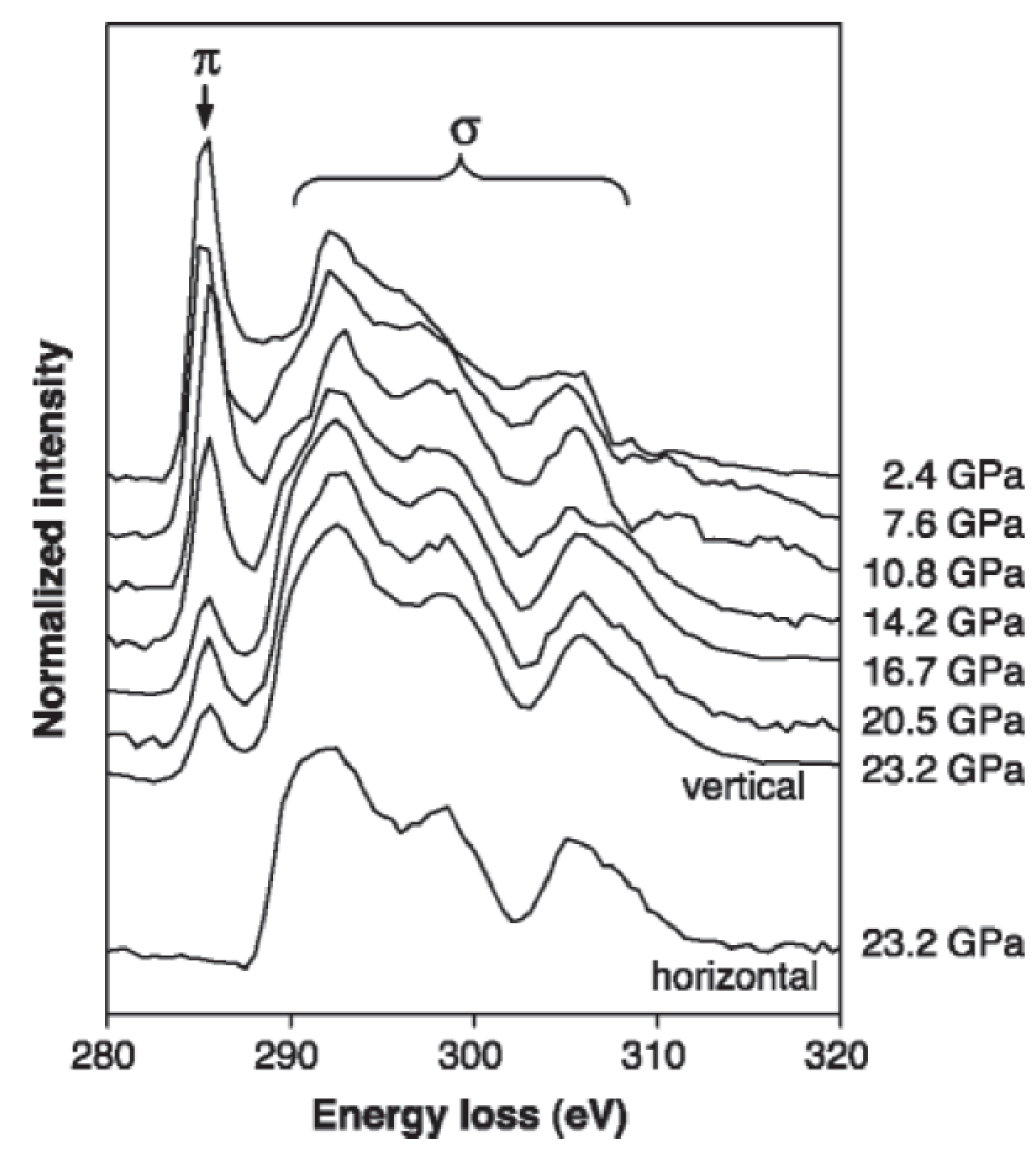}
\caption[C K-edge in graphite under pressure]{C K-edge in graphite under pressure measured by x-ray Raman scattering (XRS). From~\textcite{Mao2003}.}
\label{fig:graphite_HP}
\end{figure}

An emblematic example of pressure-induced bonding change is graphite, one of the simplest 2D materials. Under pressure graphite undergoes a metal-insulator transition around 15 GPa which is signaled by a significant drop in reflectivity, a broadening of the vibrational modes and a change of the x-ray diffraction pattern. The observation of the C-K edge in compressed graphite has clarified the mechanism of this transition. Fig.~\ref{fig:graphite_HP} shows the variation of the C K-edge in compressed graphite at ambient temperature. The narrow $\pi^*$ features at low energy are related to the $2p_z$ antibonding orbital while the broad humps on the high energy side corresponds to $\sigma^*$ in-plane bonds. With increasing pressure, about half of the $\pi$-bonds transforms to $\sigma$-bonds~\cite{Mao2003} as deduced from the transfer of spectral weight under pressure. The conversion implies a partial change from $sp^2$ carbon to an $sp^3$ form, though without a full transformation to diamond like structure. This apparently contrasts with fullerene, another form of $sp^2$ carbon at ambient conditions, which was reported to convert fully to $sp^3$ diamond structure under pressure~\cite{Kumar2007}. One has to remain cautious however with the quantitative interpretation since the $\pi^*$ and  $\sigma^*$ C near edge features consist of excitonic excited states~\cite{Bruhwiler1995} whose binding energy and localization will be strongly affected by pressure.

\begin{figure}[htb]
\includegraphics[width=0.90\linewidth]{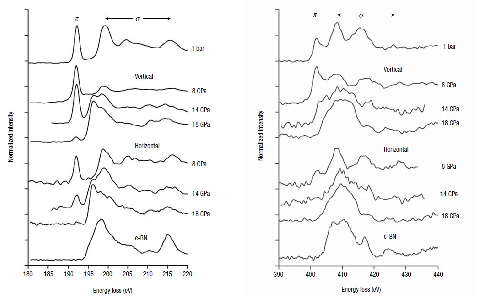}
\caption[B and N K-edge in BN under pressure]{B (left) and N (right) K-edges in BN under pressure measured by x-ray Raman scattering (XRS). From~\textcite{Meng2004}.}
\label{fig:BN_HP}
\end{figure}

That the compressed graphite presents a remarkable hardness is a clear indication that a new form of carbon is synthesized at high pressure. Boron shows a comparable sensitivity to the compressed lattice leading to an change of the local coordination and structural transformations, aided in compounds by the hybridization with the ligand $p$ states. In-situ measurements of the B and O K-edges in B$_2$O$_3$ glass by XRS~\cite{Lee2005a} for instance has proven the conversion of tri-coordinated B to tetra-coordinated B under pressure while the O-$\pi^*$ bonds are progressively transformed into $\sigma^*$ type. Figure~\ref{fig:BN_HP} illustrate the case of hexagonal BN under pressure investigated by XRS.
The spectral changes at the B-K edge denote a transformation of the B coordination shell from $sp^2$ to $sp^3$ above 14 GPa. The transition is accompanied by a structural transformation of h-BN to the hexagonal close-packed structure (w-BN), another polymorph of BN. Simultaneously, the N K-edge spectra reflect the diminution of $\pi$ bonds in the compressed BN to the depend of $\sigma$ ones. The observed chemical changes provide a mechanism for the densification of BN under pressure.  

\subsection{Structure of water and ice: Hydrogen bonding}
Compared to the 2D materials, water exhibits a far greater complexity. Both liquid and solid phases forms a three dimensional network of H$_2$O molecules linked by hydrogen bonds that lead to a rich variety of phases under specific temperature and pressure conditions. Beside infrared spectroscopy and x-ray diffraction, soft-x ray XES at the O-K$\alpha$ line \cite{Guo2002} and XAS at the O K-edge \cite{Wernet2004} have been suggested as probes of the local bonding configurations in water. When performed via the XRS process, the latter gives readily access to high pressure phases of water that are not attainable by soft x-ray techniques. 

\begin{figure}[h!]
\includegraphics[width=0.80\linewidth]{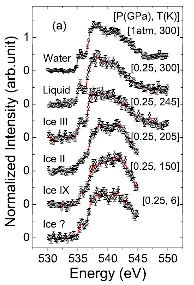}
\caption[O K-edge in water and ice]{(Color online) O K-edge in water and ice phases as a function of pressure and temperature. From \textcite{Cai2005}.}
\label{fig:H2O_XRS_Cai}
\end{figure}

Figure~\ref{fig:H2O_XRS_Cai} shows XRS spectra of water and ice measured at the O K-edge in a pressure cell as a function of temperature at a pressure of 0.25 GPa. Starting from liquid water at high temperature, the experiment explores successively the phases III, II and IX of ice upon cooling. Discernible pressure-dependent effects can be observed, especially in the pre-edge and post-edge regions. 
According to density functional calculations for liquid water and ice \cite{Wernet2004}, the strength of the near-edge structure can be related to the number of uncoordinated hydrogen bonds. From the liquid phase to ice III at 0.25 GPa, for instance, the decrease of the pre-edge and main edge intensities is understood as a consequence of the ordering of the oxygen framework which reduces the number of uncoordinated hydrogen bonds. The trend seemingly continues from ice III onwards. But the validity of such an interpretation has been questioned recently by among others \textcite{Prendergast2006}. Using ab-initio calculations, the authors found that the pre-edge structures is mostly of excitonic nature, thus with little bearing on the local environment.

\section{Summary and perspectives}
\label{sec:conclusion}
In this article we have reviewed the spectroscopy of electronic properties of materials under extreme conditions from the point of view of inelastic x-ray scattering. IXS, whether in the resonant or non-resonant mode has several useful features. It is an all photon technique with bulk sensitivity and superior penetration depth, therefore compatible with difficult sample environments; RIXS shares chemical selectivity with first order spectroscopic techniques but in contrast to them can furnish improved resolution better than the core-hole lifetime while enhancing the signal from electronic excitations through resonant and spectral sharpening effects; momentum conservation provides a means to study the dispersion of these excitations while spin conservation gives a handle to local magnetic properties; finally in the non resonant mode, XRS offers the opportunity to measure the K edges of light elements with a high energy probe.  These features can be fully exploited in the new generation of x-rays sources which are tunable, extremely brilliant and highly focused, down to the micron level, well within typical sample sizes in pressure cells.

The versatility of the IXS technique allows one to address a rich variety of physico-chemical phenomena in materials under pressure. RIXS for instance has been applied to various strongly correlated $d$-electron compounds and Kondo-like $f$-electron systems while XRS is well-suited to the study of light elements in materials such as graphite or water. As it turns out, the behavior of compressed matter, especially in the presence of strong electronic correlations, is far more complex than that expected from a simplistic picture of electron delocalization. Indeed, spectroscopic results reveal unusual behavior in the electronic degrees of freedom brought up by increased density under pressure, changes in the charge-carrier concentration, overlapping between orbitals and hybridization. Many of these have been discussed in this review: Magnetic collapse and the metal-insulator transition in transition-metal oxides that are coupled to strong magnetovolumic effects especially important for their geophysical implications; mixed valent behavior and Kondo screening of the magnetic moment in compounds with narrow $f$-electron bands as a result of their interaction with the conduction electrons; and finally, change of coordination, local structure and chemical bonding with pressure in covalently bonded or hydrogen bonded compounds. 

Expanding these investigations to still high pressure, extremely low or high temperatures or high magnetic field, is the next step. In the following, we suggest one line of research which could profit from such an improvement.

\subsection{Quantum critical points}
\label{sec:QCP}
\begin{figure}[htb]
\includegraphics[width=0.80\linewidth]{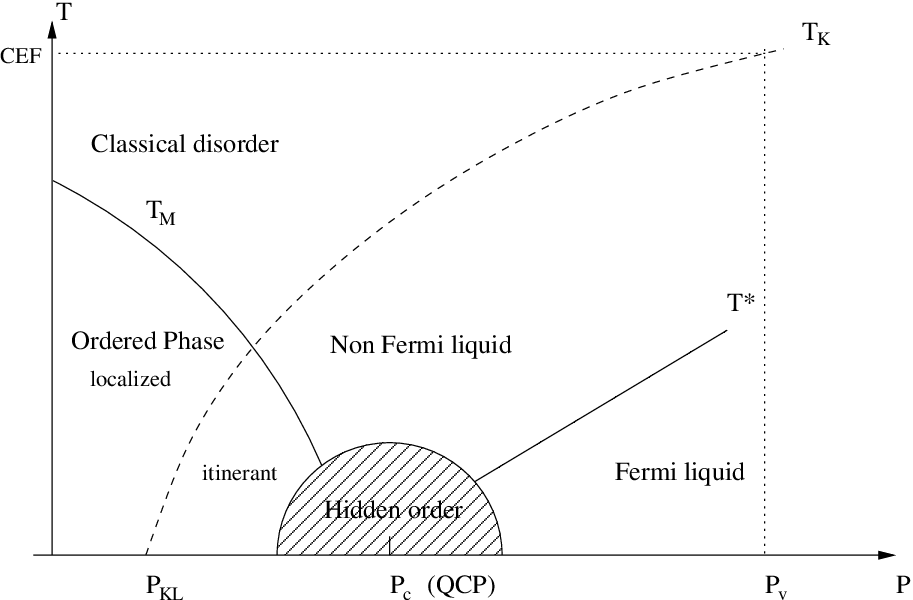}
\caption[QCP]{Phase diagram in the vicinity of a quantum critical point.}
\label{fig:QCP}
\end{figure}

The discovery of non-conventional superconductivity close to a quantum critical point (QCP) and the deviation from Fermi liquid behavior are among the challenging features observed in heavy fermions. A generalized phase diagram around the QCP is illustrated in Fig.~\ref{fig:QCP} as a function of a non-thermal control parameter $P$. The latter has no unequivocal meaning but in the present context can be associated to pressure. In this framework, one can derive three relevant pressures which characterize the $f$ magnetism and hybridized state following the definitions of~\textcite{Flouquet2005a}: $P_{KL}$ denotes the onset of itinerant magnetism, $P_c$ (the critical pressure) marks the disappearance of long range magnetism and the onset of the Fermi liquid behavior, while at higher pressure $P_v$ indicates the regime where the angular momentum $J$ is quenched by the Kondo coupling. At low temperature, the system undergoes a transition from a classically ordered state to a quantum disordered phase where the electrons behaves as a Fermi liquid below a characteristic temperature $T^*$. The two regions are separated by the QCP, a singularity marking the divergence at $P_c$ of the quantum coherence length. It is believed that a novel ordered phase is reached as the systems approach $P_c$. Above this region, non Fermi-liquid behavior prevails~\cite{Custers2003}. 
Hints of quantum criticality have been found in many heavy fermion compounds. In particular, it seems that the vicinity of a QCP is fundamental for superconductivity in Ce and U compounds. This includes recently the discovery of superconductivity under pressure well within the AFM (CeIn$_3$~\cite{Mathur1998}) and FM (UGe$_2$~\cite{Saxena2000a}) phase domains. But the detailed knowledge of the electronic properties in the vicinity of the QCP is still hampered by experimental difficulties. Exploring the QCP phase diagram requires high-pressure measurements at low temperatures and thus mastering technical difficulties such as fine and stable remote control of a pressure cell inside a cryostat. Recent developments in pressure setups open up new perspectives for such studies \cite{Rueff2009}. Although temperatures of few K are beyond reach at the moment, the influence of the QCP that is expected to encompass a wide region of the phase-space could be investigated. 

\subsection{Theoretical developments}
Finally, from the theoretical point of view, the parameterized multiplet approach has shown its limits and insufficiencies for the description of the strongly correlated state, although it remains one of the most efficient calculating methods for spectroscopy. One step toward an improved scheme is to implement realistic density of states, computed from first principles, in a cluster model. Such a method has found a perfect testing ground in $1s2p$-RXES thanks to the wealth of information it provides, e.g. in cuprates~\cite{Shukla2006}. Alternatively, new ab-initio theoretical frameworks, such as dynamical mean field theory, have proven to be extremely useful for describing the Kondo state of $f$-electrons~\cite{Medici2005,Amadon2006}. These calculations can well reproduce quantities such as the XPS spectral function, and the hope is that they could equally well describe second order processes including RIXS. 

\section*{Acknowledgments}
The variety of results explored in this review, the complexity of the experimental techniques and the subsequent theoretical work is the ultimate product of a large scale scientific effort. We would like to warmly thank all colleagues whose work has been cited in this review, especially in various institutes worldwide C. Hague, J.-M. Mariot (LCPMR, Paris), G. Loupias, J. Badro, F. Guyot, G. Fiquet, M. d'Astuto, J.-C. Chervin (IMPMC, Paris), J.-P. Kappler (IPCMS, Strasbourg), S. Raymond, D. Braithwaite, J.-P. Sanchez (CEA, Grenoble), C. Dallera, L. Braicovich (Politecnico di Milano), M. Grioni (EPFL, Lausanne), H. K. Mao, R. Hemley, V. Struzhkin (Geophyiscal Lab., Washington D.C.), M. Abd-Elmeguid (K\"{o}ln University), M. Acet (Duisburg University), A. Mattila, K. H\"{a}m\"{a}l\"{a}inen (University of Helsinki), M. Taguchi (RIKEN), A. Yaresko (MPI Dresden), F. M. F. de Groot (Utrecht University) and A. Kotani (RIKEN), colleagues in synchrotron light sources, F. Sette, M. Krisch, G. Monaco, G. Vank\'{o}, S. Huotari, P. Glatzel, M.Hanfland, M. Mezouar (ESRF), F. Baudelet, J.-P. Iti\'{e} (SOLEIL), Y. Cai, H. Ishii, I. Jarrige (Spring-8) and G. Shen (APS), and G. Lander.  Finally, we would like to acknowledge C. C. Kao (NSLS) for his seminal intuition in successfully combining inelastic x-ray scattering with high pressure physics.   


\end{document}